\documentclass{emulateapj}
\usepackage{apjfonts}

\usepackage{graphicx}

\slugcomment{AJ in press}

\begin{document}
\title{Stability of Distant Satellites of the
Giant Planets in the Solar System}

\shorttitle{Dynamics of Distant Satellites}

\shortauthors{SHEN \& TREMAINE}

\author{Yue Shen\altaffilmark{1} and Scott Tremaine\altaffilmark{2}}
\affil{$^1$Princeton University Observatory, Princeton, NJ 08544;
\quad $^2$Institute for Advanced Study, Princeton, NJ 08540}

\begin{abstract}
We conduct a systematic survey of the regions in which distant
satellites can orbit stably around the four giant planets in the
solar system, using orbital integrations of up to $10^9$ yr. In
contrast to previous investigations, we use a grid of initial
conditions on a surface of section to explore phase space
uniformly inside and outside the planet's Hill sphere (radius
$r_{\rm H}$; satellites outside the Hill sphere sometimes are also
known as quasi-satellites). Our confirmations and extensions of
old results and new findings include the following: (i) many
prograde and retrograde satellites can survive out to radii $\sim
0.5r_{\rm H}$ and $\sim 0.7r_{\rm H}$, respectively, while some
coplanar retrograde satellites of Jupiter and Neptune can survive
out to $\sim r_{\rm H}$; (ii) stable orbits do not exist within
the Hill sphere at high ecliptic inclinations when the semi-major
axis is large enough that the solar tide is the dominant
non-Keplerian perturbation; (iii) there is a gap between $\sim
r_{\rm H}$ and $2r_{\rm H}$ in which no stable orbits exist; (iv)
at distances $\gtrsim 2r_{\rm H}$ stable satellite orbits exist
around Jupiter, Uranus and Neptune (but not Saturn). For Uranus
and Neptune, in particular, stable orbits are found at distances
as large as $\sim 10r_{\rm H}$; (v) the differences in the stable
zones beyond the Hill sphere arise mainly from differences in the
planet/Sun mass ratio and perturbations from other planets; in
particular, the absence of stable satellites around Saturn is
mainly due to perturbations from Jupiter.  It is therefore likely
that satellites at distances $\gtrsim 2r_{\rm H}$ could survive
for the lifetime of the solar system around Uranus, Neptune, and
perhaps Jupiter.
\end{abstract}

\section{Introduction}
\label{sec:intro}

Most of the satellites of the four giant planets in the solar
system can be divided into two groups, usually called the regular
and irregular satellites. Regular satellites orbit close to the
planet (within $\sim 0.05r_{\rm H}$, where $r_{\rm H}$ is the Hill
radius\footnote{The Hill radius is defined as $r_{\rm
H}=a_p(\mu/3)^{1/3}$, where $a_p$ is the semi-major axis of the
planet orbit and $\mu\equiv m_p/(m_p+M_{\odot})$ with $m_p$ the
planet mass.}), and move on nearly circular, prograde orbits that
lie close to the planetary equator. Irregular satellites are found
at distances $\sim 0.05r_{\rm H}-0.6r_{\rm H}$, with large orbital
eccentricities and inclinations, on both prograde and retrograde
orbits. An alternative division between regular and irregular
satellites is given by the critical semi-major axis (e.g.,
Goldreich 1966; Burns 1986), $a_{\rm crit}=(2\mu
J_2R^2a_p^3)^{1/5}$; those with $a>a_{\rm crit}$ are classified as
irregular satellites. Here $J_2$ is the planet's second zonal
harmonic coefficient (augmented by any contribution from the inner
regular satellites) and $R$ is the planet's radius. This critical
radius marks the location where the precession of the satellite's
orbital plane is dominated by the Sun rather than by the planet's
oblateness. The current number ratios of irregular to regular
satellites are $55/8$ for Jupiter, $35/21$ for Saturn, $9/18$ for
Uranus, and $7/6$ for Neptune (e.g., Jewitt \& Haghighipour 2007).
The regular satellites are likely to have formed within a
circumplanetary disk of gas and solid bodies. The kinematic
differences between regular and
irregular satellites suggest that the latter must have formed
through a quite different mechanism, most likely capture from the
circumstellar disk (for a recent review, see Jewitt \&
Haghighipour 2007).

The search for irregular satellites of the giant planets has been
fruitful in recent years, owing mainly to modern high-sensitivity,
large-scale CCDs (e.g., Gladman et al.\ 1998, 2000, 2001; Holman et
al.\ 2004; Kavelaars et al.\ 2004; Sheppard \& Jewitt 2003; Sheppard et
al.\ 2005, 2006). Up to 2007, 106 irregular satellites of the giant
planets had been discovered, compared to 53 regular satellites. Two
features stand out in the distributions of orbital parameters of these
irregular satellites. First, retrograde irregular satellites extend to
larger semi-major axes than prograde ones ($\sim 0.6r_{\rm H}$
compared to $\sim 0.4r_{\rm H}$); second, satellites with orbital
inclination in the range $\sim 60^\circ-130^\circ$ relative to the ecliptic
are absent.

A number of authors have shown that these features can be
explained reasonably well by the requirement that the satellite
orbits be stable. H\'{e}non (1969, 1970) studied the planar
circular restricted three-body problem in Hill's (1886)
approximation, where the mass ratio $\mu\rightarrow 0$ while the
radii of interest shrink to zero as $\mu^{1/3}$. He showed that
prograde satellite orbits are stable up to a mean distance from
the planet $\sim 0.4r_{\rm H}$, while retrograde satellite orbits
can be stable at much larger distances from the planet. Thus it is
not surprising that retrograde satellites are found at larger
distances than prograde ones. Hamilton \& Krivov (1997) studied
the dynamics of distant satellites of asteroids in heliocentric
orbits using a ``generalized Tisserand constant'' and, among other
conclusions, confirmed that retrograde orbits are more stable than
prograde ones. Carruba et al.\ (2002) used a combination of
analytic arguments and numerical integrations to show that
high-inclination orbits inside the Hill sphere exhibit large eccentricity oscillations
(Kozai oscillations; Kozai 1962) due to secular solar
perturbations. They found that orbits with inclinations (relative
to the planetary orbital plane) between $55^\circ$ and $130^\circ$
are generally unstable, thus explaining the absence of irregular
satellites on high-inclination orbits. Nesvorn\'{y} et al.\ (2003)
performed detailed orbital integrations of the four giant planets
plus a grid of test-particle satellites for intervals of
$10^6$--$10^8$ yr. They confirmed that retrograde satellites can
be stable at larger radii than prograde ones, and that highly
inclined orbits are unstable. They argued that the largest
semi-major axes at which satellites of the four giant planets
could survive for times comparable to the lifetime of the solar
system were $\sim 0.7r_{\rm H}$ for retrograde satellites and
$\sim 0.4r_{\rm H}$ for prograde ones, and that these upper limits
were achieved only for nearly circular orbits close to the plane
of the ecliptic.

Other authors have examined the possibility that stable satellite
orbits exist with mean distance from the planet $\gtrsim r_{\rm
H}$. In the planetocentric frame, the dominant force on such
satellites is due to the Sun, rather than the
planet\footnote{Hence these are sometimes called
``quasi-satellites'' (Lidov \& Vashkov'yak 1994a,b; Mikkola \&
Innanen 1997).}. Nevertheless, the satellite remains close to the
planet because it is in a 1:1 resonance in the sense that its
heliocentric mean longitude librates around that of the planet;
the resulting orbit relative to the planet is a retrograde ellipse
with axis ratio 2:1, the short axis pointing towards the Sun, and
synodic period equal to the planet's orbital period. The
analytical theory of such orbits is described by Jackson (1913),
Lidov \& Vashkov'yak (1994a,b), Mikkola \& Innanen (1997), Namouni
(1999), Mikkola et al.\ (2006), and others. H\'{e}non's (1970)
numerical analysis of Hill's approximation to the planar circular
restricted three-body problem suggests that stable retrograde
satellites can exist at arbitrarily large distance from the
planet. Benest (1971) confirmed that stable retrograde orbits at
large distances persist in the elliptic restricted three-body
problem, where the mass ratio and eccentricity were chosen to
match those of Jupiter. Wiegert et al.\ (2000) demonstrated that
retrograde satellites of Uranus and Neptune could be stable for up
to $10^9$ yr at distances up to $\sim 10r_{\rm H}$, suggesting
that primordial objects of this type could still exist in the
solar system although none are currently known.

Despite the number and quality of these investigations, there are several
unanswered questions that lead us to revisit the problem of
orbital stability of satellites at large distances from the host
planet. (i) Wiegert et al.\ (2000) found stable satellite orbits
beyond the Hill radius only for Uranus and Neptune, not Jupiter or
Saturn. What is the reason for this difference? The possibilities
include differences in the planetary masses and orbital
eccentricities, or different perturbations from neighboring planets. (ii)
Wiegert et al.\ (2000) explored orbits outside the Hill radius, while
Nesvorn\'y et al.\ (2003) explored orbits inside the Hill radius
(indeed, in the former paper the integrations were terminated when the
particles entered the Hill sphere of radius $r_{\rm H}$ around the
planet, while in the latter paper the integrations were terminated
when the particles exited the Hill sphere). Are there stable satellite
orbits that cross the Hill sphere? (iii) As we shall describe further
in \S\ref{sec:method}, the grids of initial conditions used by
Nesvorn\'y et al.\ and Wiegert et al.\ do not provide a complete
exploration of the phase space in which stable satellite orbits
exist.

The primary goal of this paper is to map out the entire stability
region in phase space---both inside and outside the Hill sphere---in
which satellite orbits that can survive around the four giant planets
for times comparable to the age of the solar system (our main
integrations last for up to 100 Myr). We describe our setup in
\S\ref{sec:method}, and present the results in \S\ref{sec:result}. We
conclude and discuss our results in \S\ref{sec:con}.

Following Fabrycky (2008), we shall define a ``satellite'' of a
planet to be a small body whose distance from the planet never
exceeds the semi-major axis of the planet, $a_p$. This definition
excludes bodies on Trojan orbits around the triangular Lagrange
points, bodies on horseshoe orbits, and objects such as asteroid
2003 YN107 (Connors et al.\ 2004), which oscillates between a
horseshoe orbit and an orbit centered on Earth. This definition
seems simple and reasonable to us, but other definitions are
common in the literature. Many authors define
``satellite'' to be an object that always remains within the Hill
sphere of the planet or whose Jacobi constant constrains it to
remain within the last closed zero-velocity surface around the
planet. Benest (1971) defines a satellite to be a body whose
heliocentric orbital frequency is the same as the planet's, but
whose synodic frequency around the planet is non-zero. Wiegert et
al.\ use the term ``quasi-satellite'' for an object that remains
outside the Hill sphere but whose heliocentric longitude
difference from the planet never exceeds $120^\circ$ and regularly
passes through zero. However, the term ``quasi-satellite'' is
confusing because it is also
used for objects such as 2003 YN107 that spend part of their time on
horseshoe orbits and thus are only temporarily satellites in our sense.

\section{Methods}
\label{sec:method}

Although all of our results are based on direct numerical integrations
of the N-body problem (Sun, one or four giant planets, plus a test
particle orbiting one planet), we shall find it useful to interpret
our results in terms of the coordinates and notation used by H\'{e}non
(1970) in the exploration of satellite orbits in Hill's approximation.

\subsection{Hill's approximation}\label{subsec:Hill_equation}
When studying satellite motions near a planet ($r\ll r_{\rm H}$)
it is conventional to employ a {\em non-rotating} planetocentric
coordinate system, which we denote as $(xyz)$. However, in Hill's
approximation to the circular restricted three-body problem, it is
more convenient to use a {\em rotating} planetocentric coordinate
system $(\xi\eta\zeta)$, where $\xi$, $\eta$ and $\zeta$ are
scaled coordinates in the rotating frame in which the planet is at
the origin, the $\xi$ axis is along the direction opposite the Sun
and the $\zeta$ axis is perpendicular to the Sun-planet orbital
plane. In Hill's formulation the unit of length is $\mu^{1/3}a_p$,
and the unit of time is $n^{-1}$ where $n\equiv
[G(M_\odot+m_p)/a_p^3]^{1/2}$ is the mean motion of the planet. As
usual, the orbit of the planet in the inertial frame is
counter-clockwise as viewed from the positive $z$ or $\zeta$ axis.
In Hill's coordinate system the collinear Lagrangian points $L_1$
and $L_2$ are located at $\eta=\zeta=0$, $\xi=\pm 3^{-1/3}\simeq
0.6934$, and the Hill radius is $r_{\rm H}=3^{-1/3}$. Similar
definitions are used in this paper when the planet orbit is
eccentric and/or perturbed by other planets; in this case the
$\xi$ axis points away from the instantaneous position of the Sun,
the $\zeta$ axis is perpendicular to the instantaneous orbital
plane of the planet around the Sun, and $a_p$ is the initial
semi-major axis of the planet.

In the circular restricted three-body problem, Hill's approximation
is achieved by taking the limit $\mu\rightarrow 0$, where the
equations of motion reduce to (e.g., H\'{e}non 1974; Murray \&
Dermott 1999):
\begin{eqnarray}
&&
\ddot{\xi}=2\dot{\eta}+3\xi-\frac{\xi}{(\xi^2+\eta^2+\zeta^2)^{3/2}}\
,\ \
\ddot{\eta}=-2\dot{\xi}-\frac{\eta}{(\xi^2+\eta^2+\zeta^2)^{3/2}}\ ,\nonumber\\
&&\qquad\qquad\qquad\ddot{\zeta}=-\zeta-\frac{\zeta}{(\xi^2+\eta^2+\zeta^2)^{3/2}}\
.
\end{eqnarray}

There exists an integral of motion for these equations,
\begin{equation}
\Gamma=3\xi^2+\frac{2}{(\xi^2+\eta^2+\zeta^2)^{1/2}}-\zeta^2-(\dot{\xi}^2+
\dot{\eta}^2+\dot{\zeta}^2)\ ,
\end{equation}
which corresponds to the Jacobi constant in the circular
restricted three-body problem.

\begin{figure*}
\centering
\includegraphics[width=0.45\textwidth]{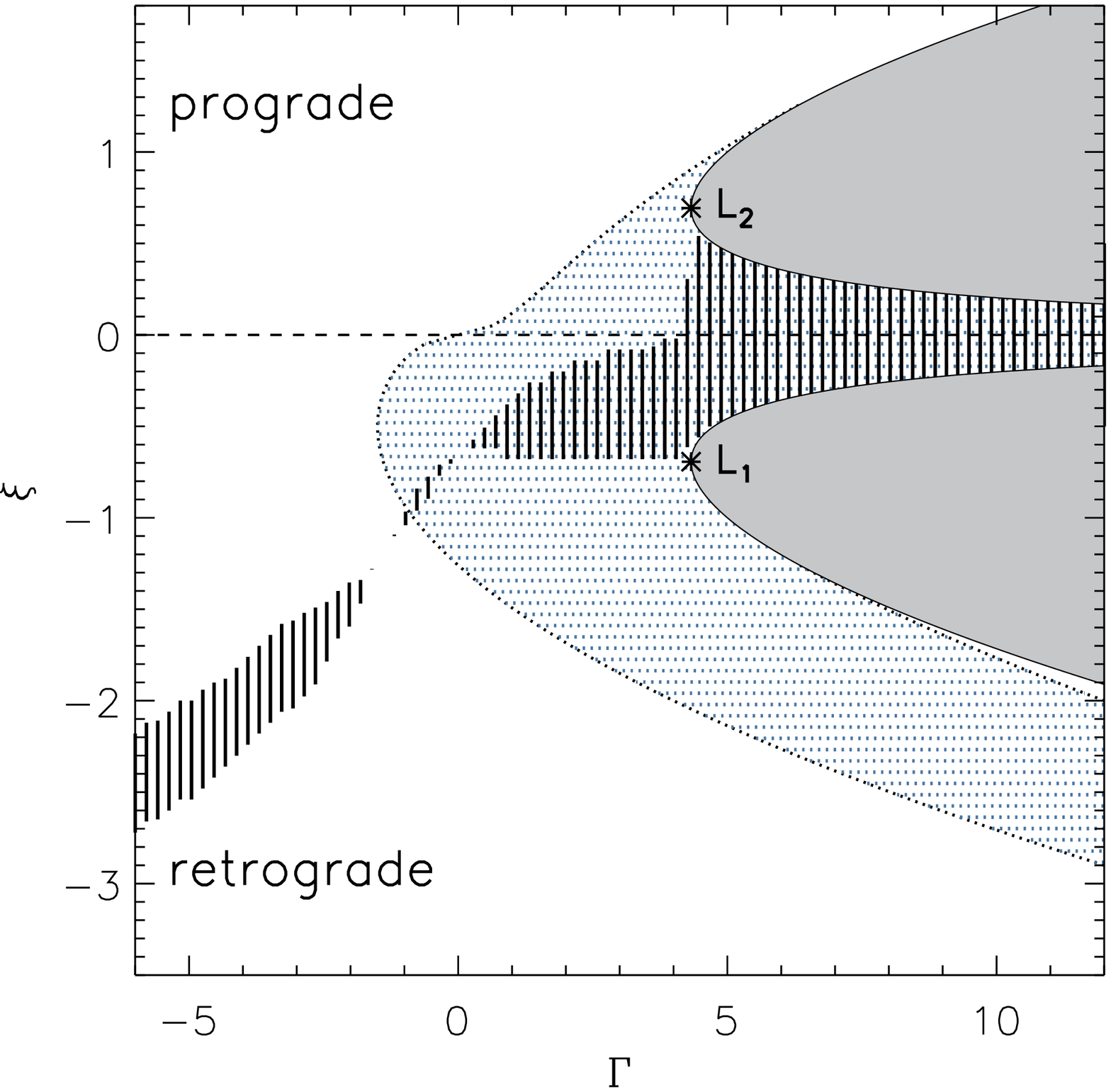}
\includegraphics[width=0.45\textwidth]{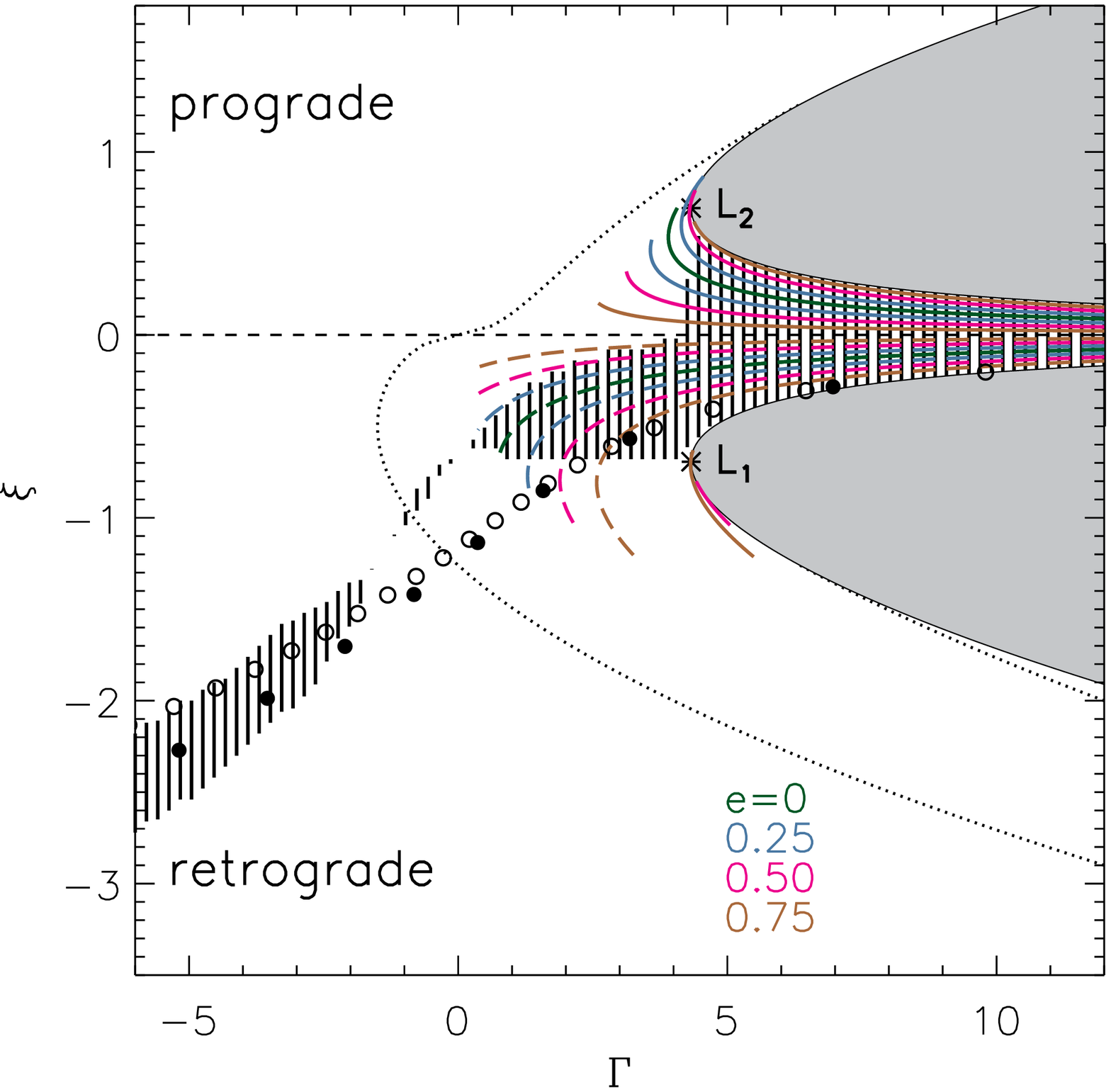}
\caption{Sampling of initial conditions in terms of the H\'{e}non
diagram. The gray-shaded regions are forbidden.  The Lagrange points
$L_1$ and $L_2$ are marked by $*$. The horizontal dashed line at
$\xi=0$ separates retrograde orbits from prograde ones (as defined in
the rotating planetocentric frame). {\em Left}: The region shaded by
vertical lines is an approximate reproduction of the stable region as
estimated in H\'{e}non (1970). The dotted region is where the
osculating Kepler elements correspond to bound elliptical orbits
($a>0$, $e<1$). {\em Right}: Curved lines represent the initial
conditions derived using osculating elements in the high-resolution
survey of Nesvorn\'{y} et al.\ (2003), color-coded according to
eccentricity. Solid and long-dashed lines represent orbits which are
prograde and retrograde in the {\em non-rotating} planetocentric frame
respectively. There are two sets of lines for each eccentricity
corresponding to argument of pericenter $\omega=0^\circ$ and
$180^\circ$ respectively. The short segments at the lower right below $L_1$ are
extensions of the $e=0.50$ and $0.75$ branches which are prograde in
the non-rotating frame.  Thus orbits that are retrograde in the
rotating frame can be prograde in the non-rotating frame. The initial
conditions for zero-inclination orbits sampled by Wiegert et al.\
(2000) are shown as filled circles for Uranus and open circles for
Jupiter.}\label{fig:check_Henon}
\end{figure*}

For the moment let us restrict ourselves to motion in the Sun-planet
orbital plane, so $\zeta=\dot\zeta=0$ at all times. Then to study
orbital motions we may use a surface of section defined by
$\eta=0$, $\dot\eta>0$. The trajectory in the four-dimensional
$(\xi,\eta,\dot\xi,\dot\eta)$ phase space is then represented by
a set of points in the $(\xi,\dot\xi)$ plane, and for a given
value of the Jacobi constant $\Gamma$ the other two phase-space coordinates
can be derived from
\begin{equation}
\eta=0,\qquad
\dot\eta=\left(3\xi^2-\dot\xi^2+{2\over|\xi|}-\Gamma\right)^{1/2}.
\end{equation}
We define ``prograde'' and ``retrograde'' in the rotating frame
unless otherwise noted. Thus retrograde orbits have $\xi<0$ in
this surface of section and prograde orbits have $\xi>0$.

A drawback of this surface of section is that a different plot is
needed for each value of the Jacobi constant $\Gamma$. To obtain a
global view of the dynamics, we use a different surface of section
defined by $\eta=0$, $\dot\xi=0$,
$\dot\eta=(3\xi^2+2/|\xi|-\Gamma)^{1/2}$. A trajectory is
represented by a point in the $(\Gamma,\xi)$ plane. This surface
of section was introduced by H\'{e}non (1969; 1970), and we shall
call it the H\'enon surface of section or H\'enon diagram. The
H\'enon diagram, like any surface of section, will not show orbits
that do not cross it; the usefulness of the H\'enon diagram
derives from the observation that most stable orbits periodically
pass close to the point $\eta=\dot\xi=0$---for example, this
occurs for nearly Keplerian orbits close to the planet when their
line of apsides precesses past the Sun-planet line. The orbits
not shown on the H\'enon diagram include those confined to some
resonant islands, which should occupy a small
fraction of phase space, and escape orbits, which we are not
interested in anyway.

Fig. \ref{fig:check_Henon} ({\em left panel}) is a H\'enon diagram
modeled on Figure 12 of H\'enon (1970). The Lagrange points are at
$(\Gamma,\xi)=(3^{4/3},\pm 3^{-1/3})=(4.32675,\pm0.69336)$.
Forbidden regions, in which $\dot\eta^2=3\xi^2+2/|\xi|-\Gamma$
would be negative, are shaded in gray. The stable regions of phase
space, as estimated by H\'enon, are denoted by vertical stripes.
The diagram shows that retrograde satellites ($\xi<0$) have a
larger stable region than prograde satellites ($\xi>0$), a
conclusion consistent with the numerical studies described in
\S\ref{sec:intro}. Moreover, the stable band in this diagram that
begins at $(\Gamma,\xi)=(-1.4,-1.2)$ and stretches downward to the
left shows that retrograde satellites can be stable at distances
much larger than the Hill radius; in fact, this band continues to
arbitrarily large negative values of $\Gamma$ and $\xi$ (see
Figure 13 of H\'enon 1970), so retrograde satellites can be stable
at arbitrarily large distances from the planet, at least in Hill's
approximation to the planar circular restricted three-body
problem.

In future discussions we divide the stable regions in Fig.\
\ref{fig:check_Henon} into three branches: the inner prograde
branch ($\xi>0$), the inner retrograde branch ($\xi<0$ and
$\Gamma>0$), and the outer retrograde branch ($\xi<0$ and
$\Gamma<0$).

A simple and rather complete way to sample initial conditions in the
planar three-body problem is to use the H\'{e}non diagram, i.e., to
sample uniformly in the $(\Gamma,\xi)$ plane. As described above, this
approach is based on the assumption that most stable orbits
periodically have their apocenter or pericenter on the Sun-planet
line. Note that even without invoking Hill's approximation the
question of which initial conditions on the H\'enon diagram correspond
to stable orbits is well-posed. Accordingly, we may present our
stability results in terms of the H\'{e}non diagrams, even though our
orbit integrations do not use Hill's approximation.

We may compare this approach to the grids of initial conditions used
in other investigations of the stability of satellite orbits.  The
initial conditions for Nesvorn\'{y} et al.'s ``high-resolution
survey'' were chosen from a grid of planet-centered osculating
Keplerian orbital elements, with semi-major axis $a$ given typically
by $a/r_{\rm H}=0.1$--$1$, eccentricity $e=0$--$0.75$, inclination
$i=0^\circ$--$180^\circ$, argument of pericenter
$\omega=0^\circ,90^\circ$, and the other elements distributed
uniformly between $0^\circ$ and $360^\circ$.  The right panel of Fig.\
\ref{fig:check_Henon} shows similar initial conditions on the H\'enon
diagram ($i=0^\circ$ or $180^\circ$ and $\omega=0^\circ$ or
$180^\circ$).  The conversions from osculating elements to
$(\Gamma,\xi)$ were done using equations (8) and (10) in H\'{e}non
(1970). It is clear that the initial conditions sampled in
Nesvorn\'{y} et al.\ do not provide a complete exploration of the
phase space in which stable satellite orbits could exist; in
particular, they completely missed the stable region that extends
beyond the Hill sphere (of course, such orbits are also excluded from
their study by their artificially imposed escape criterion $r>r_{\rm
H}$). In fact, most of the stable orbits beyond the Hill sphere have
hyperbolic osculating elements. In Fig.~\ref{fig:check_Henon} we plot
the boundaries that separate regions of hyperbolic osculating elements
from those with elliptical osculating elements, where the latter are
shaded by a dotted pattern. The functional forms of these boundaries
are:
\begin{eqnarray}
\Gamma&=&2\xi^2-2^{3/2}|\xi|^{1/2}\ ,\qquad (\xi<0)\ ,\\
\Gamma&=&2\xi^2+2^{3/2}|\xi|^{1/2}\ ,\qquad (0\le\xi\le 2^{1/3})\
,\\
\Gamma&=&2\xi^2+2^{3/2}|\xi|^{1/2}\ ,\qquad (\xi<-2^{1/3})\ .
\end{eqnarray}

The initial conditions explored by Wiegert et al.\ (2000) were
chosen from a grid of heliocentric osculating Keplerian elements,
these being the same as the elements of the host planets except
for the eccentricity and inclination. The eccentricity was
typically chosen in the range $e=0$--$0.5$ and inclination in the
range $0^\circ$--$30^\circ$. With this procedure, zero-inclination orbits
appear in the H\'enon diagram along the locus
\begin{equation}
   \Gamma=2/|\xi|-\xi^2 \qquad\hbox{with}\qquad \xi=-e\mu^{-1/3}\
   ,
\end{equation}
where the expression for $\Gamma$ is evaluated using Hill's
approximation. The grid sampled by Wiegert et al.\ for $i=0$ is
also shown in the right panel of Fig.\ \ref{fig:check_Henon},
converted from the heliocentric frame using Hill's units (but
without Hill's approximation); for clarity, only Jupiter and Uranus
are shown. Although Wiegert et al.'s initial conditions do probe the
stability region found by H\'enon beyond the Hill sphere,
the coverage is far from complete.

\begin{figure*}
\centering
\includegraphics[width=0.45\textwidth]{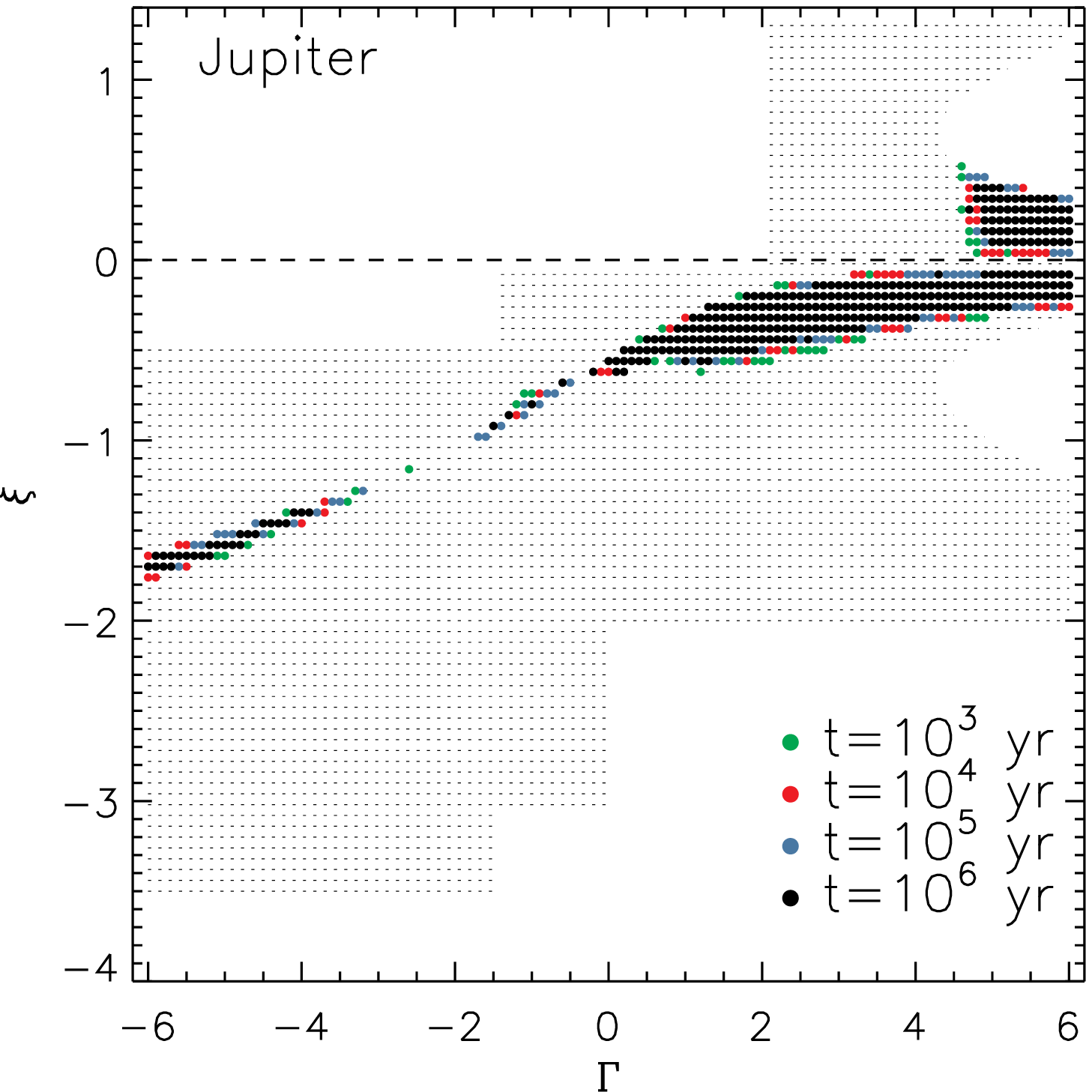}
\includegraphics[width=0.45\textwidth]{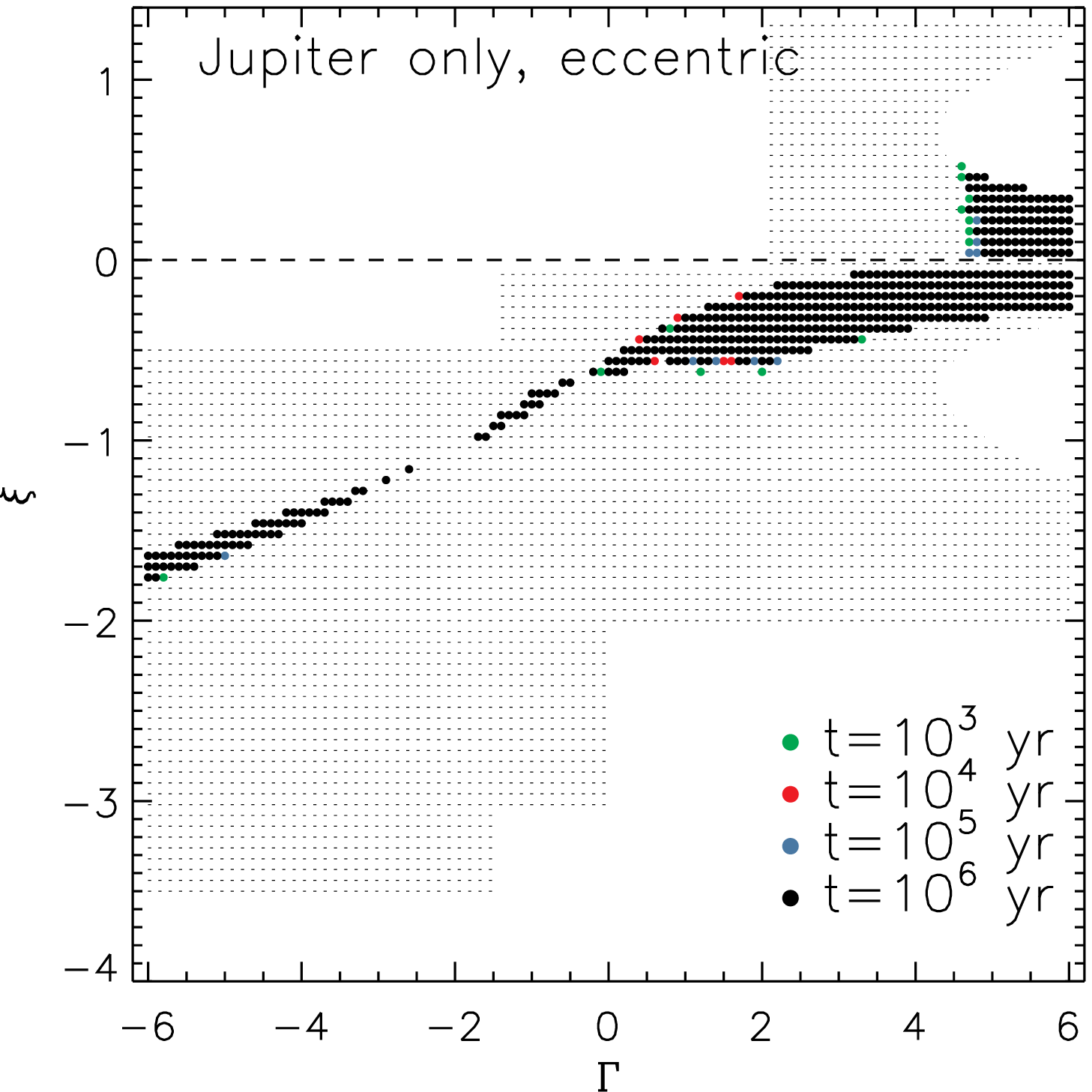}
\includegraphics[width=0.45\textwidth]{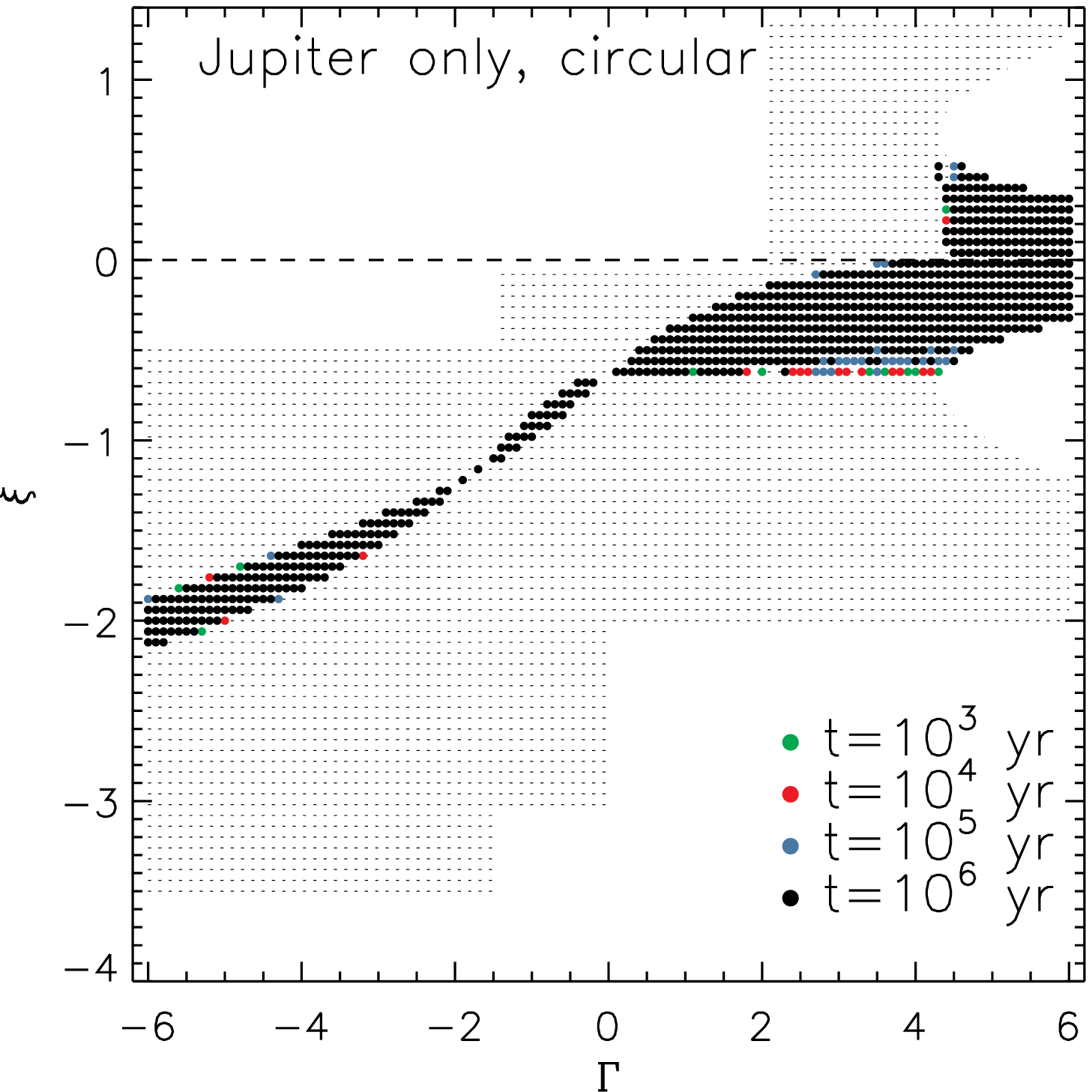}
\includegraphics[width=0.45\textwidth]{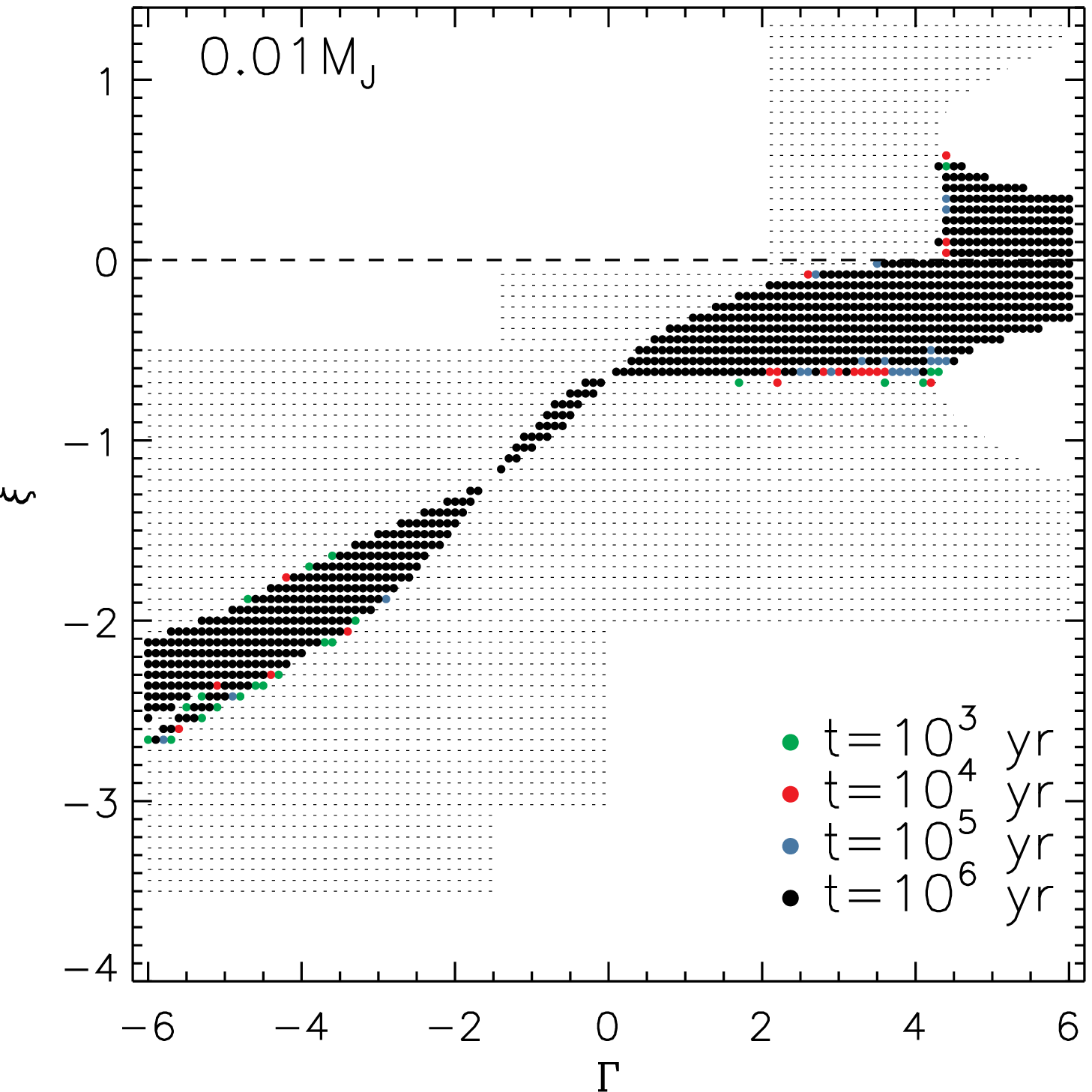}
\caption{H\'{e}non diagrams for Jupiter under various
circumstances (see \S\ref{sec:2D_result} for details). The
satellite orbital plane initially coincides with Jupiter's. In
each panel, the dotted region is the grid of initial conditions in
the $(\Gamma,\xi)$ plane. The two blank regions in the upper-right
corner are forbidden. The filled circles in different colors
represent the initial conditions of orbits that survive for
various times. }\label{fig:jupiter_henon2d}
\end{figure*}

\begin{figure*}
\centering
\includegraphics[width=0.45\textwidth]{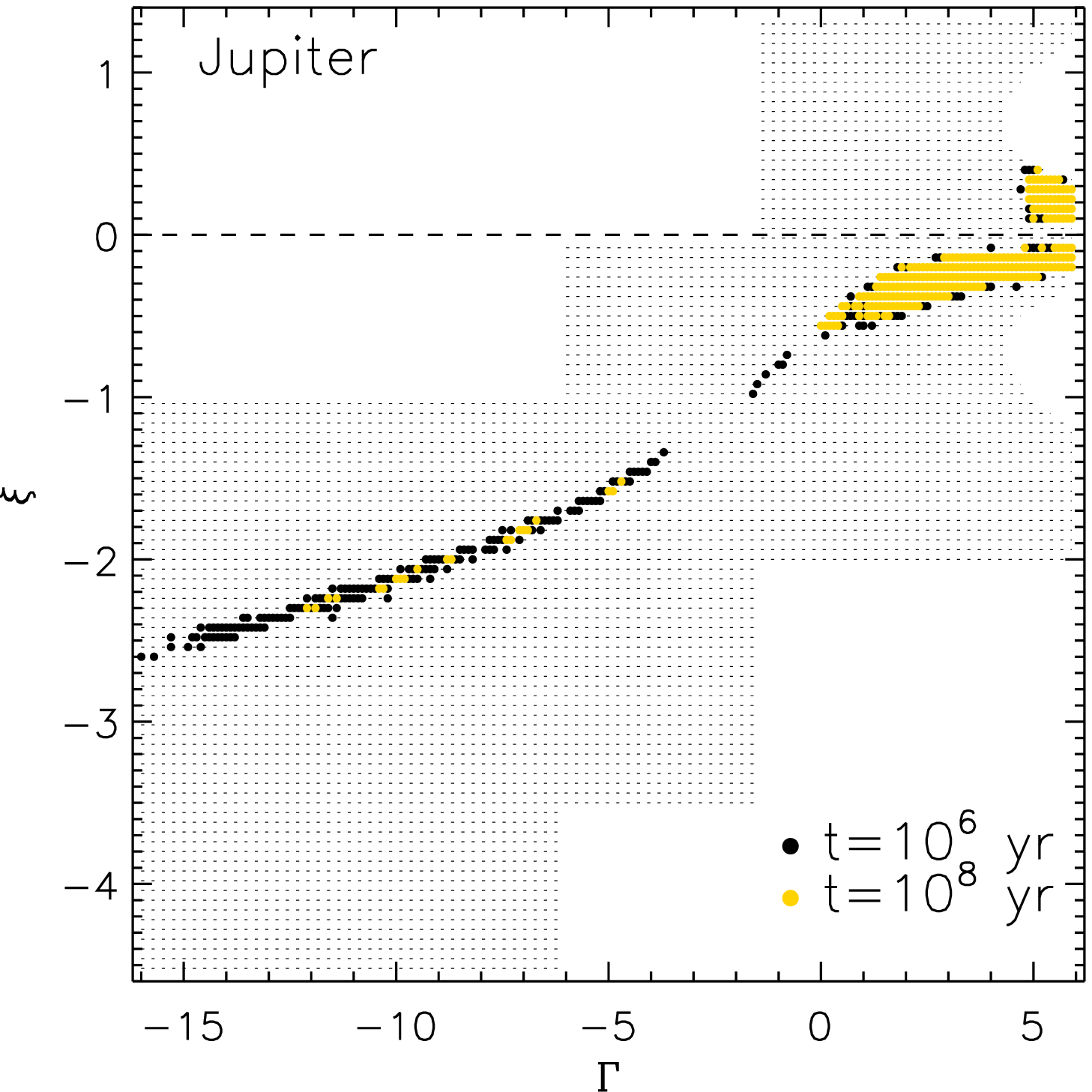}
\includegraphics[width=0.45\textwidth]{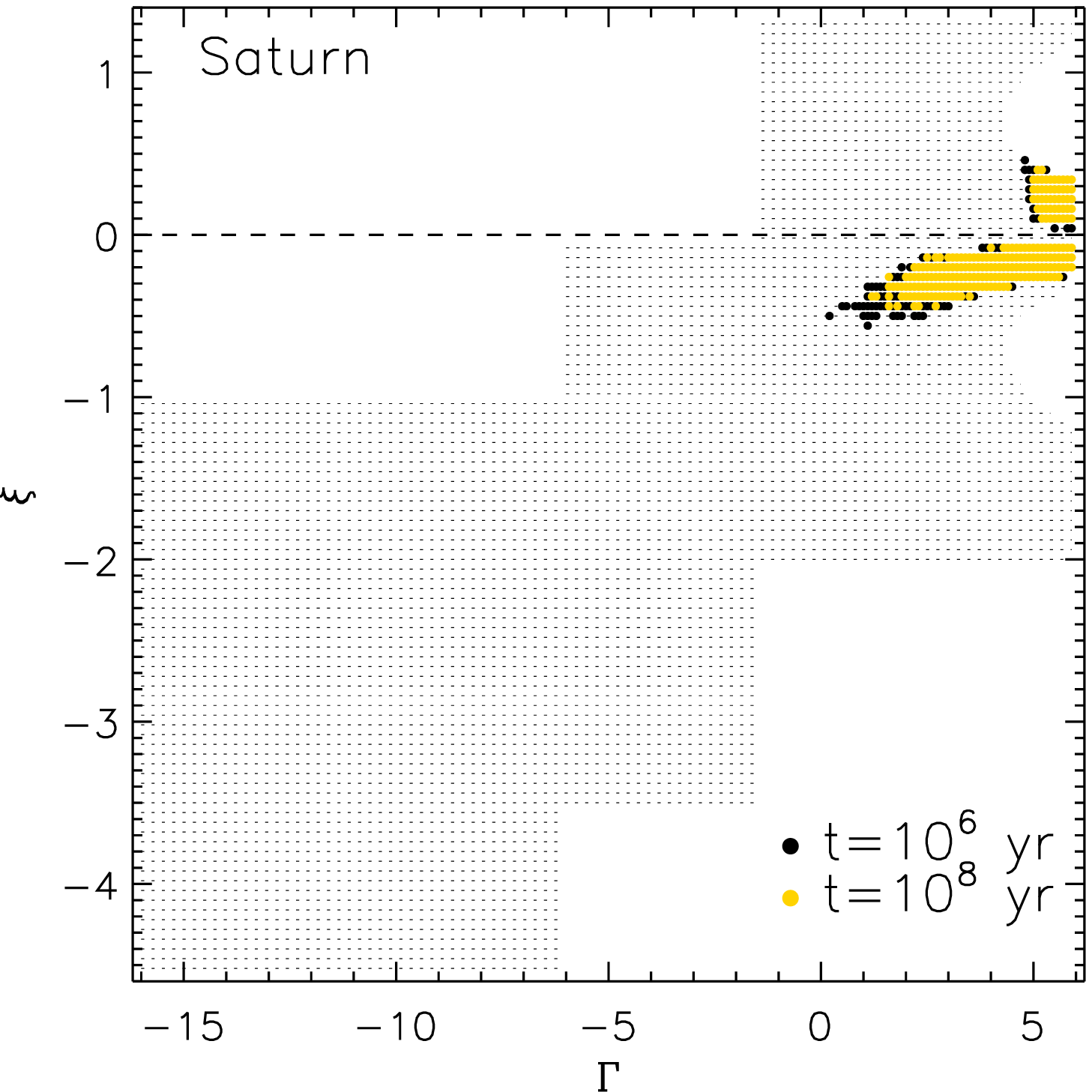}
\includegraphics[width=0.45\textwidth]{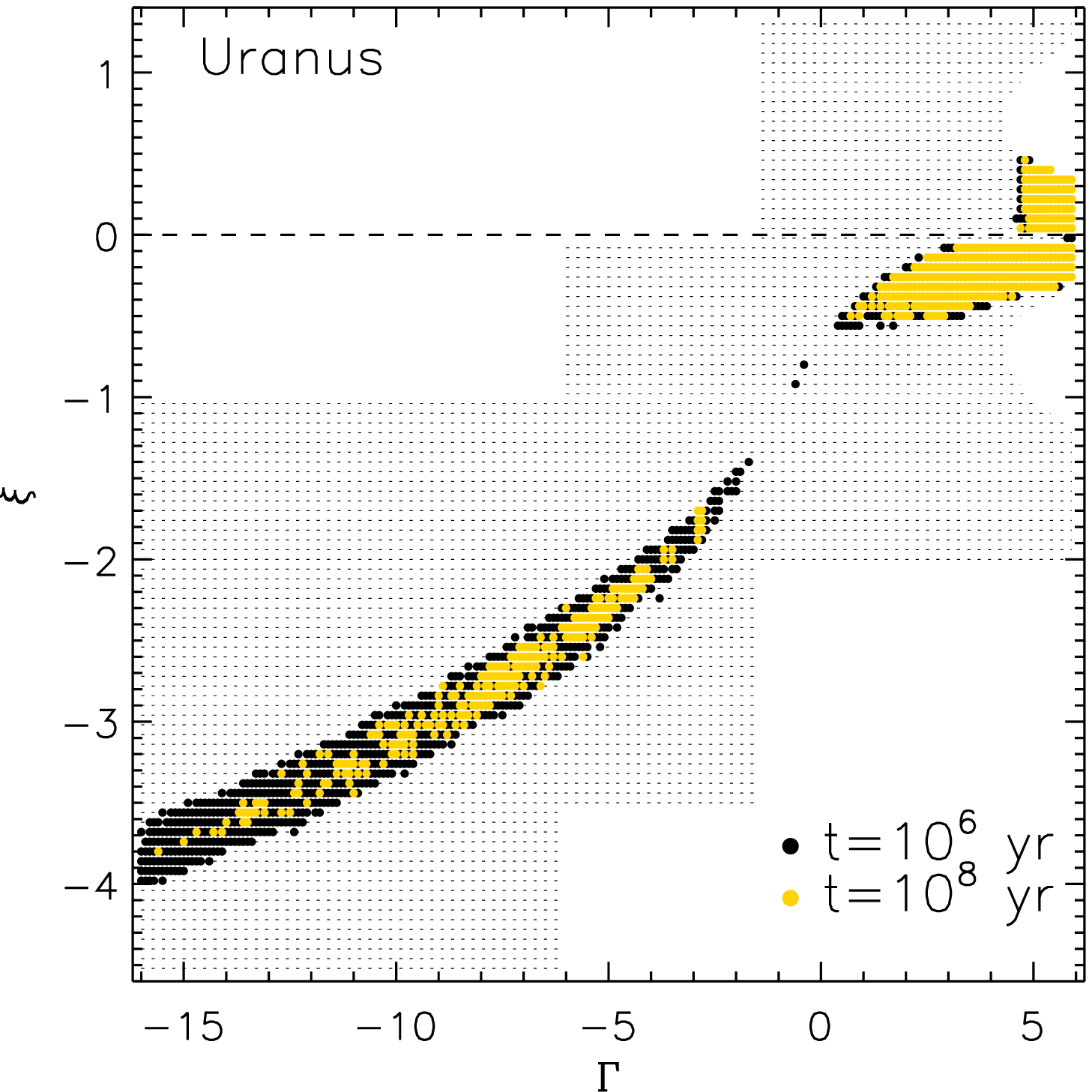}
\includegraphics[width=0.45\textwidth]{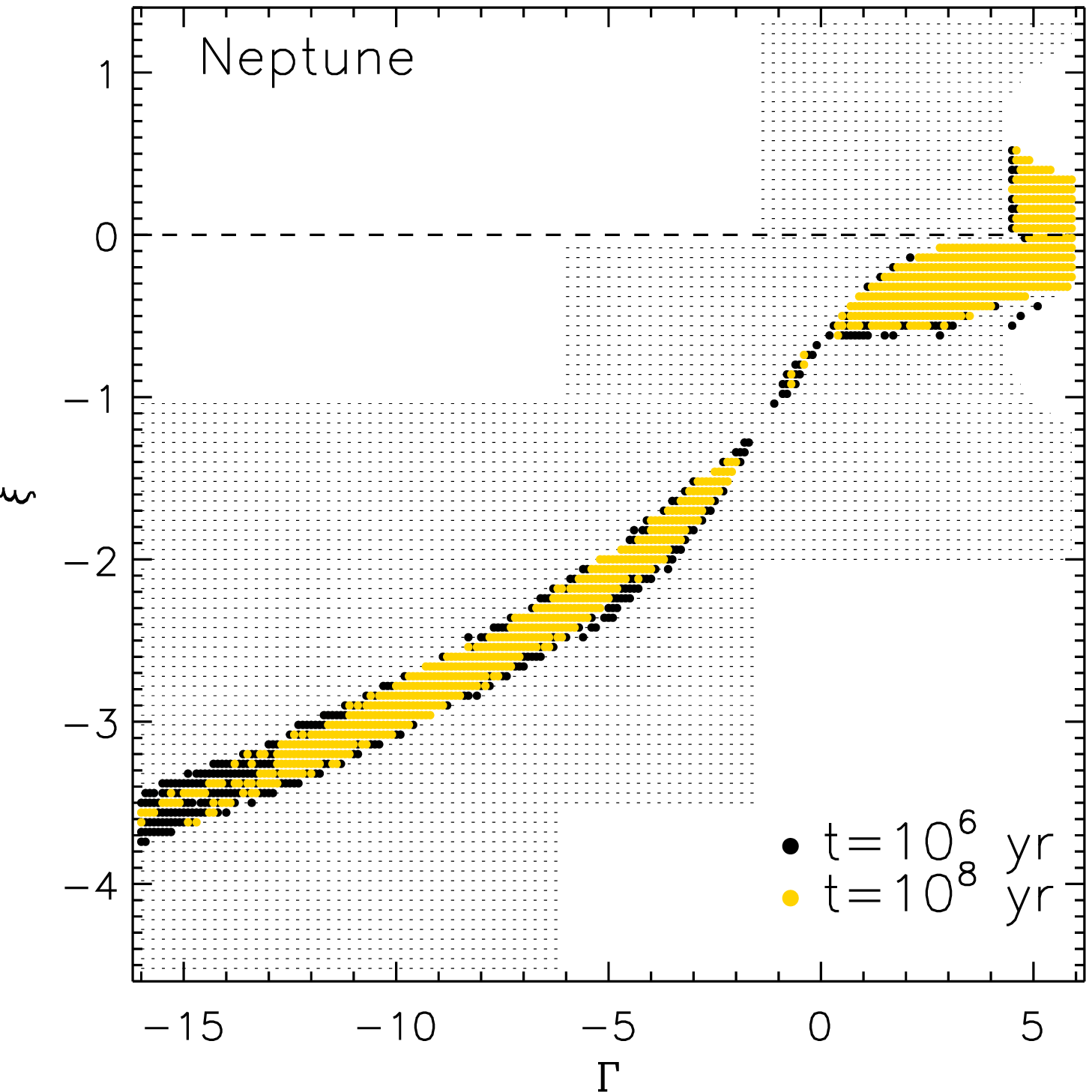}
\caption{Two-dimensional H\'{e}non diagrams for the four planets.
Each orbit is integrated up to $10^8$ yr under the gravitational
influence of its host planet, the Sun, and the other three giant
planets. Notations are the same as Fig. \ref{fig:jupiter_henon2d}.
}\label{fig:other_henon2d}
\end{figure*}

To extend our study to three-dimensional orbital motions
we use a surface of section at $\eta=\zeta=\dot{\xi}=0$, $\dot\eta
>0$. In the rotating frame, we define the initial
inclination angle $I$ by
\begin{equation}\label{eqn:inc}
\tan I=\frac{\dot{\zeta}}{\dot{\eta}}\bigg|_{t=0}\ ,
\end{equation}
such that the initial $\eta$ and $\zeta$ component velocities are
\begin{equation}\label{eqn:3d_vel}
\dot{\eta}=\cos I\bigg(3\xi^2+\frac{2}{|\xi|}-\Gamma\bigg)^{1/2}\!\!,\quad
\dot{\zeta}=\sin I\bigg(3\xi^2+\frac{2}{|\xi|}-\Gamma\bigg)^{1/2}\!\!.
\end{equation}
Since $\dot\eta>0$, the inclination is restricted to the range
$-90^\circ<I<90^\circ$.

Therefore each point in the $(\Gamma,\xi)$ plane represents a
unique set of initial conditions for a given inclination. The
usefulness of the H\'{e}non diagram in this case is based on the
assumption that most stable orbits periodically have their
line of apsides and their line of nodes simultaneously on the Sun-planet
line. This assumption is not always valid: it requires that the
argument of pericenter $\omega$ is periodically 0 or $\pi$, while a
satellite trapped in the Kozai resonance has an argument of
pericenter that librates around $\frac{1}{2}\pi$ or
$\frac{3}{2}\pi$ (Kozai 1962; Carruba et al. 2002). We estimate
the incompleteness in our survey due to such orbits in
\S\ref{sec:3D_result}.

Because
the equations of motion are symmetric around the $\zeta=0$ plane,
we may further restrict the inclination to the range $0^\circ\le
I<90^\circ$. As in the two-dimensional case, we define
``prograde'' and ``retrograde'' in the rotating frame unless
otherwise noted. Thus retrograde orbits have $\xi<0$ and prograde
orbits have $\xi>0$ at this surface of section
$\eta=\zeta=\dot{\xi}=0$, $\dot\eta>0$.

\subsection{Numerical orbit integrations}
\label{subsec:numerical_int}

Even in the two-dimensional case, we expect that the stable
regions for distant satellites of the giant planets will be
somewhat different from those derived by H\'enon (1970) and shown
in Fig.\ \ref{fig:check_Henon}, since (i) H\'enon's results are
based on Hill's approximation $\mu\to0$, while the giant planets
have $\mu$ in the range 0.00096 (Jupiter) to 0.000044 (Uranus);
(ii) H\'enon's results assume that the planet orbit is circular,
while the giant planets have eccentricities between 0.0086 and
0.056; (iii) both the satellites and their host planets are
subject to perturbations from the other planets. We must carry out
long-term numerical integrations of the satellite orbits to assess
the influence of these effects on the stability region shown in
Fig.\ \ref{fig:check_Henon}.

We sample the initial conditions using a fine grid on the
H\'{e}non diagram, with $d\Gamma=0.1$ and $d\xi=0.06$. This is
shown as the dotted grid in Fig.\ \ref{fig:jupiter_henon2d}. We
then convert them to the non-rotating $(xyz)$ planetocentric
coordinate system where we do the integrations of satellite
orbits. We require that in the rotating frame the Sun is always
located at the $-\xi$ axis, and the angular velocity of the
rotating frame equals the instantaneous angular velocity of the
Sun relative to the planet in the non-rotating planetocentric
frame; thus the angular speed of the rotating frame is
time-varying if the planet's orbit is eccentric, and the direction
of the $\zeta$ axis may vary if the planet's orbit is perturbed by
other planets. We use a unit of length $\mu^{1/3}a_p$ and unit of
time $n^{-1}$ to scale the coordinates/velocities between the two
frames, where $a_p$ is taken to be the initial semi-major axis of
the planet.

The system to be numerically integrated is composed of the four
outer giant planets (or sometimes just one of them), the Sun, and
a satellite around one of the planets; the satellite is treated as
a massless test particle. We use a second-order Wisdom-Holman
symplectic scheme (Wisdom \& Holman 1991), as implemented in the
{\bf Swift} package (Levison \& Duncan 1994). Following
Nesvorn\'y et al.\ (2003), we have modified the {\bf Swift}
code such that the integration of the planets is done in the
Jacobi coordinate system while that of the satellites is done in
the non-rotating planetocentric coordinate system. We tried
different timesteps to optimize between speed and accuracy, and
found $dt=20$ days is short enough to produce the correct results
with reasonable computational cost, for all four planets.

One potential concern is that the Wisdom-Holman symplectic scheme,
as we have implemented it, is designed for nearly Keplerian
orbits relative to the planet and might break down at large distances
from the planet, where the orbits are nearly Keplerian relative to the
Sun. However, the characteristic orbital period at large distances
is equal to the planetary orbital period, and this is much longer than the
orbital periods of satellites inside the Hill radius that the
integrator is designed to follow, so even a crude integrator should
work well. Moreover, our ability to reproduce the
H\'{e}non diagram (compare Fig.\ \ref{fig:check_Henon} and the
lower right panel of Fig.\ \ref{fig:jupiter_henon2d}), the
long-term stability of many of our orbits, and the similarity of
the characteristic orbit shapes to those found by H\'{e}non (see
\S\ref{sec:2D_result}), all indicate that even at the largest
distances probed here, the symplectic integrator seems to work
pretty well. As a further check, we have used the Bulirsch-Stoer
integrator to follow satellite orbits around Uranus for $10^6$ yr
and found almost identical results to the Wisdom-Holman integrator.

We terminate the integration if the distance of the satellite from
the planet exceeds the semi-major axis of the planet since at this
point the satellite has escaped according to our definition at the
end of \S\ref{sec:intro}, or if the distance is less than the
semi-major axis of the outermost regular satellite of each planet
(being Callisto, Iapetus, Oberon and Triton respectively), since
at this point the satellite lifetime against ejection or collision
with the regular satellite or the planet is likely to be short.
Any test particles that cross either of these two radii are
considered lost. We have experimented with including the
quadrupole moment $J_2$ of the planet (including the contribution
from the inner regular satellites) but this has no detectable
effect on our results.

\section{Results}\label{sec:result}

\begin{figure*}
\centering
\includegraphics[width=0.45\textwidth]{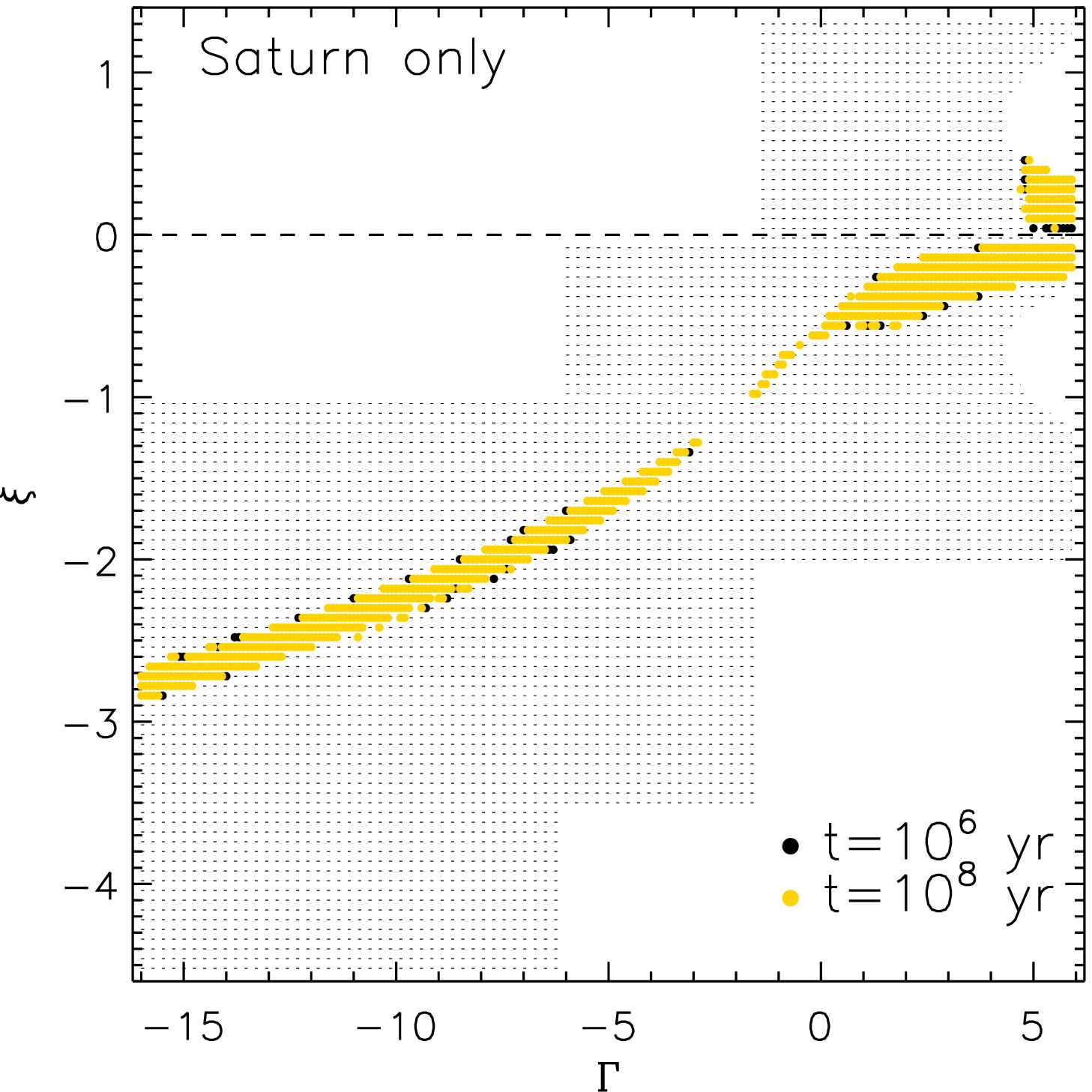}
\includegraphics[width=0.45\textwidth]{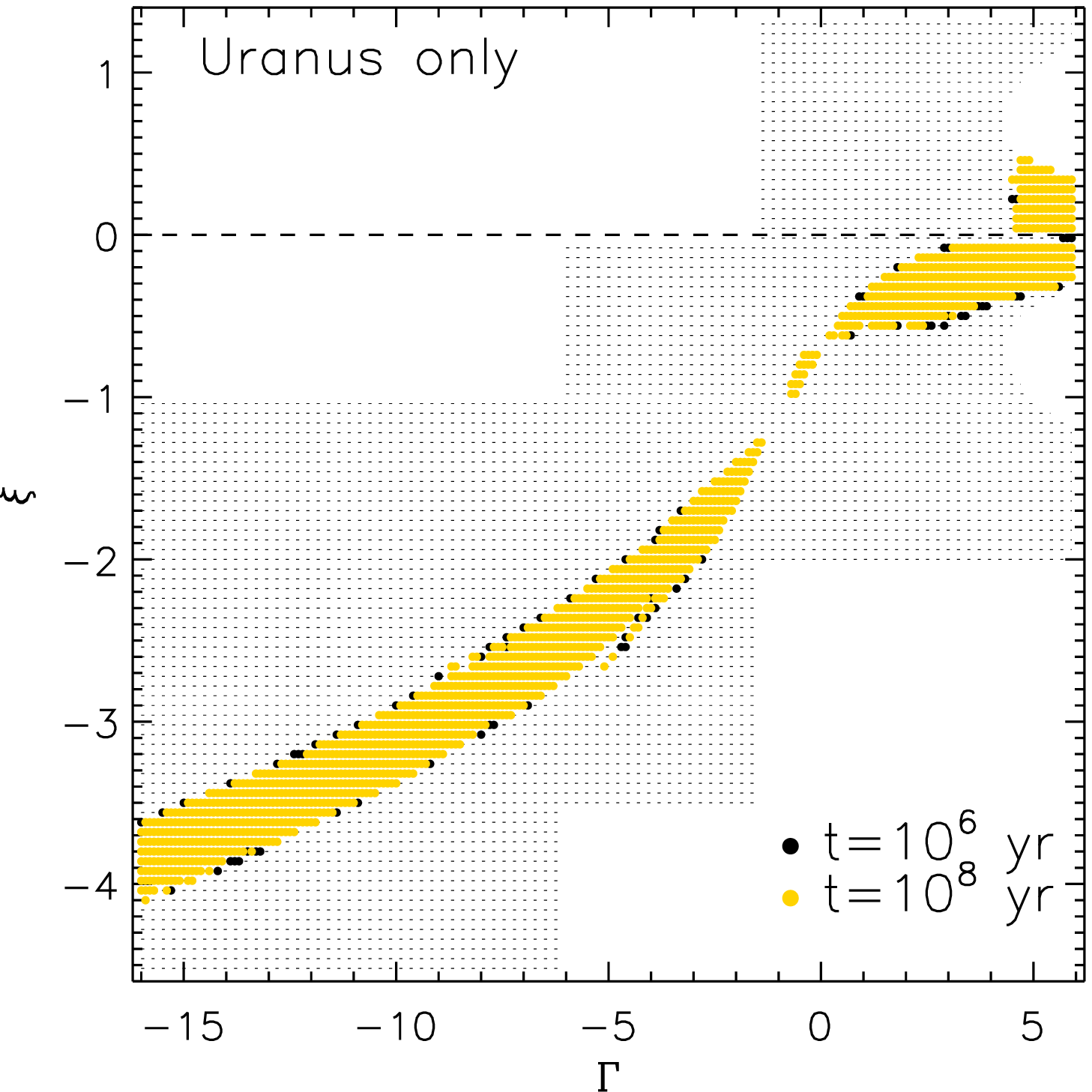}
\caption{Two-dimensional H\'{e}non diagrams for Saturn only (left)
and Uranus only (right); the effects of the other three planets
are not included in the integrations. In contrast to the
results in Fig. \ref{fig:other_henon2d}, Saturn can host stable
outer retrograde orbits, and most of the satellites that survive
for $10^6$ yr also survive for $10^8$ yr around both  planets.
}\label{fig:uranus_only_henon2d}
\end{figure*}

\subsection{Two-dimensional H\'{e}non diagrams}\label{sec:2D_result}

We first study cases in which the initial velocity vectors of
satellites lie in the planet orbital plane, i.e.,
$\zeta=\dot{\zeta}=0$. As we have described, this is different
from H\'{e}non's problem because: (i) we do not use Hill's
approximation; (ii) planets such as Jupiter have non-zero
eccentricity; and (iii) there are gravitational perturbations from
other planets.

As an illustration we show how the stable region changes under
various conditions in Fig.\ \ref{fig:jupiter_henon2d}, for
satellites around Jupiter and an integration time $10^6$ yr. We
consider four situations: (a) Jupiter moves on its actual
(slightly eccentric) orbit, including perturbations from the other
three giant planets ({\em upper left}); (b) the planar restricted
three-body problem, in which Jupiter travels on an orbit with its
current eccentricity of 0.048 and the other planets are absent
({\em upper right}); (c) the planar circular restricted three-body
problem, in which Jupiter travels on a circular orbit with its
current semi-major axis ({\em bottom left}); (d) same as (c)
except that the planet mass is $1/100$ of the Jupiter mass ({\em
bottom right}).

By comparing Figs.\ \ref{fig:check_Henon} and
\ref{fig:jupiter_henon2d} it is clear that case (d) in the bottom
right panel best reproduces the original H\'{e}non diagram; this
is not surprising since $\mu\simeq 10^{-5}$ is smallest so Hill's
approximation is satisfied best, and the other conditions assumed
in H\'enon's problem (circular planet orbit, no other planets) are
also satisfied. When using the actual Jupiter mass in (c), the
outer retrograde stable region (i.e., the lower-left branch)
shifts and shrinks. The overall stability region shrinks further---but does not vanish---when Jupiter's orbit is eccentric as in case
(b), and for the most realistic case (a).

Note that $10^6$ yr is only a small fraction of the lifetime of
the solar system, and possible erosion of the stable region over
longer times is somewhat indicated by the presence of a few red
and blue dots in the upper panels of Fig.\
\ref{fig:jupiter_henon2d}, indicating orbits that are unstable on
timescales of $10^4$ and $10^5$ yr.

These illustrative calculations show that some Jovian satellites
orbiting well outside the Hill radius can survive for at least
$10^6$ yr, although the stable region is substantially smaller
than in Hill's approximation to the circular restricted three-body
problem and appears to erode slowly with time. They also show that
the stable region is larger (relative to the Hill radius) if the
planet mass $\mu$ is smaller, suggesting that the stable regions
of the other giant planets may be larger than Jupiter's.

We now extend these calculations in the following ways: (i) we
examine satellite orbits around all four giant planets, using the
actual planetary orbits including perturbations from the three
other planets; (ii) since stable orbits are found at the most
negative Jacobi constant ($\Gamma=-6$) examined in Fig.\
\ref{fig:jupiter_henon2d}, we extend the grid of initial
conditions to $\Gamma=-16$; (iii) we extend the integration time from
$10^6$ yr to $10^8$ yr.

The results are shown in Fig.\ \ref{fig:other_henon2d}. There are
large regions of inner retrograde/prograde orbits that are stable
for $10^8$ yr. For Jupiter, there are a few outer retrograde
orbits that survive for $10^8$ yr; Wiegert et al.\ (2000) found no
orbits that survived for $\gtrsim 10^7$ yr, but this may reflect
their less complete coverage of phase space. For Saturn, the outer
retrograde stable region completely disappears in less than $10^6$
yr, a conclusion already reached by Wiegert et al.  For Uranus and
Neptune, in contrast, there is a large stable region of outer
retrograde orbits remaining after $10^8$ yr. We expect that the
stable regions around Jupiter, Uranus, and Neptune will shrink
somewhat further between $10^8$ yr and $5\times10^9$ yr, the
approximate age of the solar system, so we integrated some
outer retrograde satellite orbits around these three planets for
$10^9$ yr. We found that about a third of the Jovian orbits and
over half of the outer retrograde orbits for Uranus and Neptune
shown in Fig. \ref{fig:other_henon2d} still survive. Thus it is
very likely that Uranus and Neptune could still host primordial
satellites on such orbits to the present time. It is likely, but
not certain, that similar satellites could survive around Jupiter,
at least in small volumes of phase space.

The shrinkage of the stable region of the outer retrograde branch
between $10^6$ and $10^8$ yr, as well as the lack of stable outer
retrograde orbits around Saturn, appear to be mainly due to
perturbations from the other planets. To demonstrate this, we ran two
$10^8$ yr integrations for Saturn only and Uranus only.  The results
are shown in Fig.\ \ref{fig:uranus_only_henon2d}; in this case, Saturn
{\em can} host stable outer retrograde satellites for at least $10^8$
yr, and there is almost no difference in the size of the stable region
between $10^6$ and $10^8$ yr for either planet.

The stable region around Uranus is larger than the one around
Saturn in Fig.\ \ref{fig:uranus_only_henon2d}, and the stable
regions around Uranus and Neptune are larger than the one around
Jupiter in Fig.\ \ref{fig:other_henon2d}. These differences are
probably caused mostly by their different planet-to-Sun mass
ratios $\mu$. As $\mu$ increases, the outer retrograde stability
branch in the lower left of the H\'enon diagram shrinks, and
shifts upward (see H\'{e}non 1965; 1970, or compare the two lower
panels of Fig.\ \ref{fig:jupiter_henon2d}).

We also notice in Figs.\ \ref{fig:jupiter_henon2d}d and in the
right panel of \ref{fig:uranus_only_henon2d} that there is a
little tail or branch to the stable region around
$(\Gamma,\xi)=(-5,-2.5)$. We suspect this comes from H\'{e}non's
periodic family $g_3$, which bifurcates from the periodic
retrograde orbits at $(\Gamma,\xi)=(-2,-1.2)$ and passes close to
the point $(\Gamma,\xi)=(-5,-2.5)$ (see H\'{e}non 1970, Fig.\ 13).

\begin{figure*}
\centering
\includegraphics[width=0.4\textwidth]{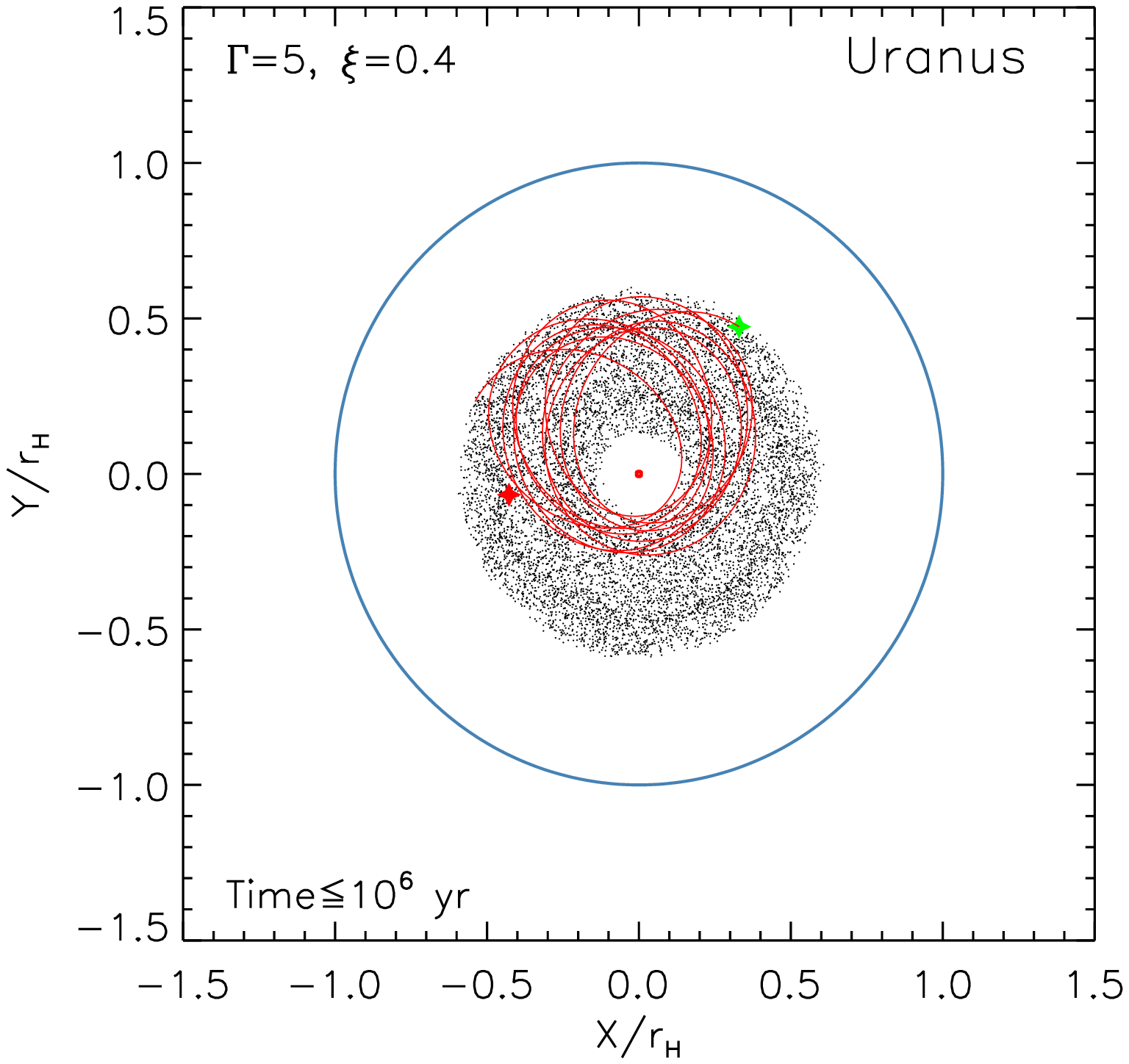}
\includegraphics[width=0.4\textwidth]{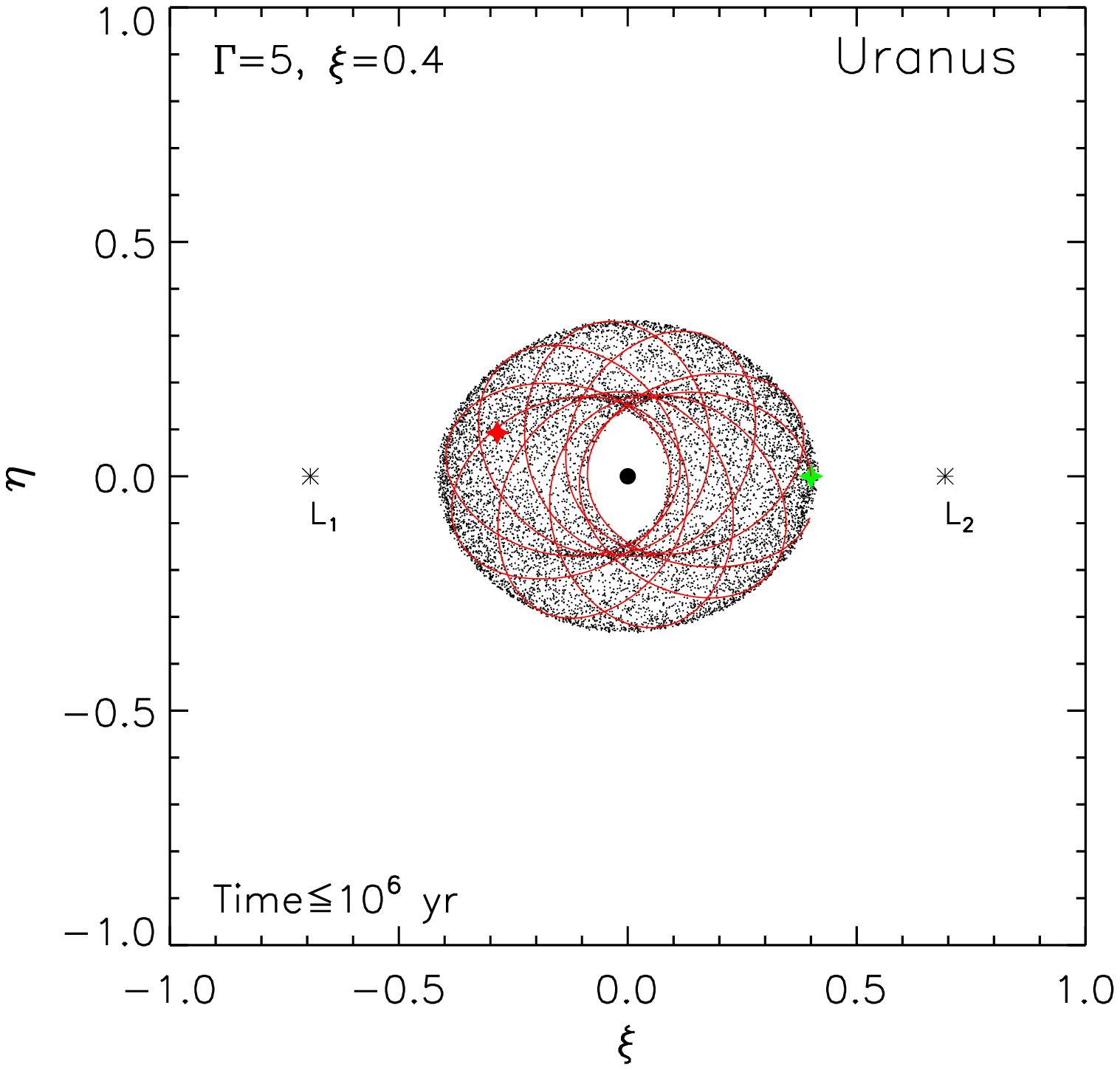}
\includegraphics[width=0.4\textwidth]{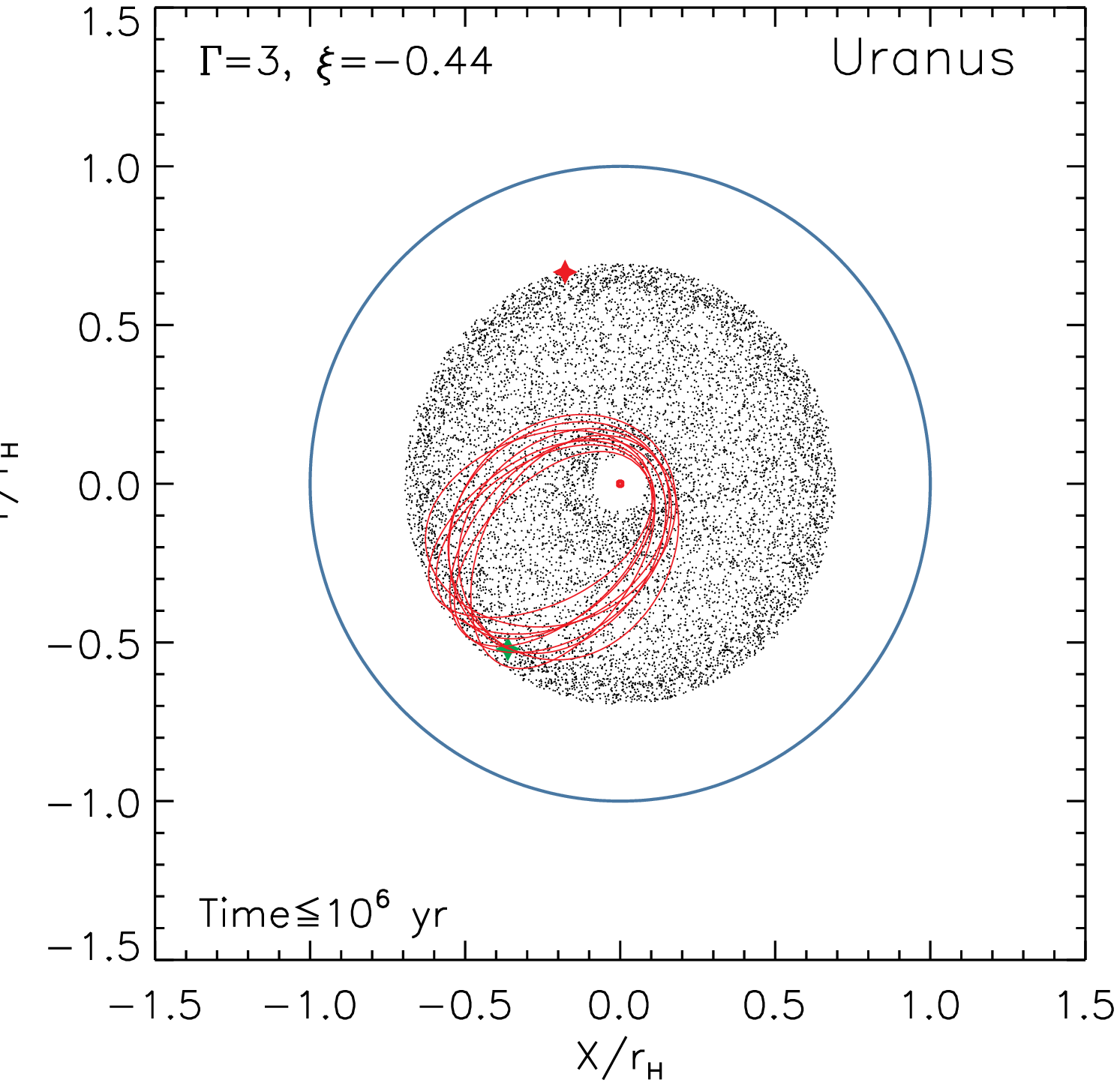}
\includegraphics[width=0.4\textwidth]{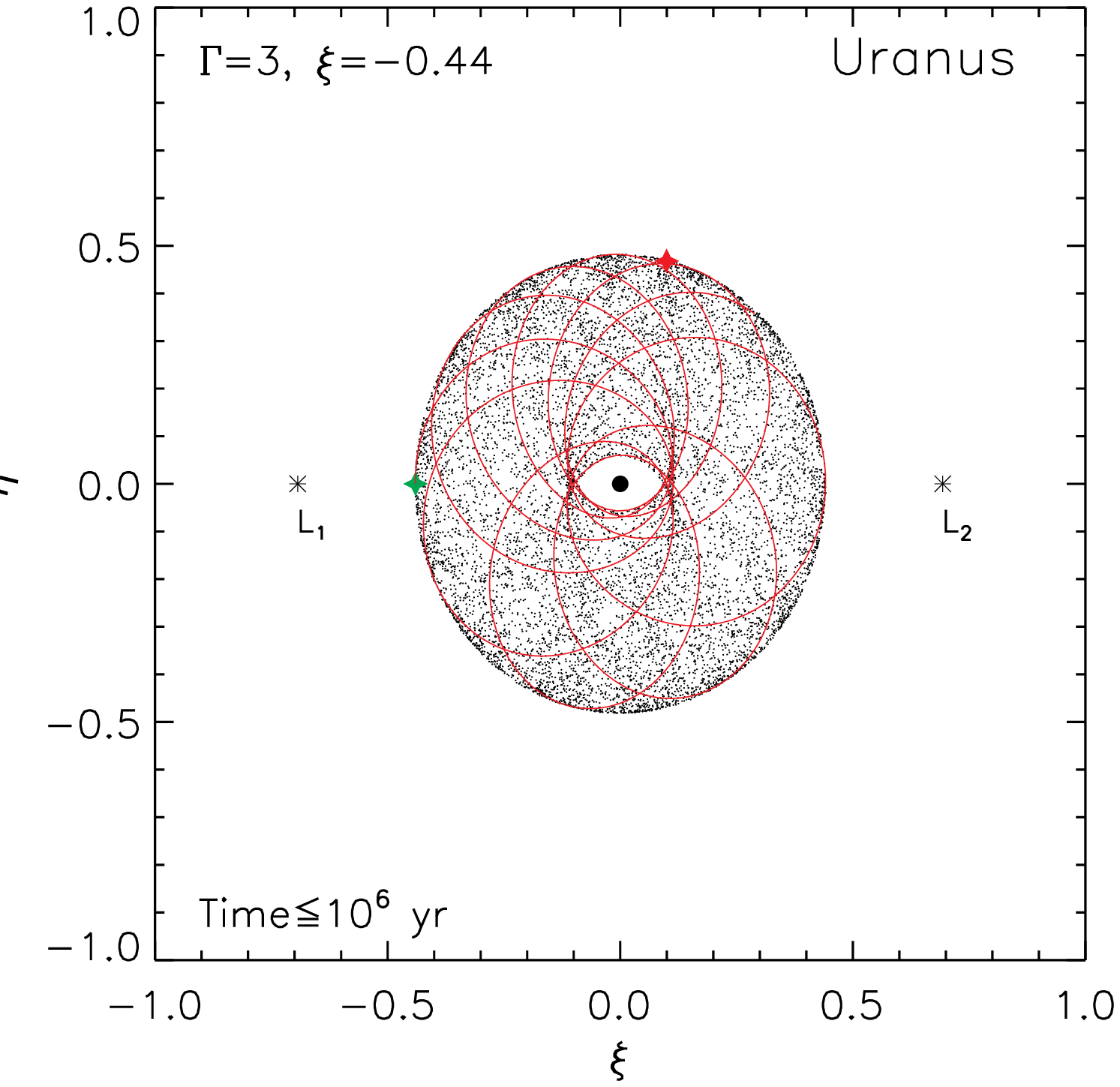}
\includegraphics[width=0.4\textwidth]{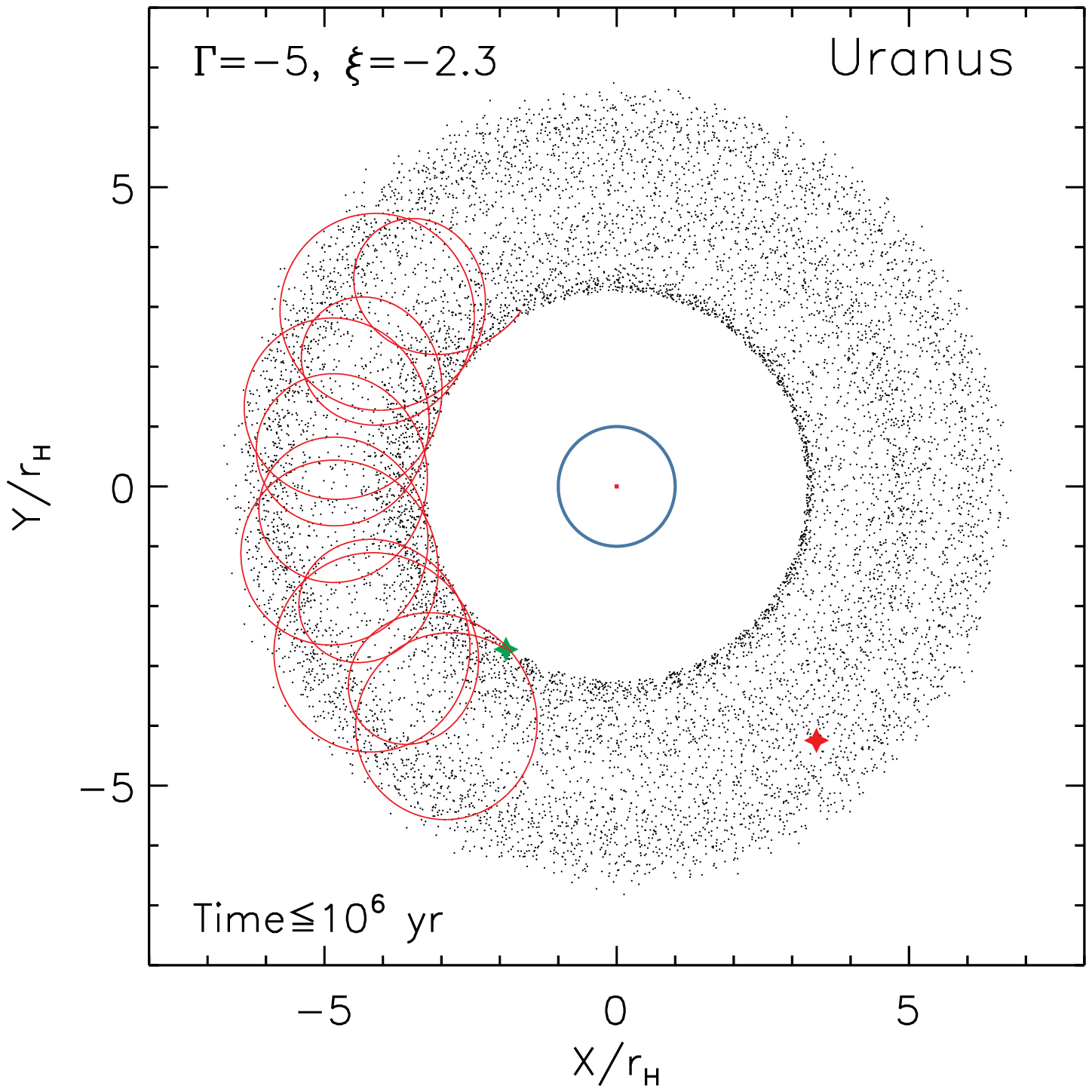}
\includegraphics[width=0.4\textwidth]{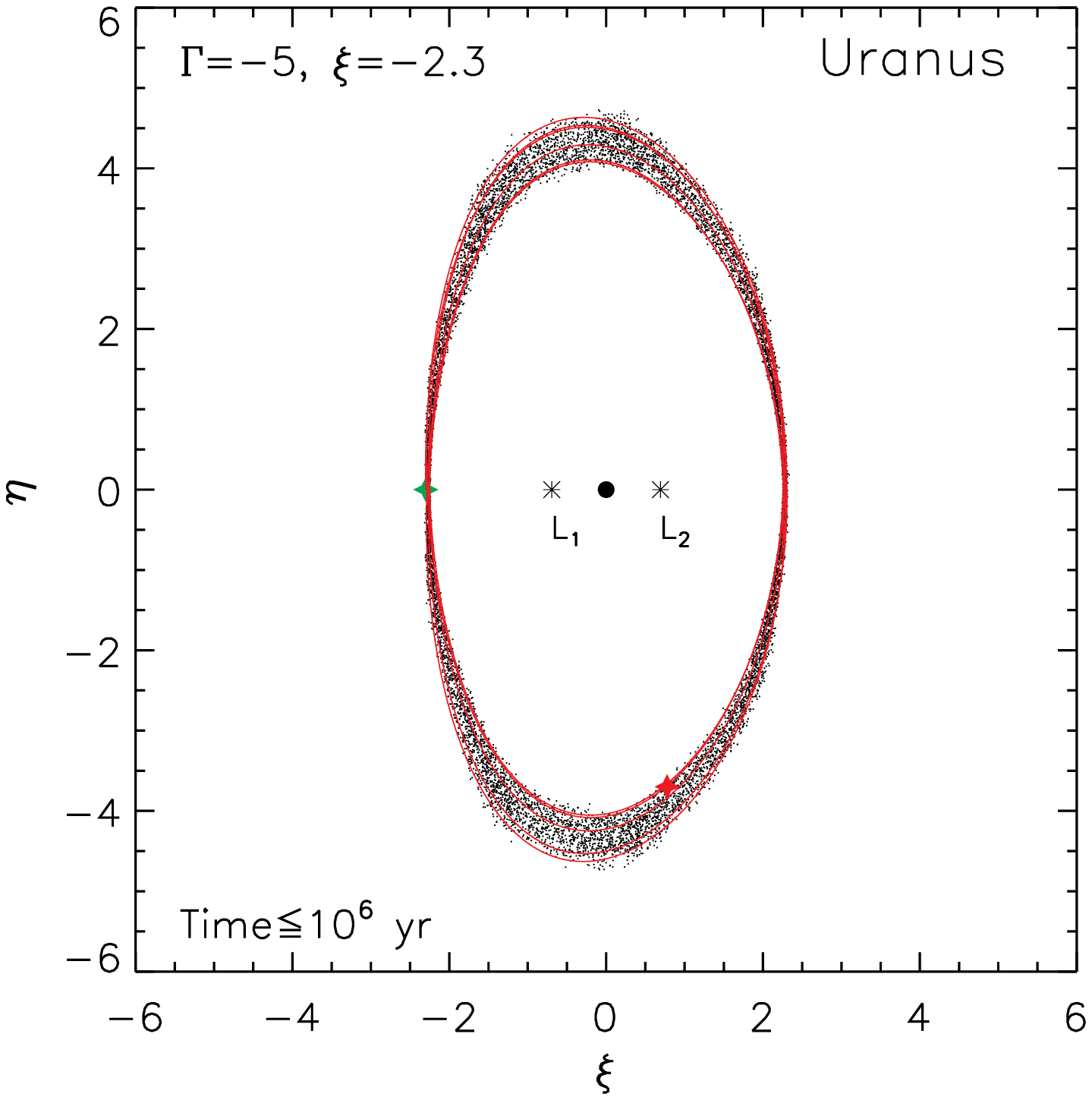}
\caption{Examples of stable orbits around Uranus. All orbits
initially lie in the orbital plane of Uranus. Dots are
instantaneous locations for the first $10^6$ yr, plotted at
intervals of 100 yr, with the green and red stars marking the
starting and ending locations. We also plot a few revolutions as
red curves. The left column is in the {\em non-rotating}
planetocentric frame and the right column is in the {\em rotating}
planetocentric frame. In the left column, the blue circles
indicate the Hill sphere. {\em Upper:} an inner prograde orbit;
{\em middle}: an inner retrograde orbit; {\em bottom}: an outer
retrograde orbit.}\label{fig:uranus_survived_TP}
\end{figure*}

\begin{figure*}
\centering
\includegraphics[width=0.45\textwidth]{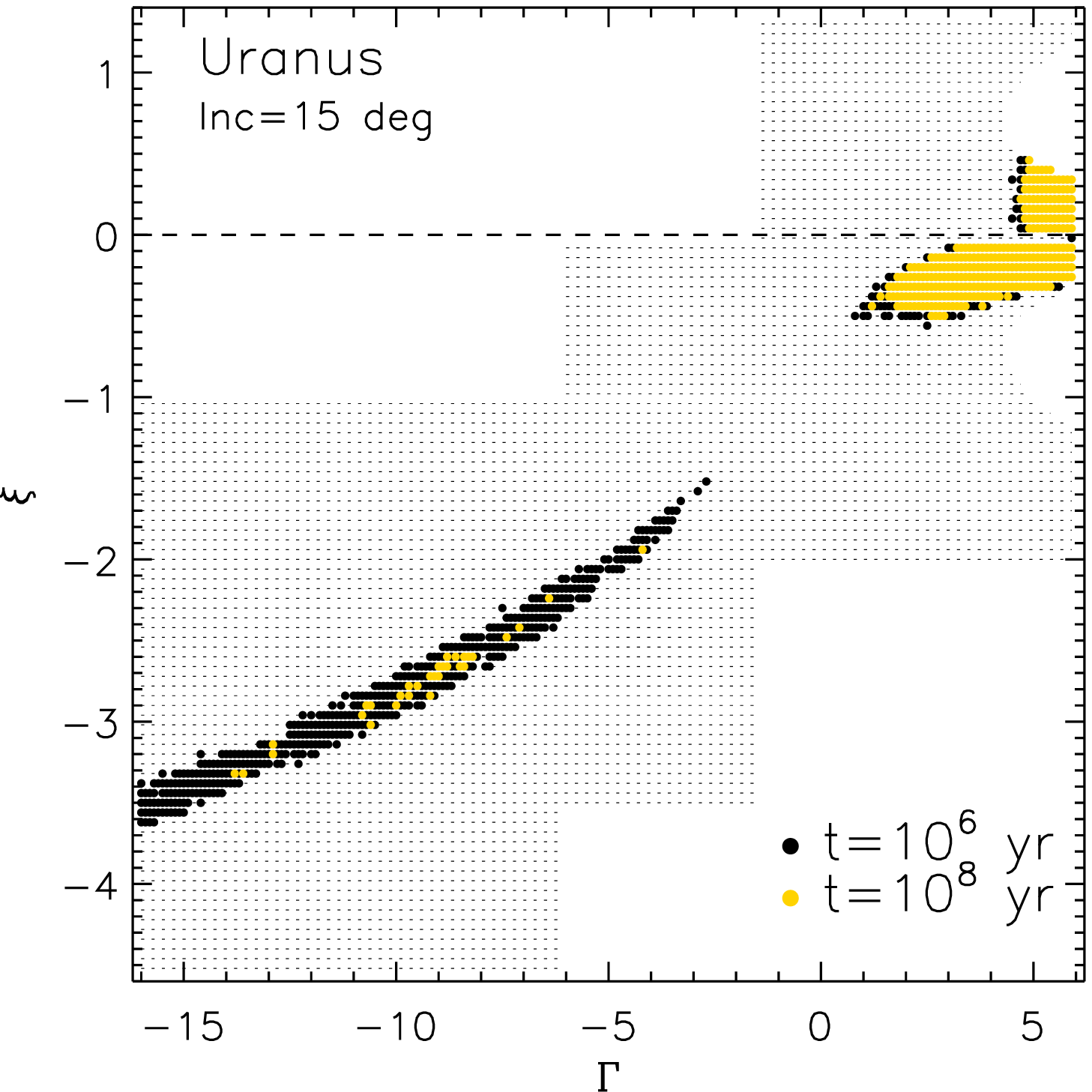}
\includegraphics[width=0.45\textwidth]{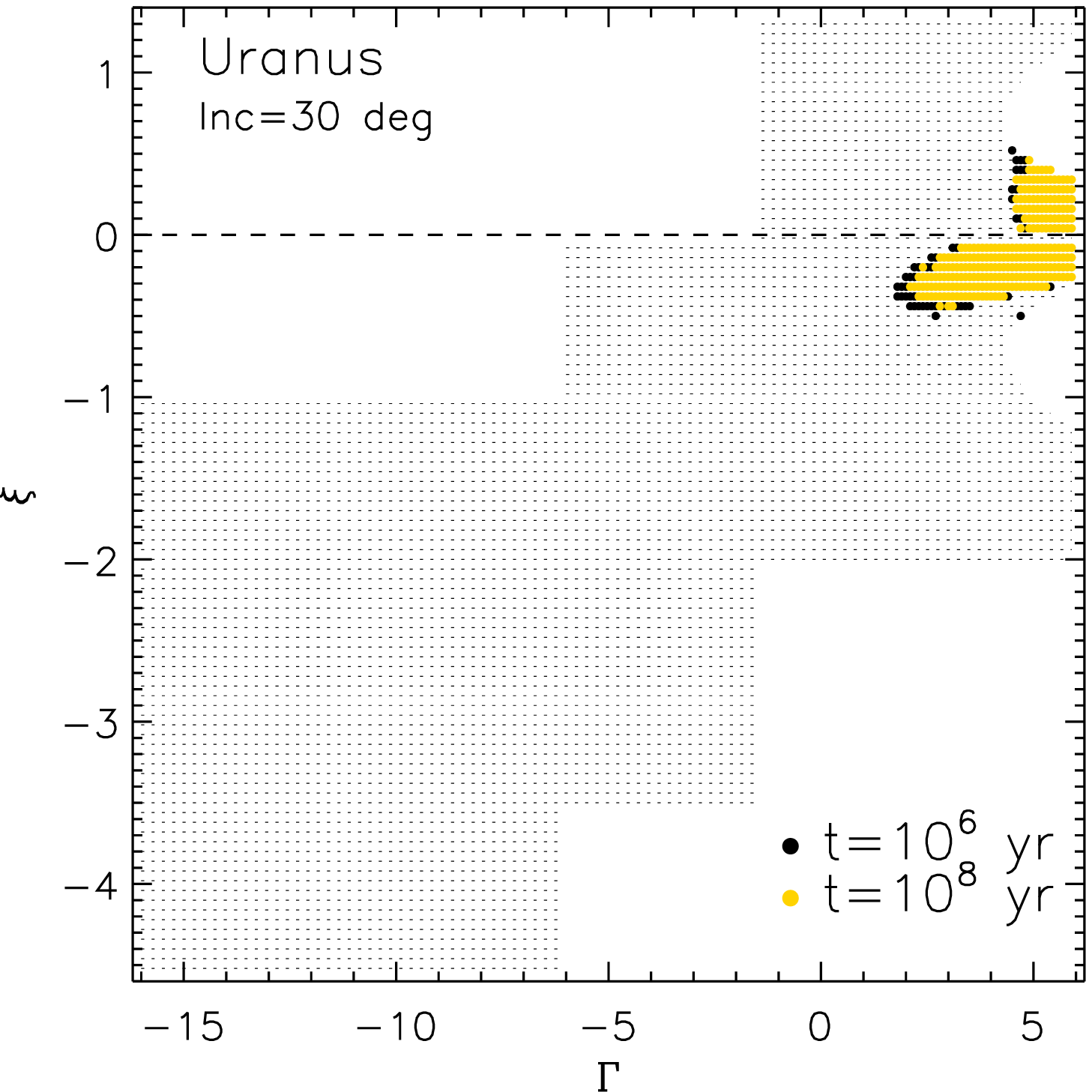}
\includegraphics[width=0.45\textwidth]{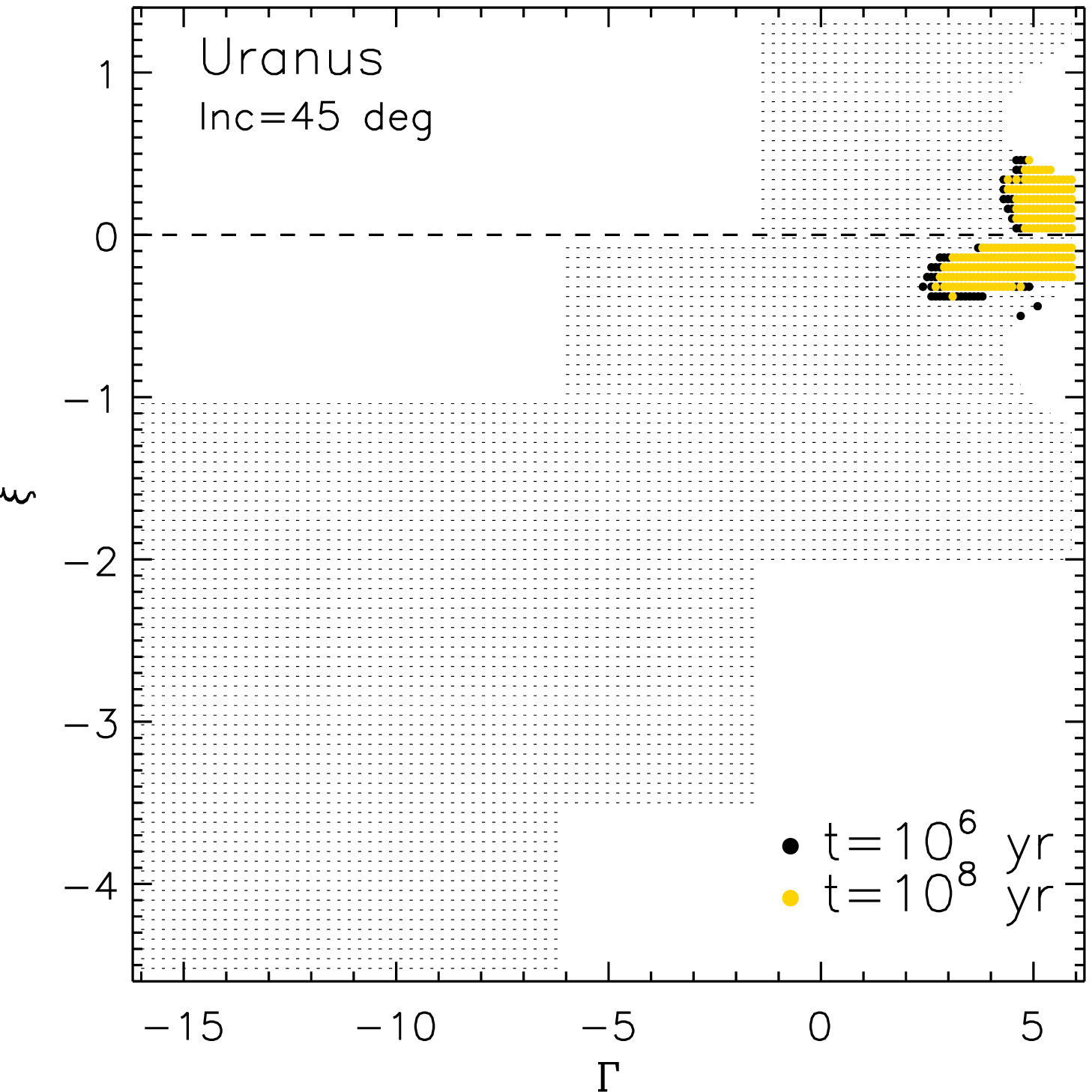}
\includegraphics[width=0.45\textwidth]{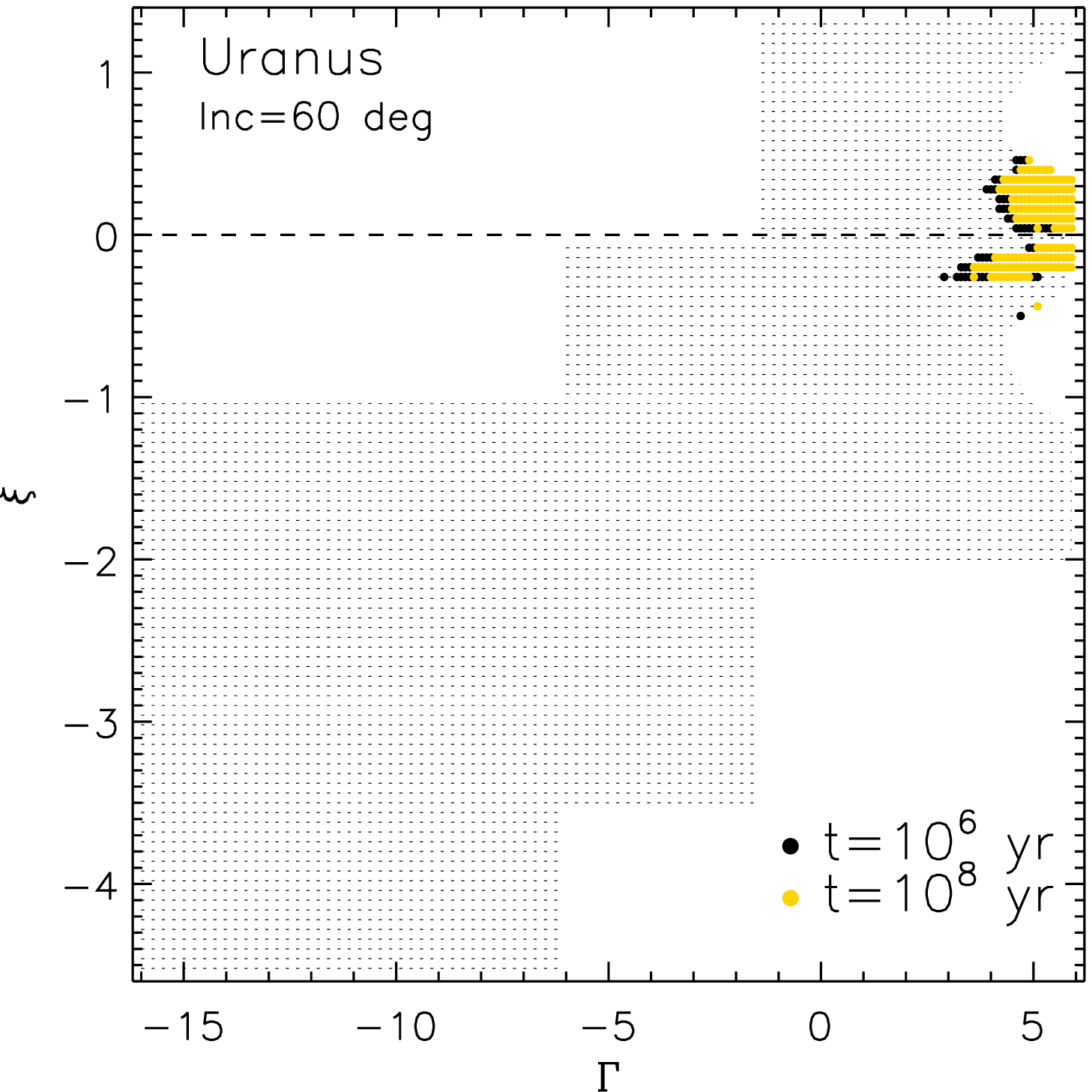}
\includegraphics[width=0.45\textwidth]{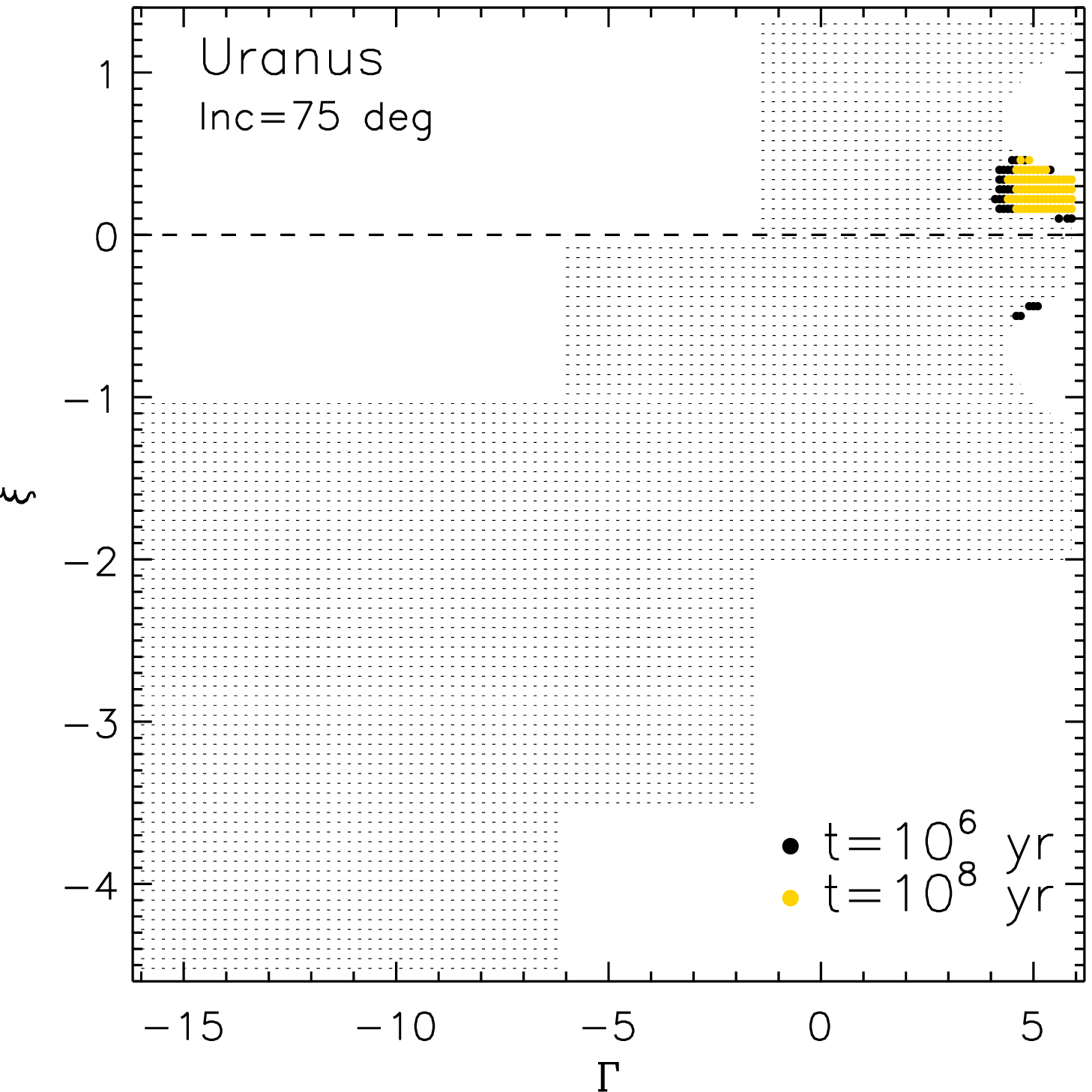}
\includegraphics[width=0.45\textwidth]{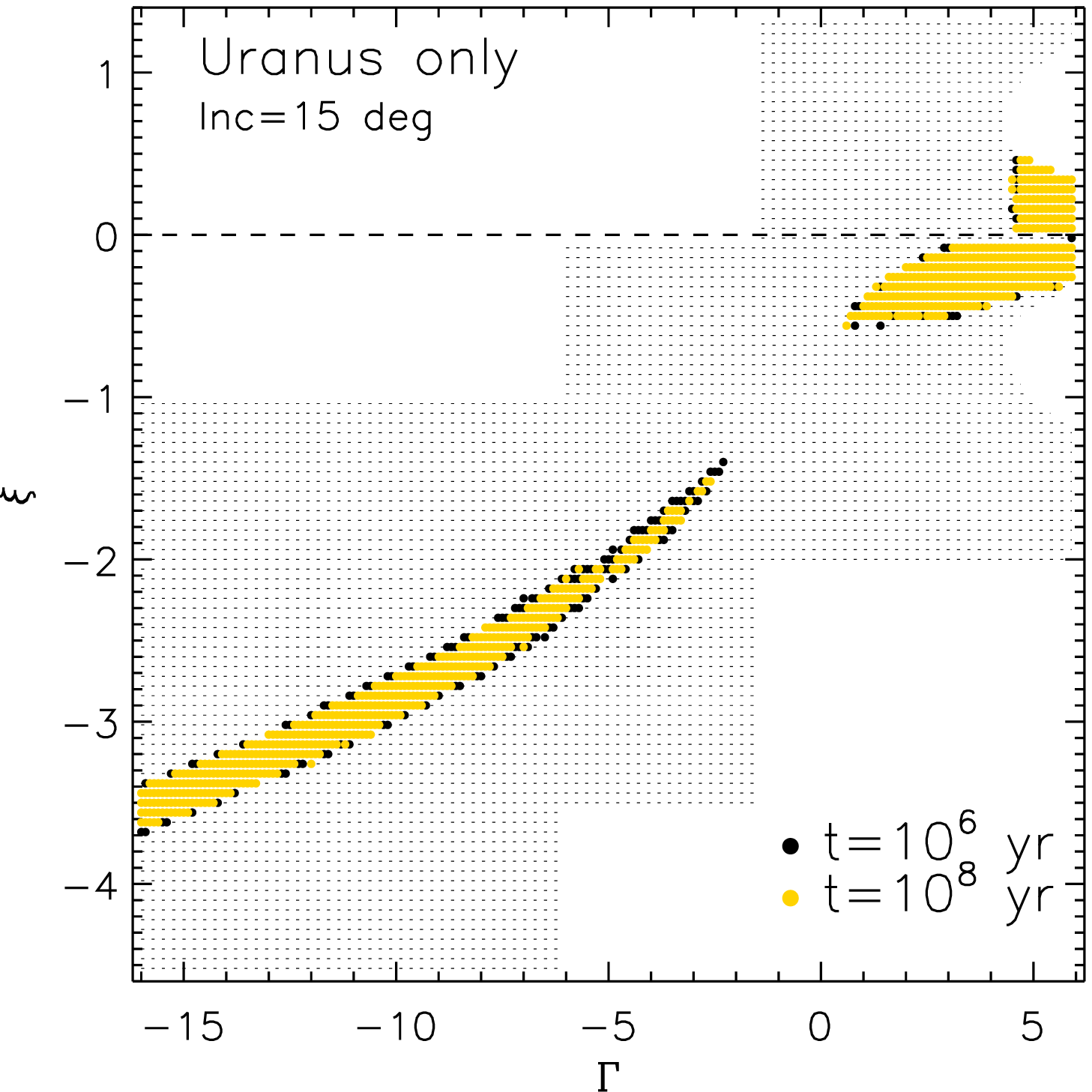}
\caption{Examples of three-dimensional H\'{e}non diagrams for
Uranus. The initial inclinations $I=15,30,45,60,75$ degrees in the
rotating frame. The notation is the same as in Figs.
\ref{fig:jupiter_henon2d} and \ref{fig:other_henon2d}. The stable
regions shrink as the inclination increases. In particular, no
stable outer retrograde orbit exists for $I\ge 30^\circ$.
}\label{fig:uranus_henon_3d}
\end{figure*}

\begin{figure*}
\centering
\includegraphics[width=0.45\textwidth]{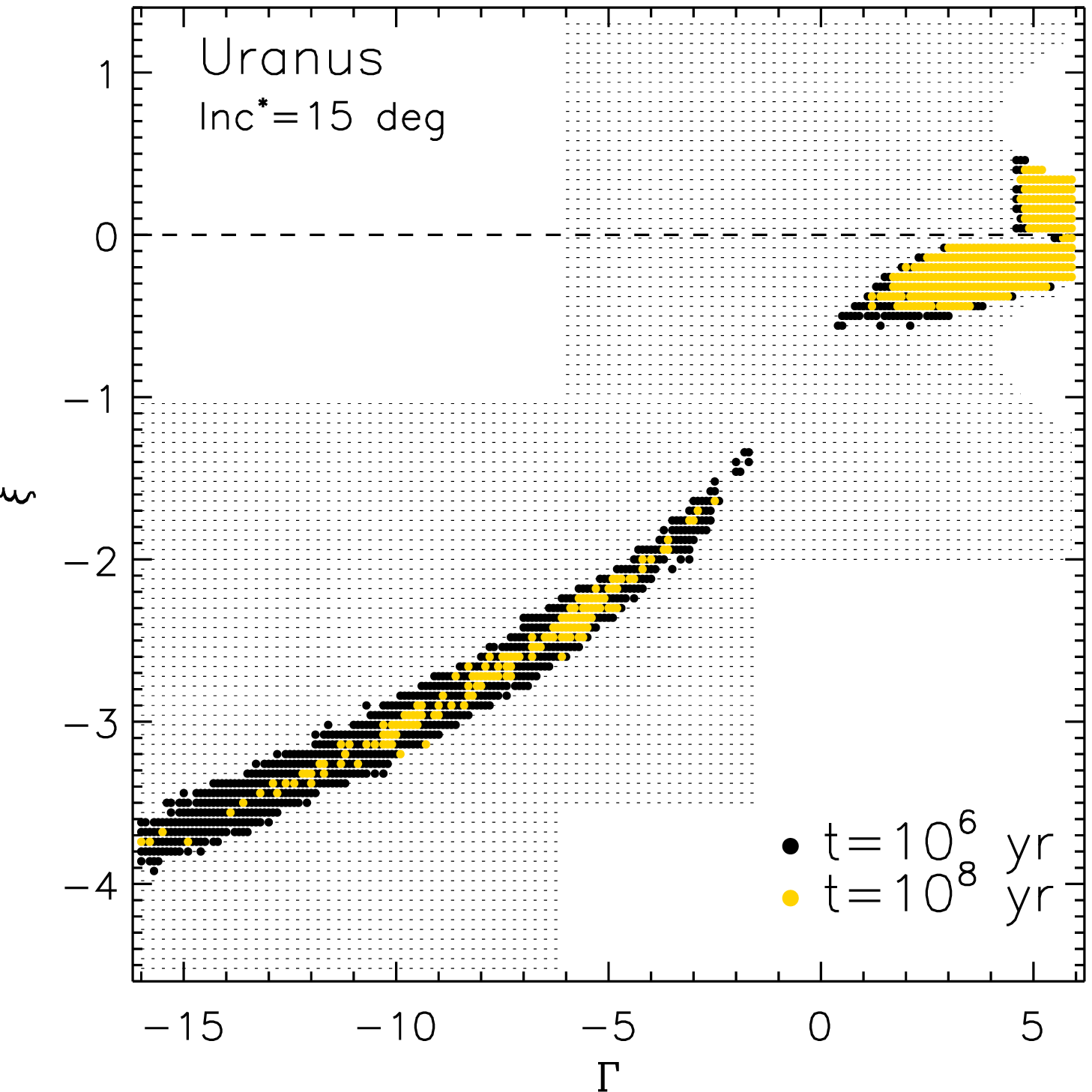}
\includegraphics[width=0.45\textwidth]{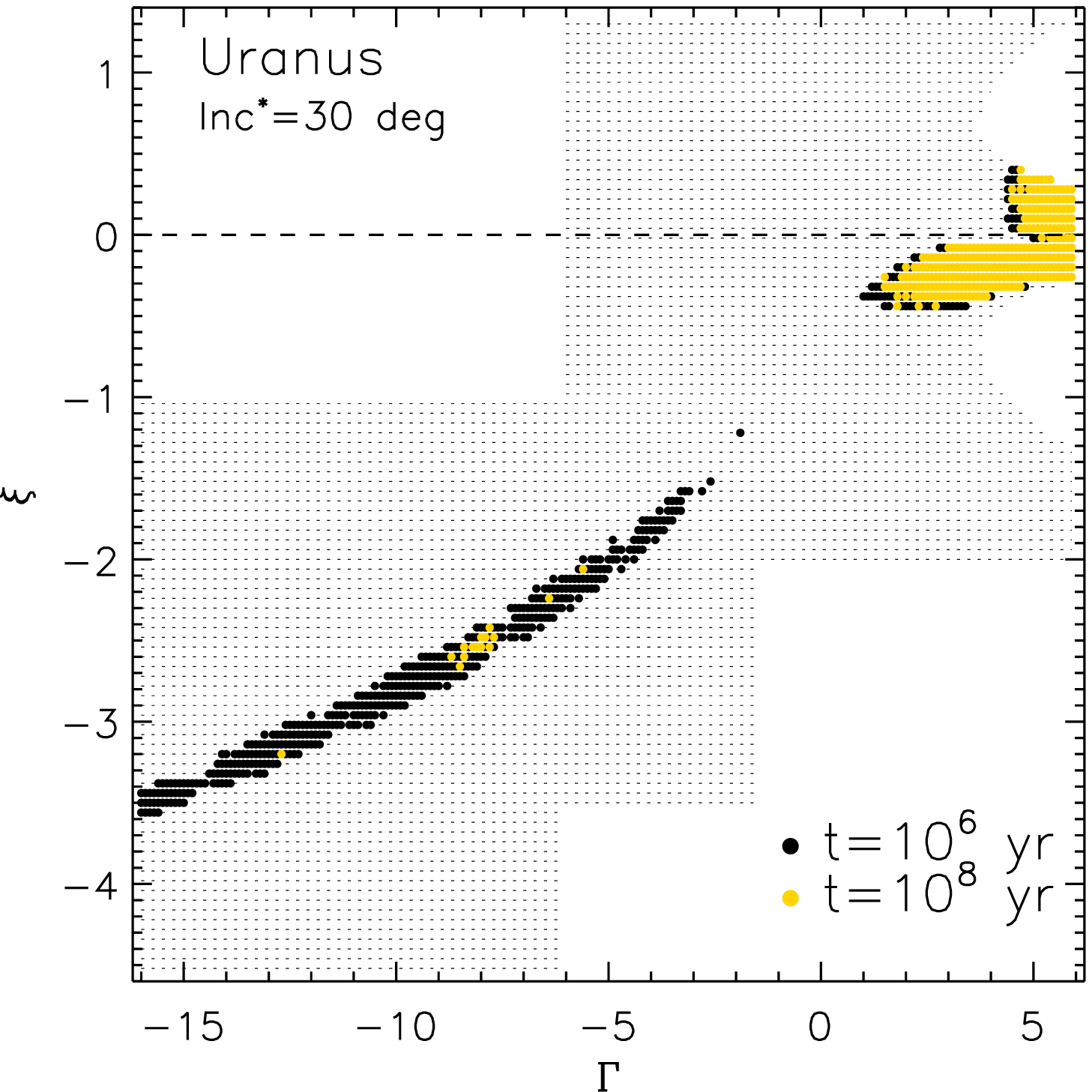}
\includegraphics[width=0.45\textwidth]{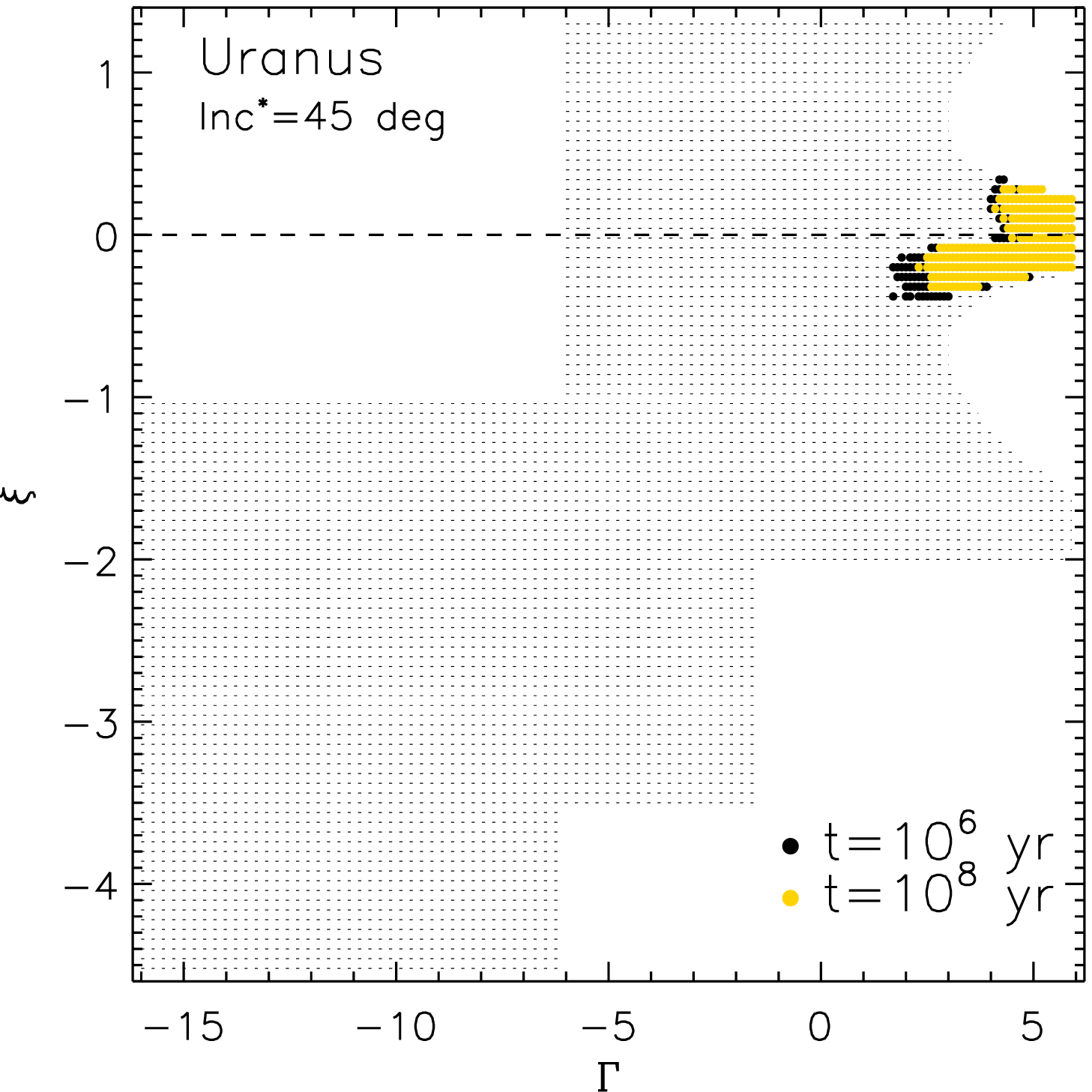}
\includegraphics[width=0.45\textwidth]{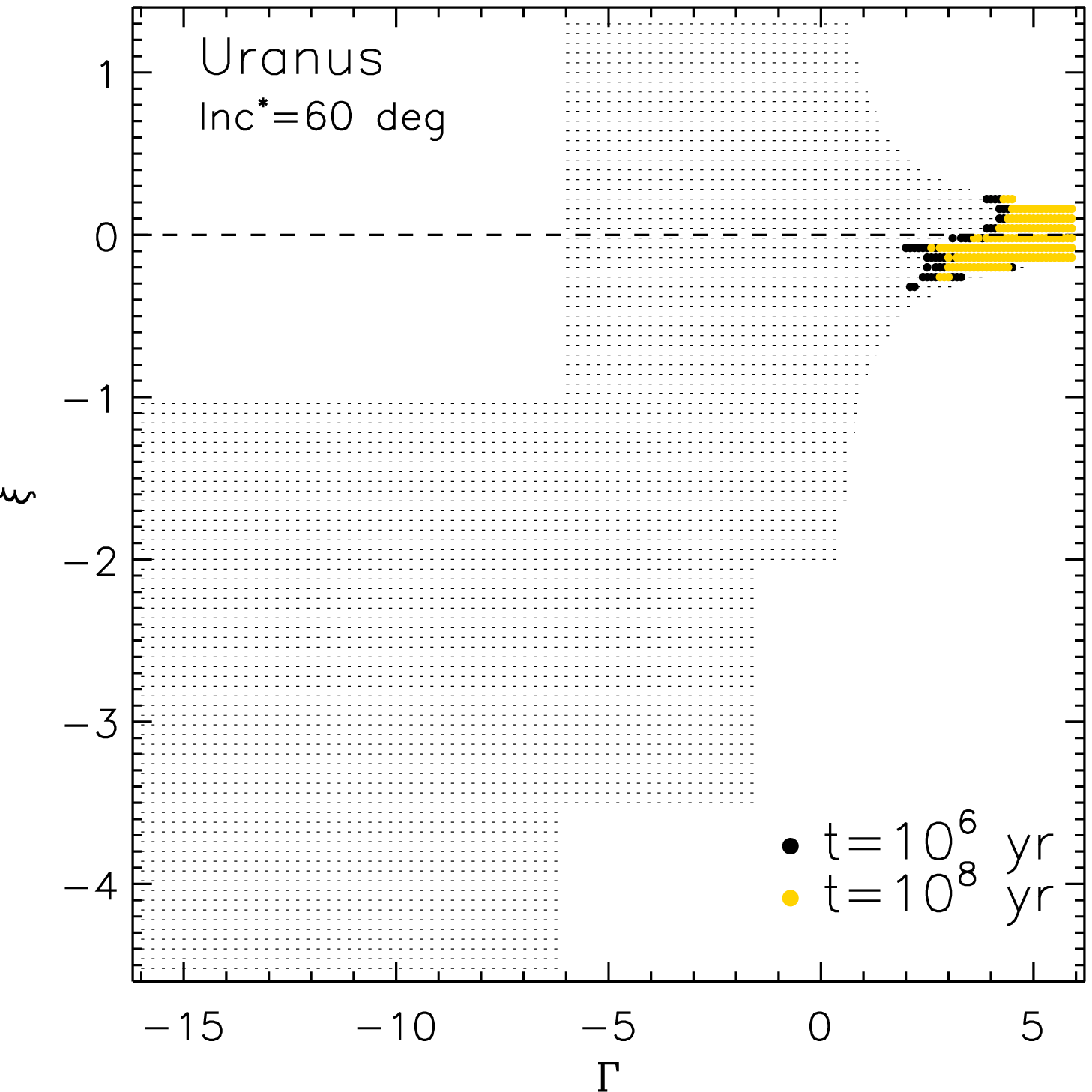}
\includegraphics[width=0.45\textwidth]{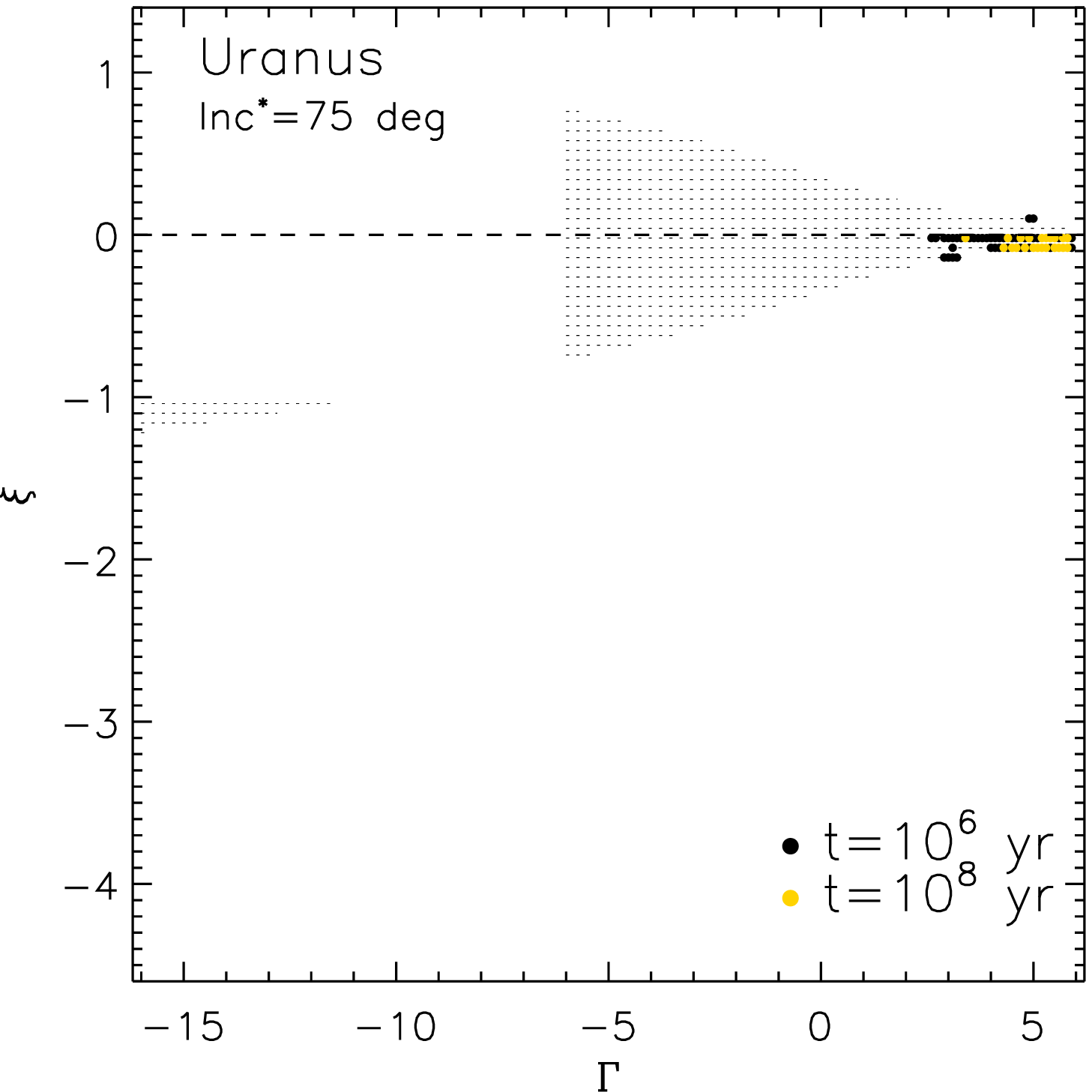}
\caption{Examples of three-dimensional H\'{e}non diagrams for
Uranus. In these diagrams the surface of section is taken when the
satellite is at maximum height above the planet's orbital plane
(in contrast to Figure \ref{fig:uranus_henon_3d} where the surface
of section is taken when the satellite crosses the plane). The
initial inclinations $I^*=15,30,45,60,75$ degrees in the rotating
frame. }\label{fig:uranus_henon_3d_alt}
\end{figure*}

\begin{figure*}
\centering
\includegraphics[width=0.45\textwidth]{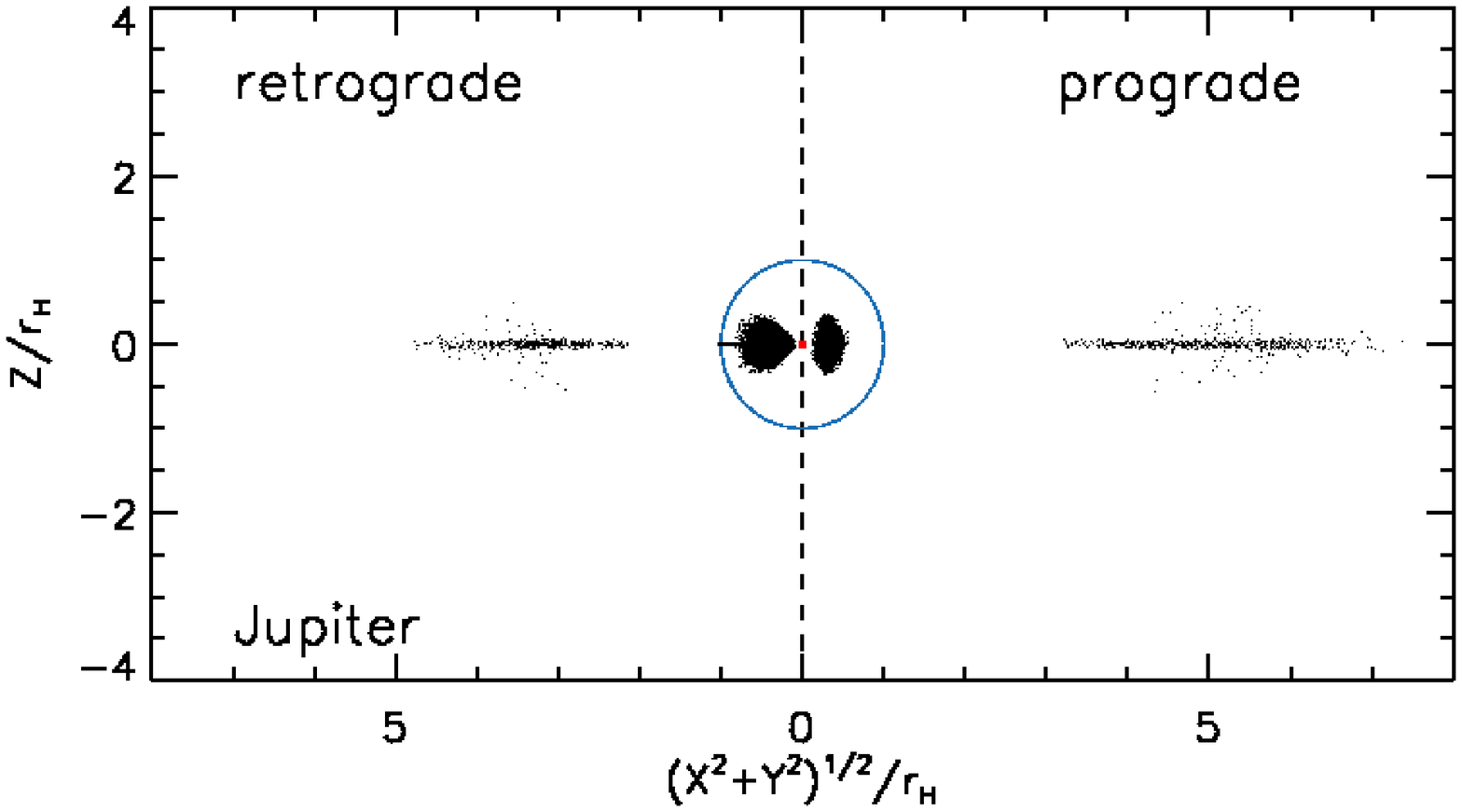}
\includegraphics[width=0.45\textwidth]{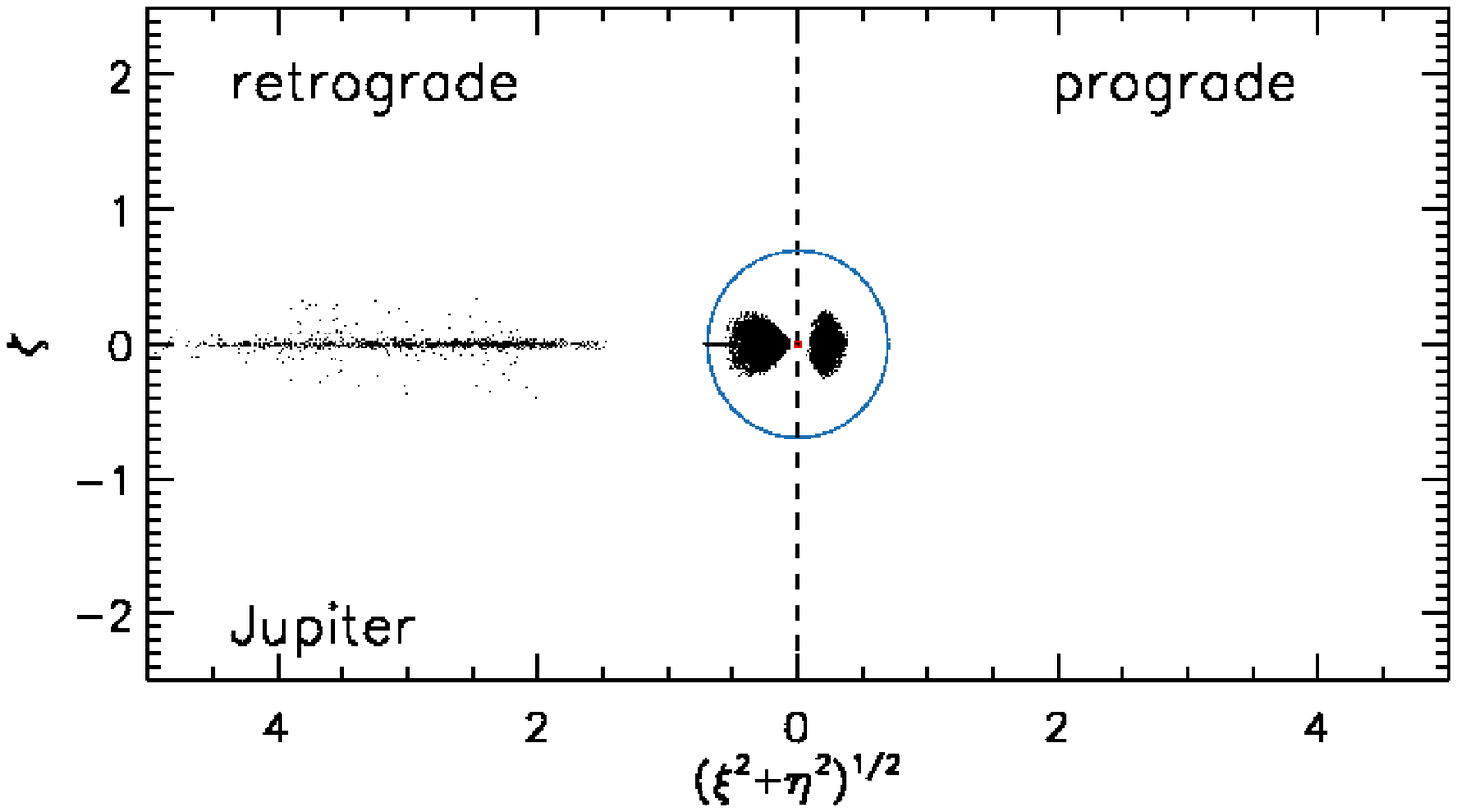}
\includegraphics[width=0.45\textwidth]{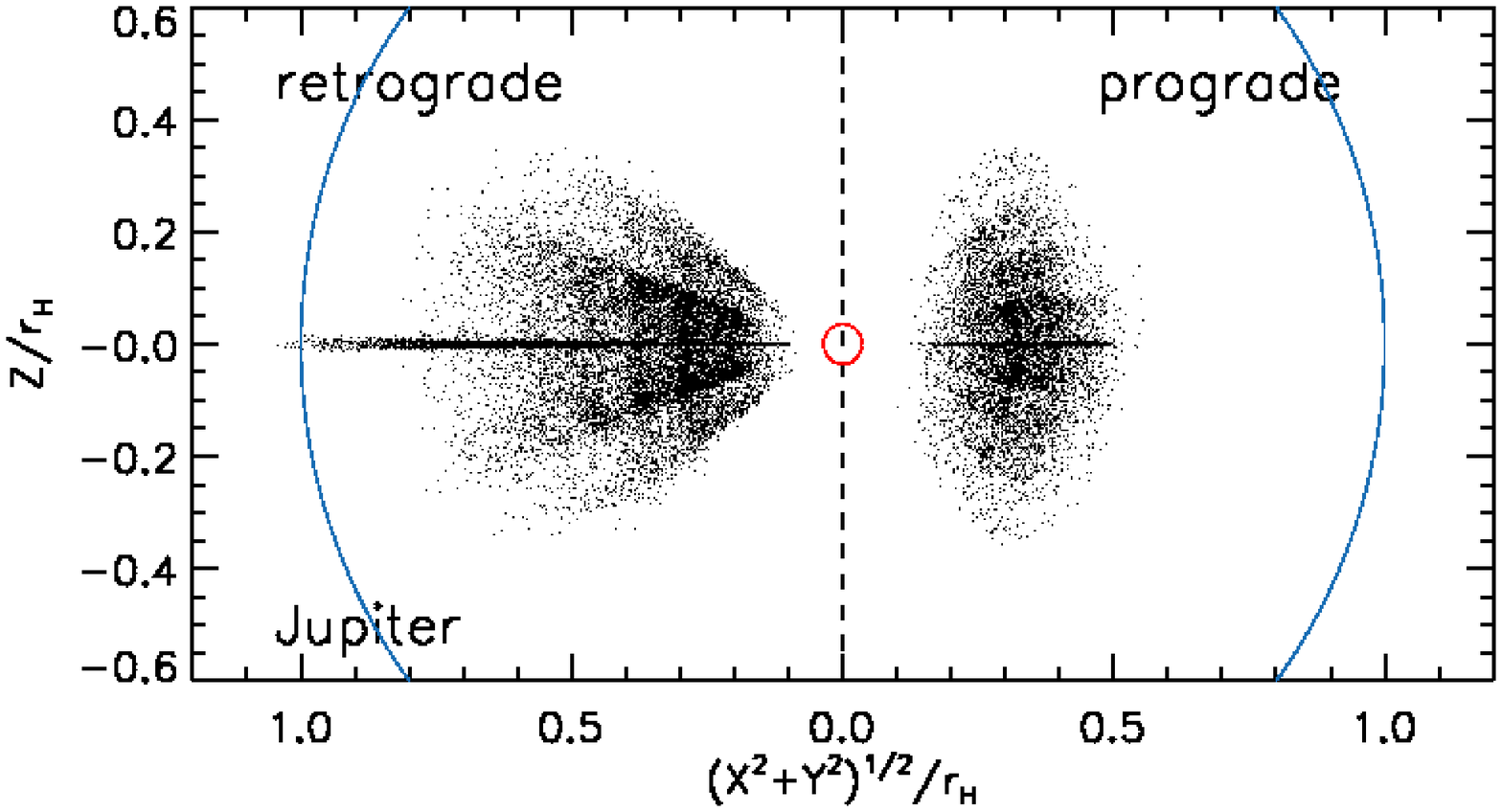}
\includegraphics[width=0.45\textwidth]{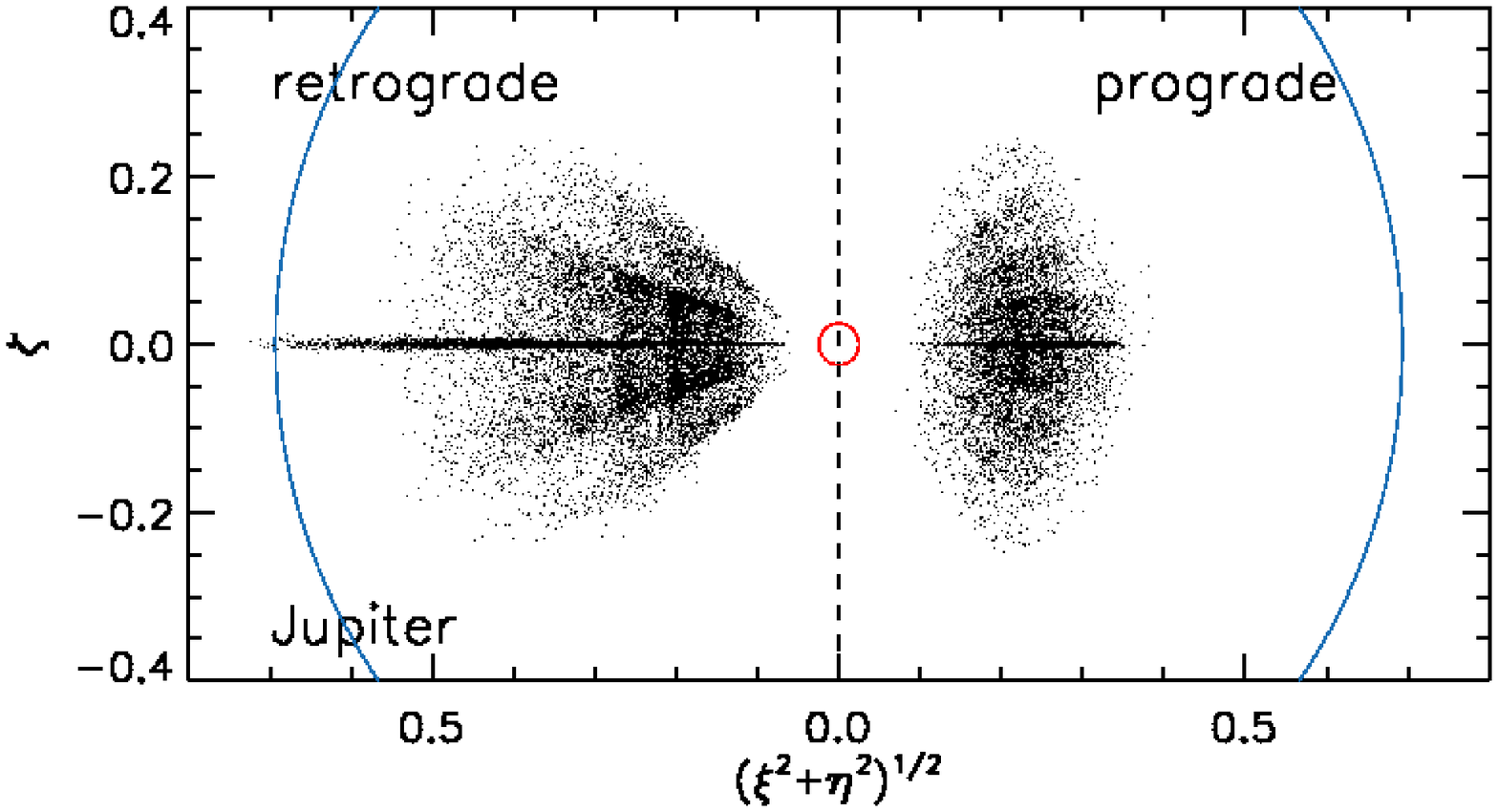}
\caption{Spatially accessible regions of stable satellite orbits
for Jupiter. The left column is in the {\em non-rotating}
planetocentric frame and the right column is in the {\em rotating}
planetocentric frame. The bottom two panels show expanded views of
the inner portions of the upper two panels. In each panel,
retrograde orbits and prograde orbits are plotted separately in
the left and right halves, with ``prograde/retrograde'' defined in
each frame used. The blue circles show the Hill sphere and the
smaller central red circles show the inner boundary in the
numerical integrations (the orbital radius of the outermost
regular satellite, in this case Callisto). The extreme thinness of
the zone of stable retrograde orbits outside the Hill sphere is an
artifact of our simulations, which sampled the initial
inclinations only at $0^\circ,15^\circ,\ldots$.
}\label{fig:stability_region_Jupiter}
\end{figure*}

\begin{figure*}
\centering
\includegraphics[width=0.45\textwidth]{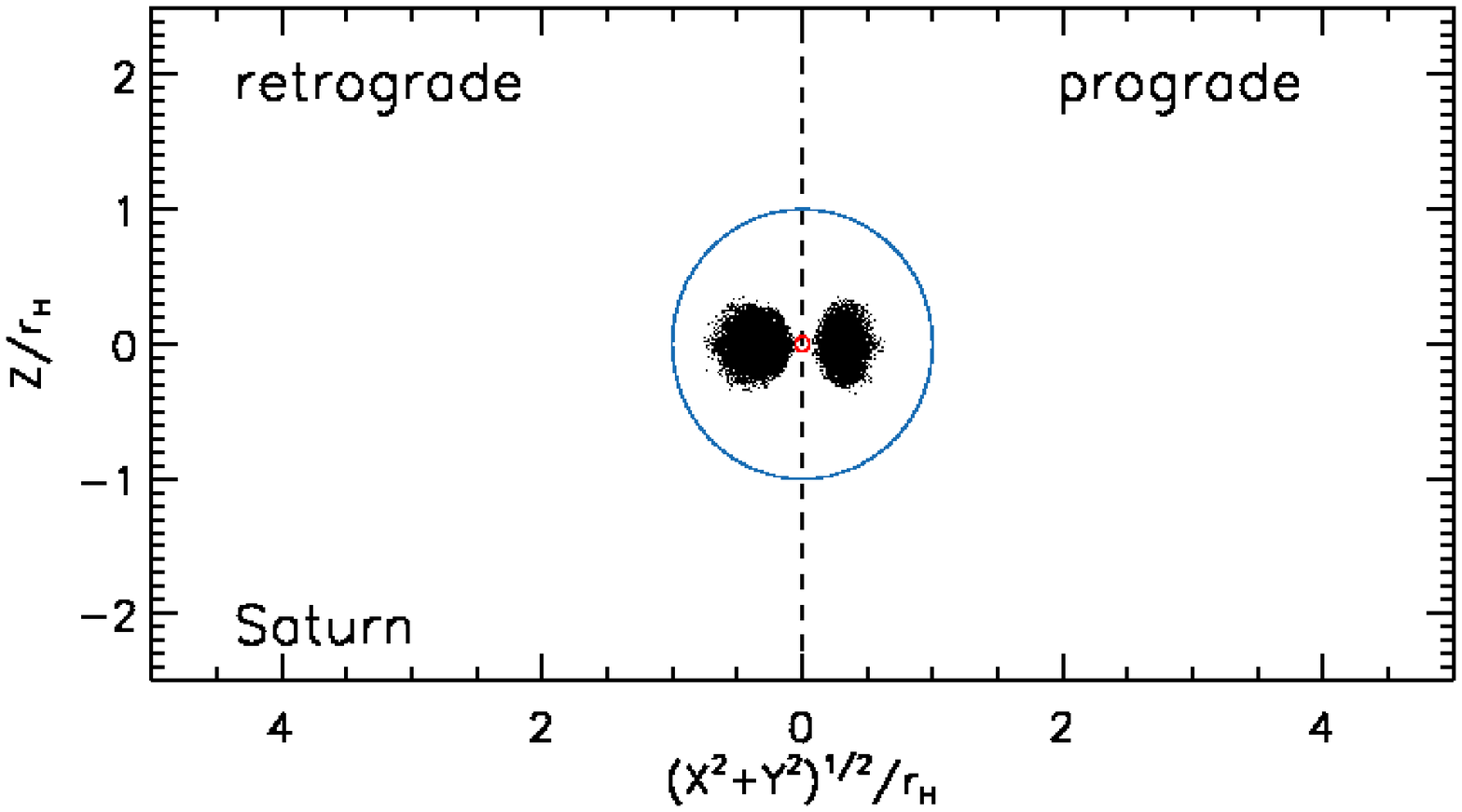}
\includegraphics[width=0.45\textwidth]{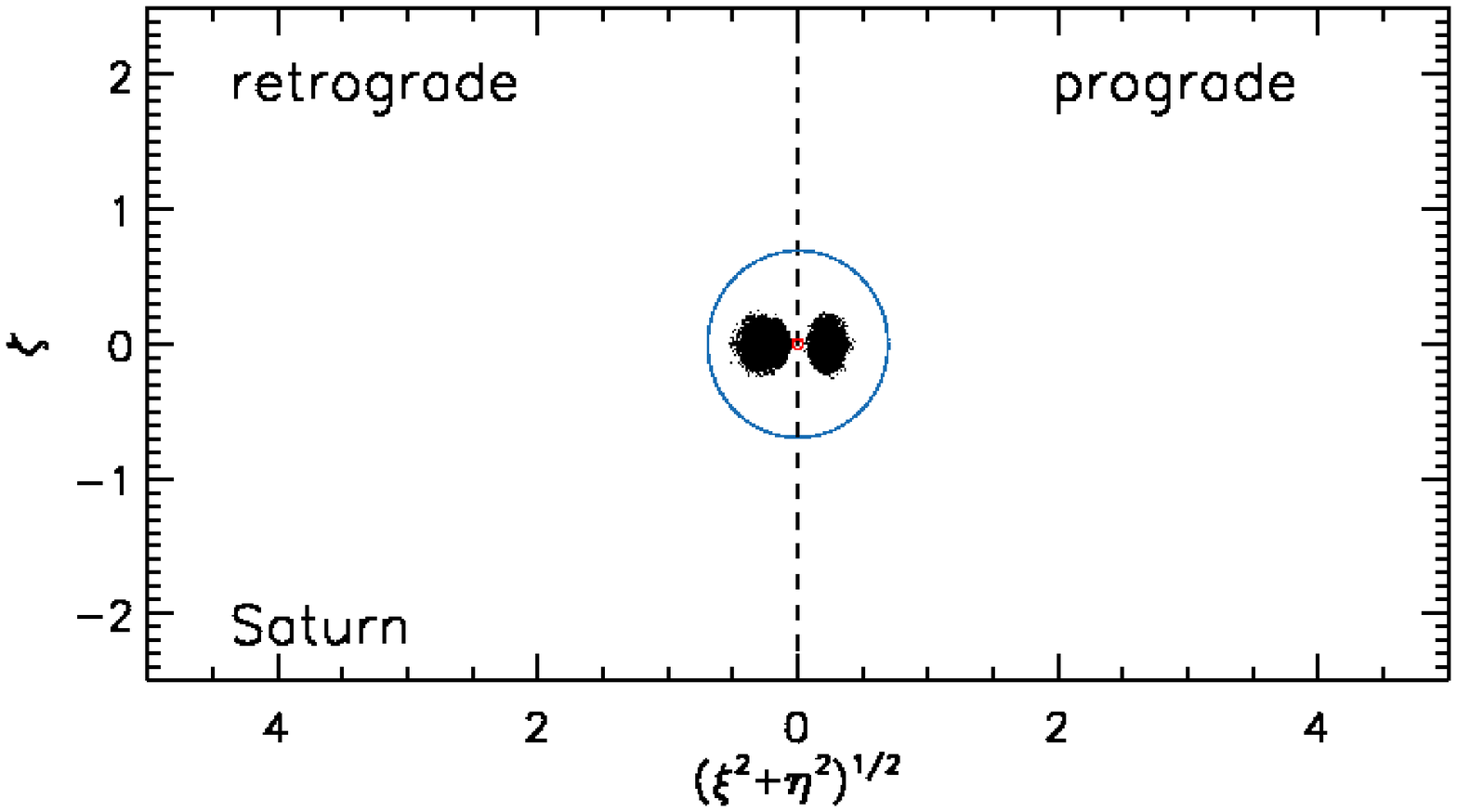}
\includegraphics[width=0.45\textwidth]{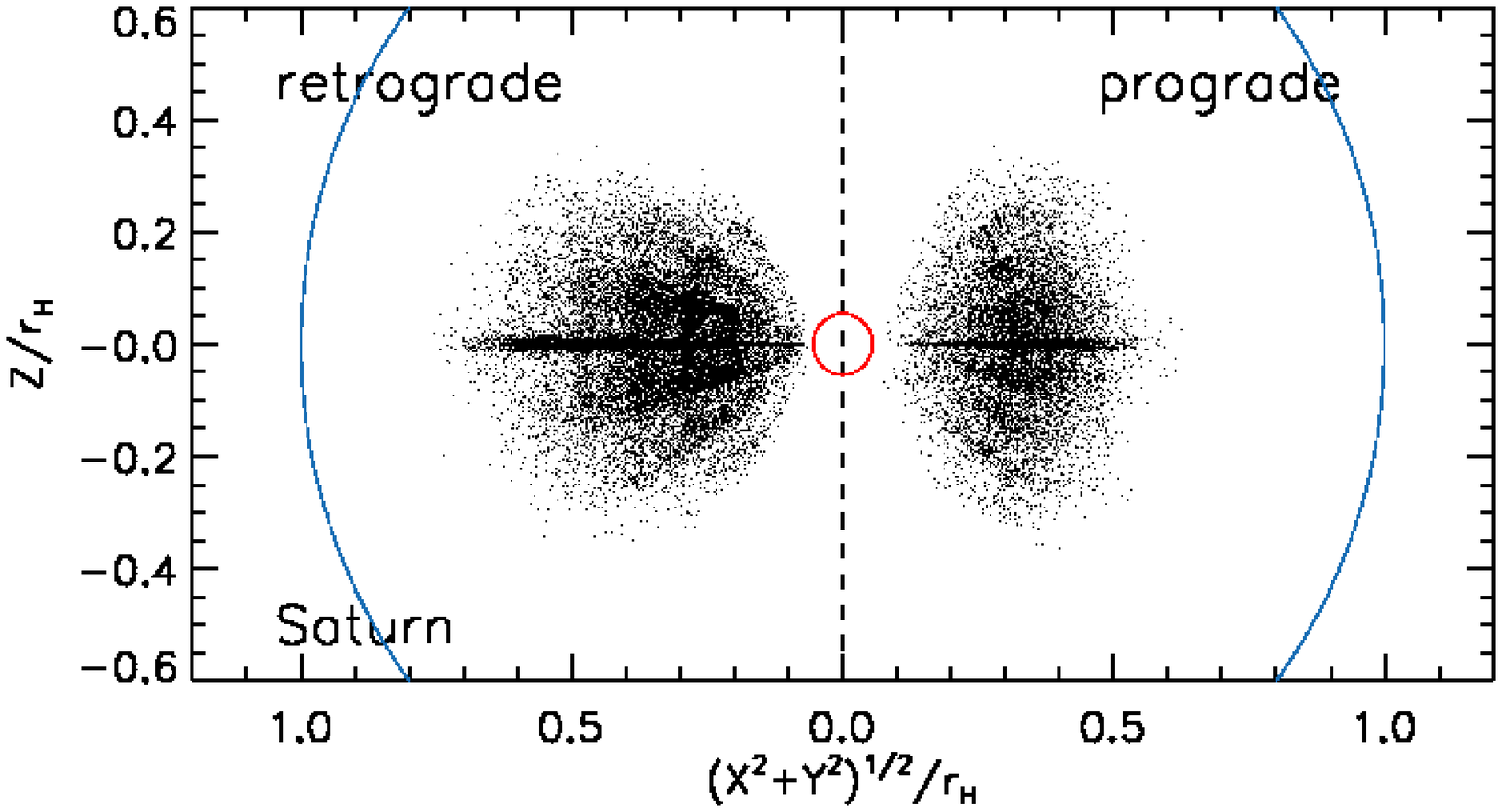}
\includegraphics[width=0.45\textwidth]{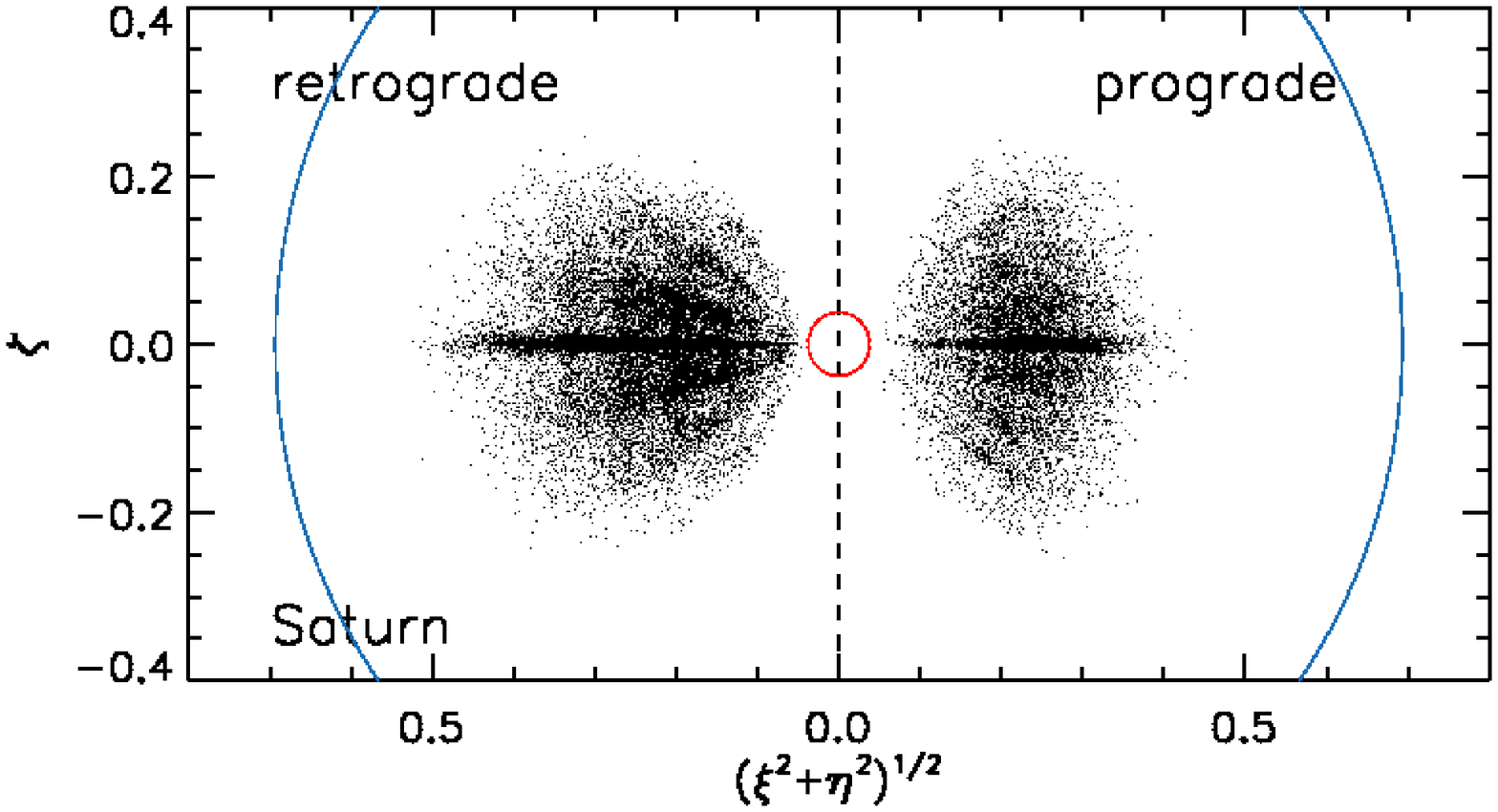}
\caption{Spatially accessible regions of stable satellite orbits for
Saturn. The notation is the same as in Fig.
\ref{fig:stability_region_Jupiter}.}\label{fig:stability_region_Saturn}
\end{figure*}

\begin{figure*}
\centering
\includegraphics[width=0.45\textwidth]{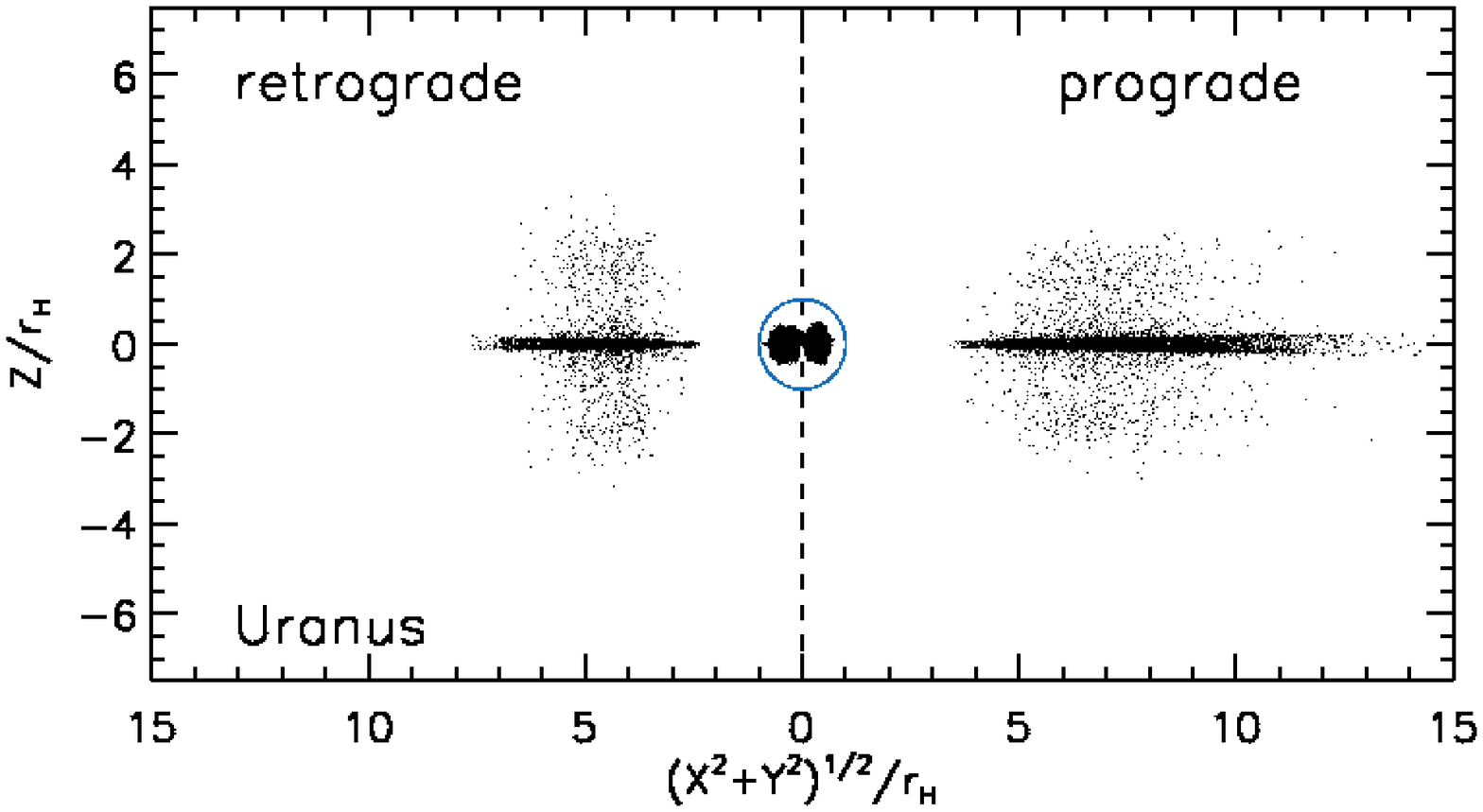}
\includegraphics[width=0.45\textwidth]{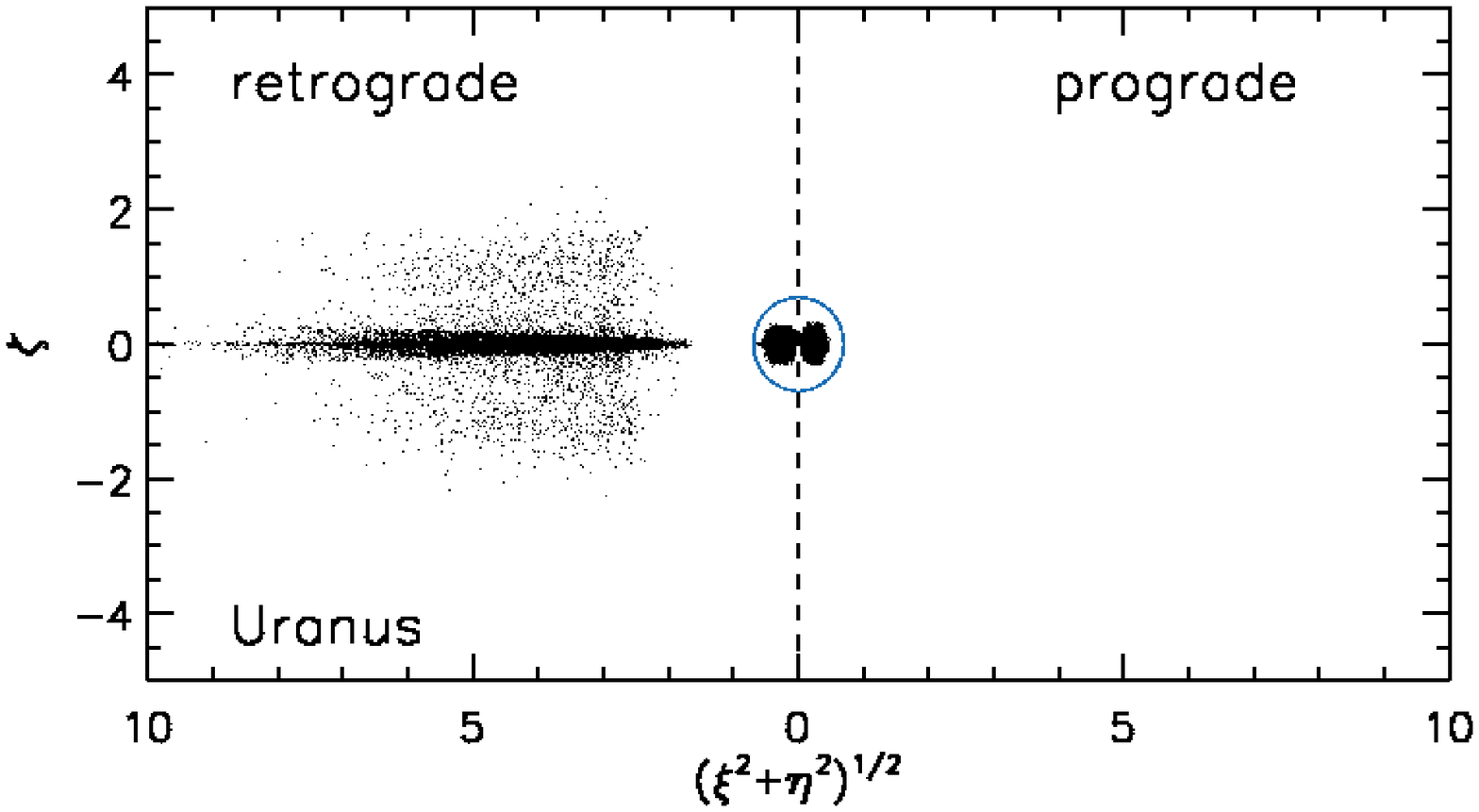}
\includegraphics[width=0.45\textwidth]{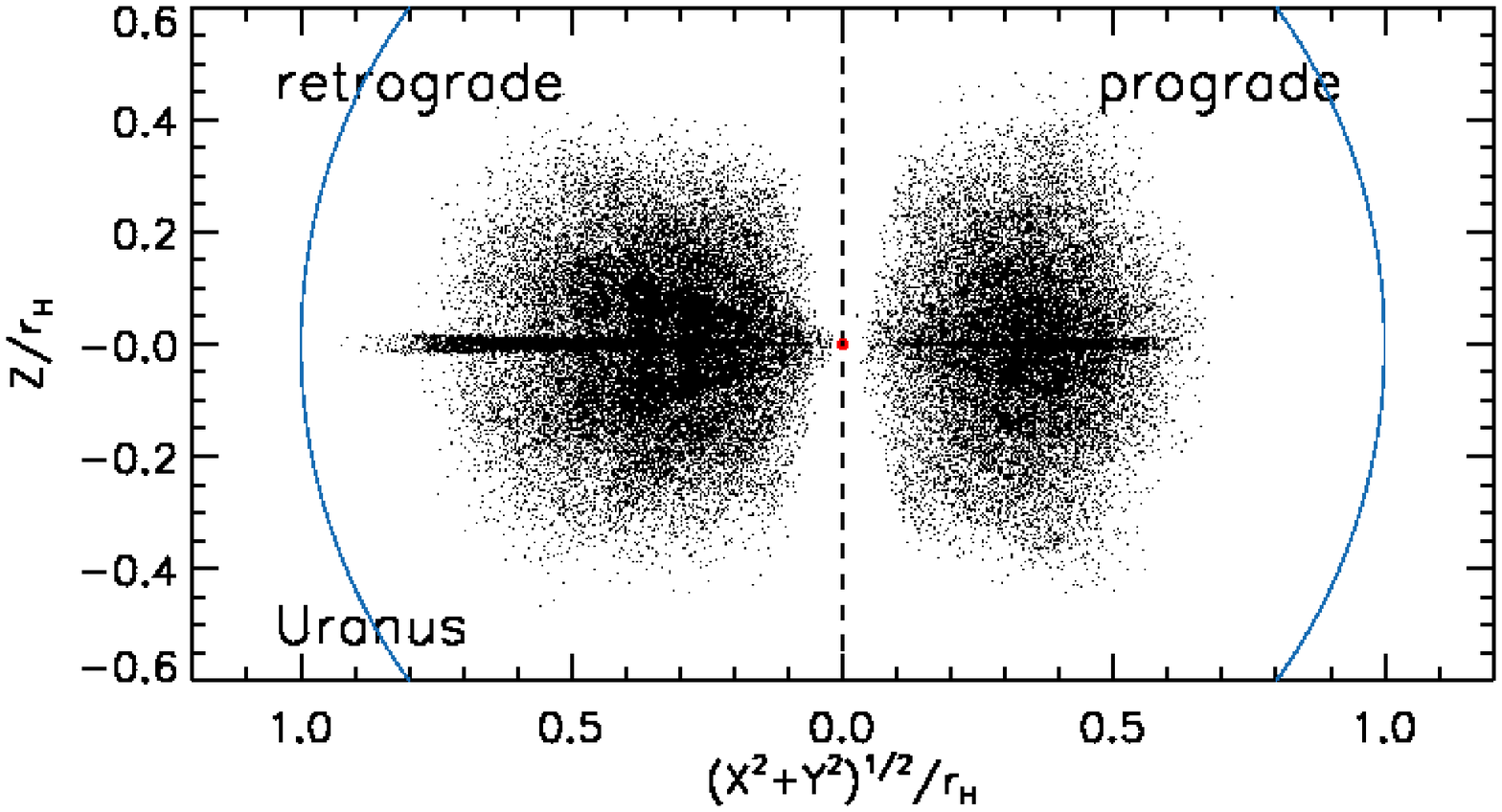}
\includegraphics[width=0.45\textwidth]{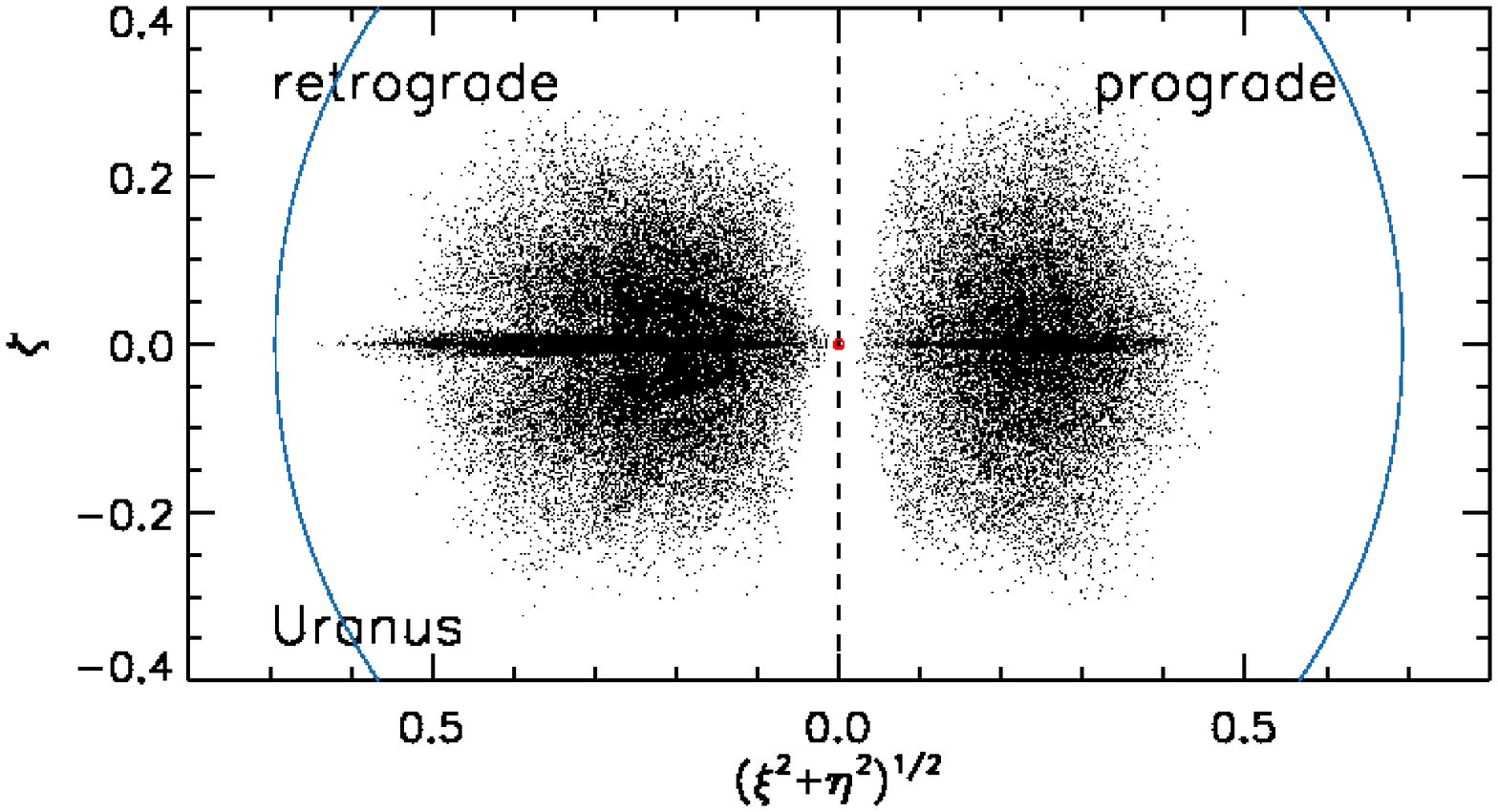}
\caption{Spatially accessible regions of stable satellite orbits
for Uranus.  The notation is the same as in Fig.
\ref{fig:stability_region_Jupiter}.}\label{fig:stability_region_Uranus}
\end{figure*}

\begin{figure*}
\centering
\includegraphics[width=0.45\textwidth]{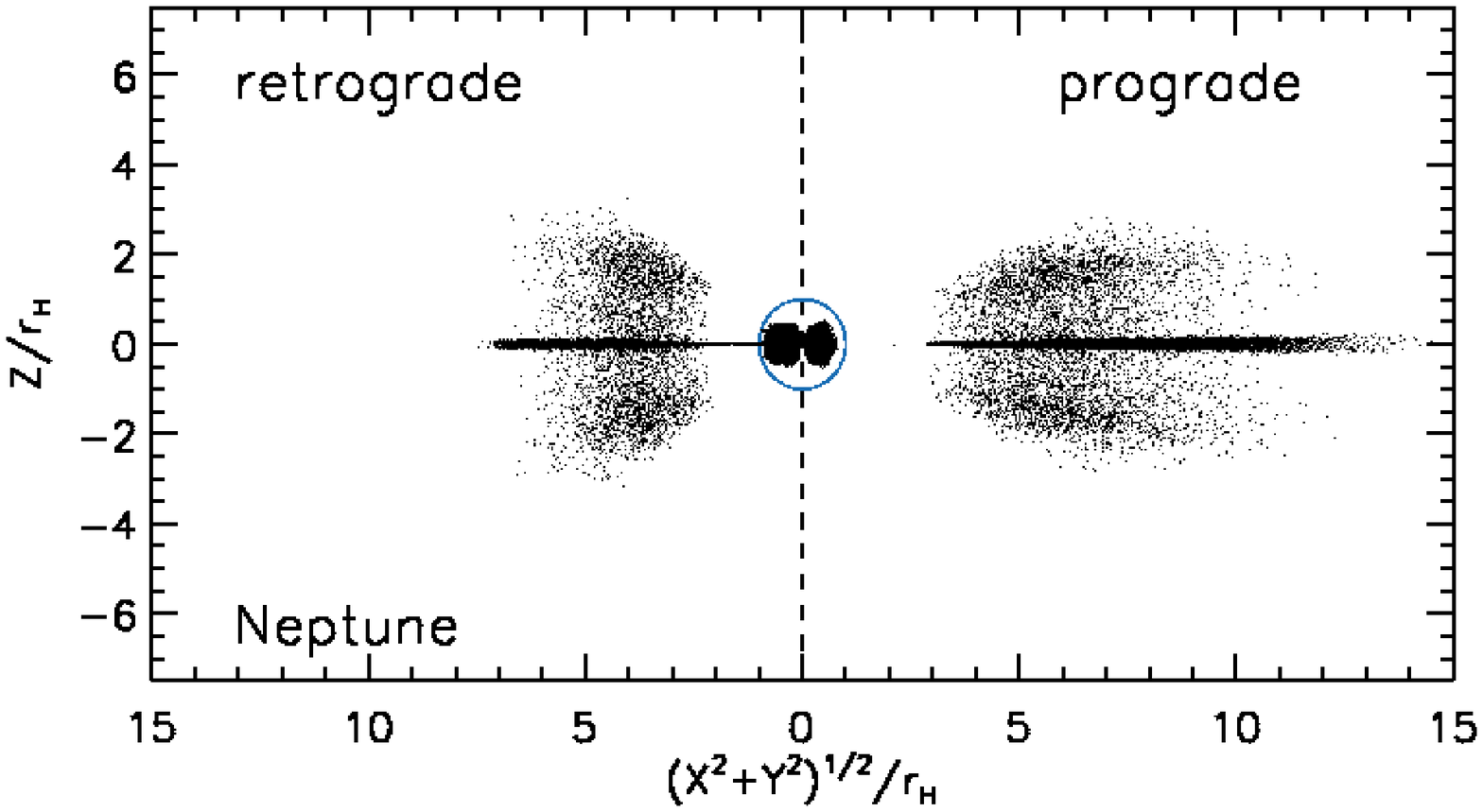}
\includegraphics[width=0.45\textwidth]{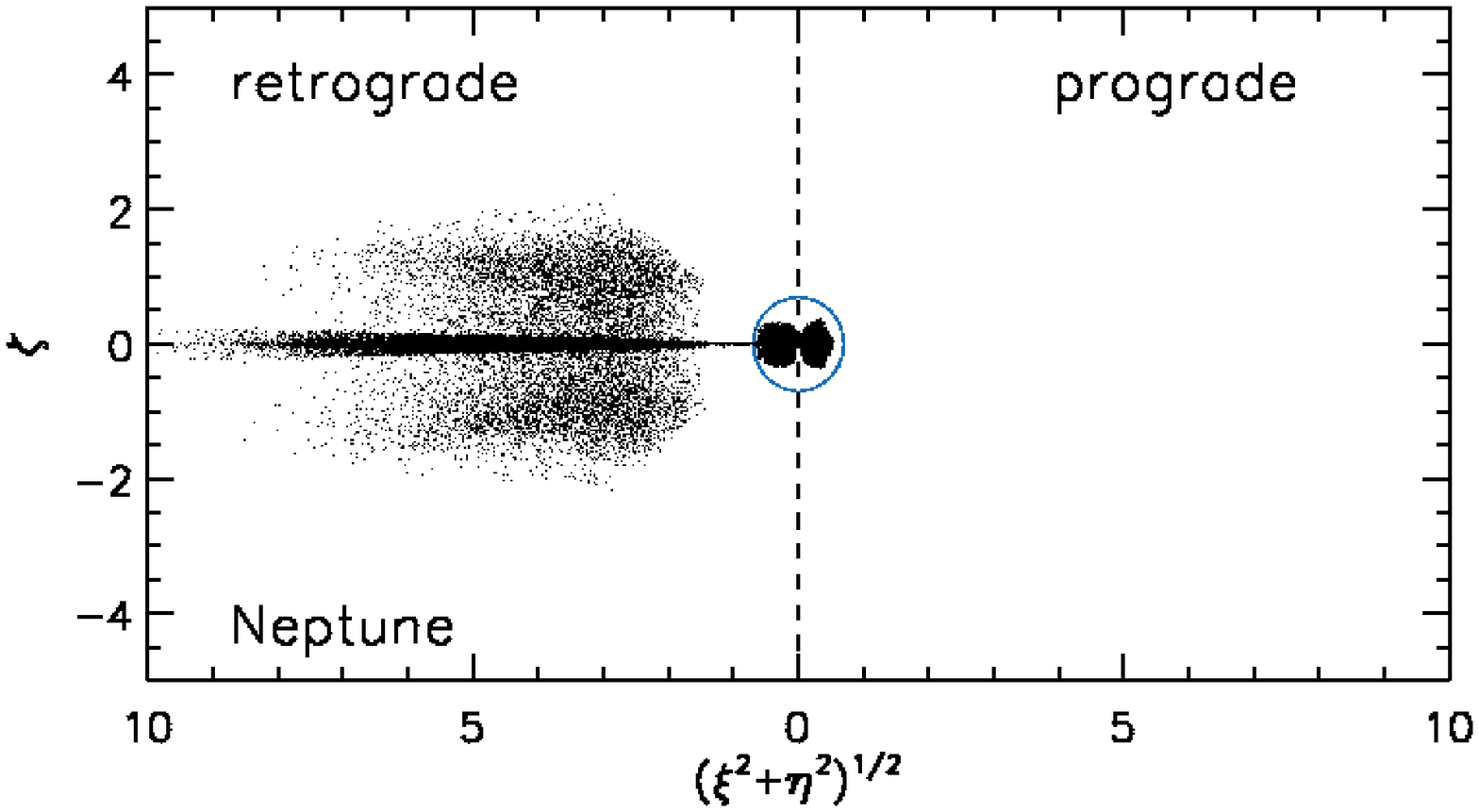}
\includegraphics[width=0.45\textwidth]{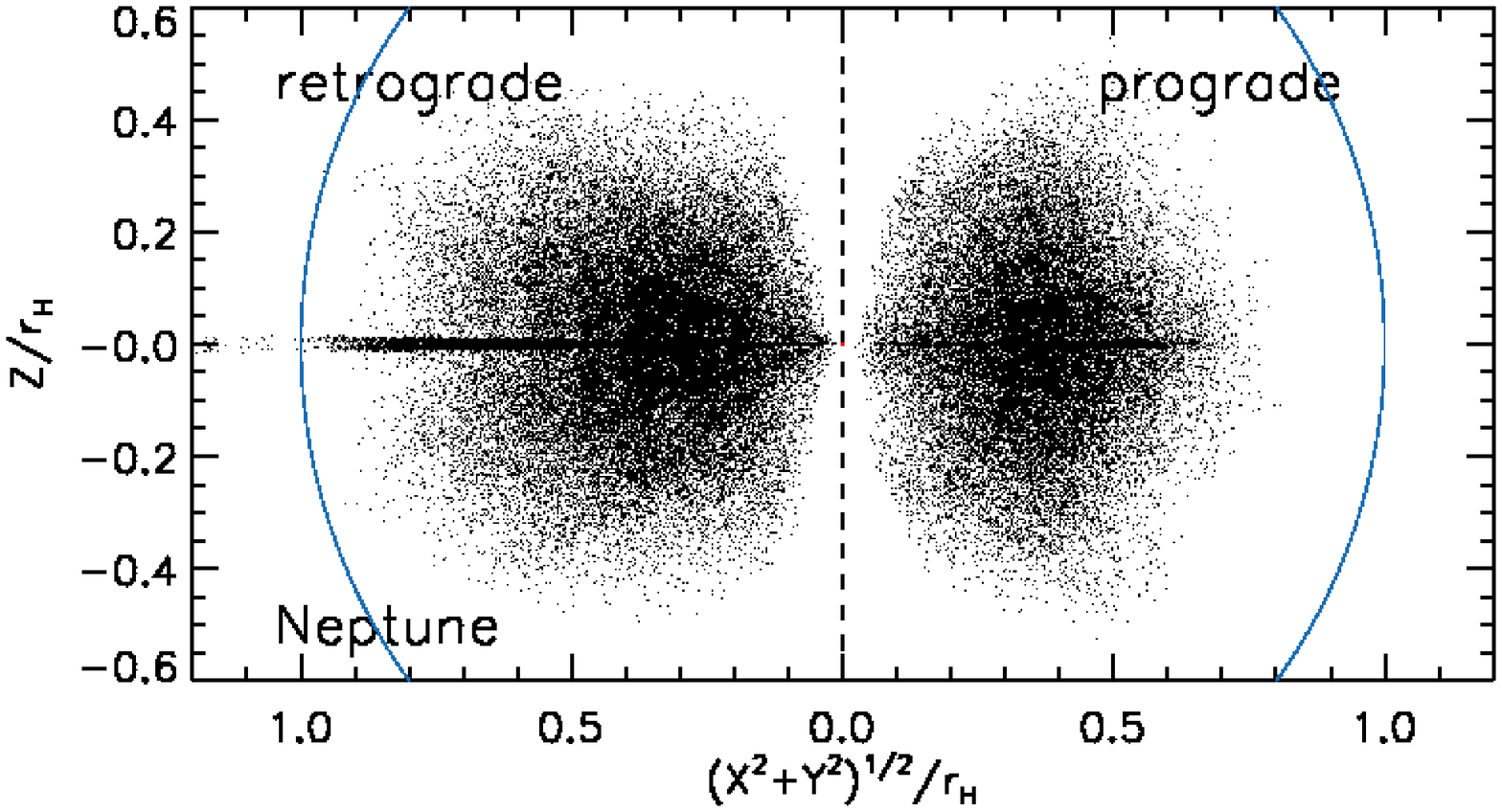}
\includegraphics[width=0.45\textwidth]{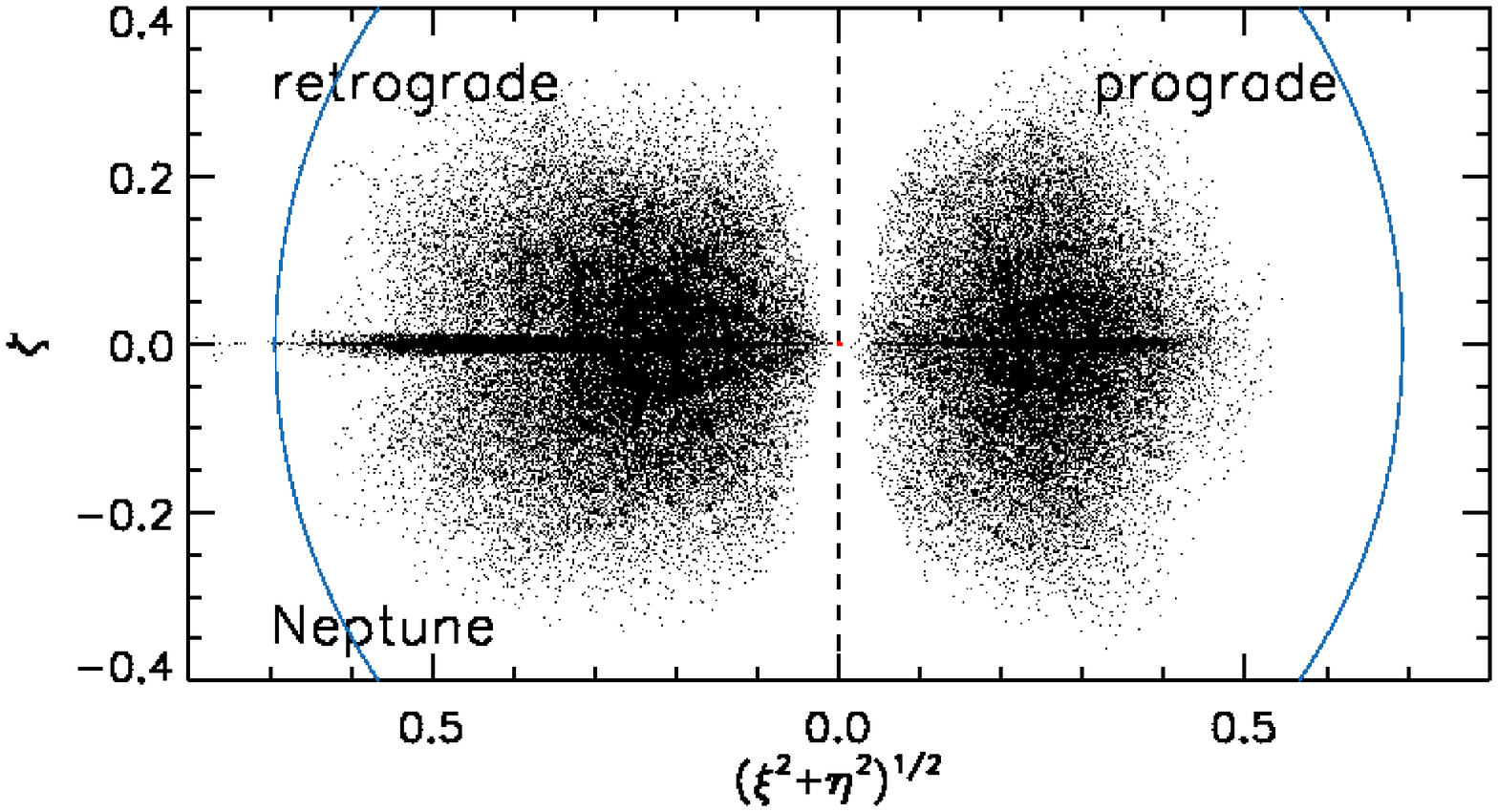}
\caption{Spatially accessible regions of stable satellite orbits for
Neptune.  The notation is the same as in Fig.
\ref{fig:stability_region_Jupiter}.}\label{fig:stability_region_Neptune}
\end{figure*}

What do these stable orbits look like? In Hill's approximation,
the stable (outer and inner) retrograde and inner prograde orbits
are generated from the periodic $f$ and $g$ families respectively
using the terminology of H\'{e}non (1969).  We show examples of
stable orbits (i.e., those that survived for $10^8$ yr) around
Uranus in Fig.\ \ref{fig:uranus_survived_TP}. For each example
orbit, we plot the instantaneous locations for the first one
million years as dots with the two stars marking the starting and
ending locations. We also plot the trajectory for several
revolutions. Each orbit is plotted in both the non-rotating
planetocentric $x$-$y$ plane (left column) and the rotating
$\xi$-$\eta$ plane (right column). In the rotating frame, the
inner prograde orbit (top panel) is elongated along the Sun-planet
axis while the inner retrograde orbit (middle panel) is elongated
perpendicular to the Sun-planet axis. The outer retrograde orbit
(bottom panel) is also elongated perpendicular to the Sun-planet
axis and oscillates about the planet as an ellipse with an axis
ratio of approximately 2:1 (compare Fig.\ 11 of H\'{e}non 1970),
as one would expect from epicycle theory.

Note that the stable retrograde orbits with $\xi$ close to $-0.6934$
in Fig.\ \ref{fig:other_henon2d} regularly cross the Hill
radius. Hence such orbits are missed by the surveys of both Wiegert et
al.\ and Nesvorn\'y et al., who terminate their integrations if
$r<r_{\rm H}$ or $r>r_{\rm H}$, respectively.

\begin{figure*}
\centering
\includegraphics[width=0.33\textwidth]{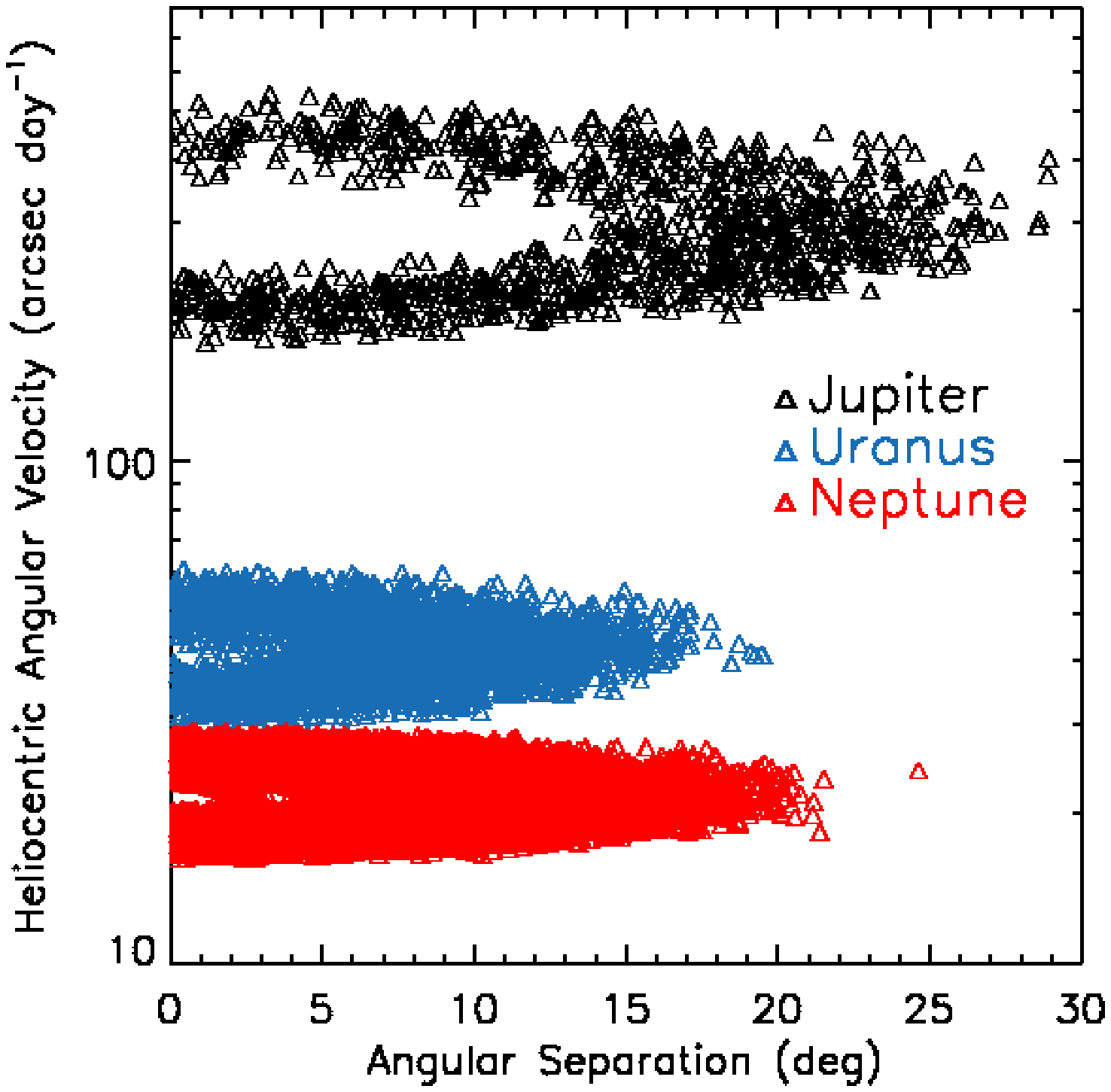}
\includegraphics[width=0.33\textwidth]{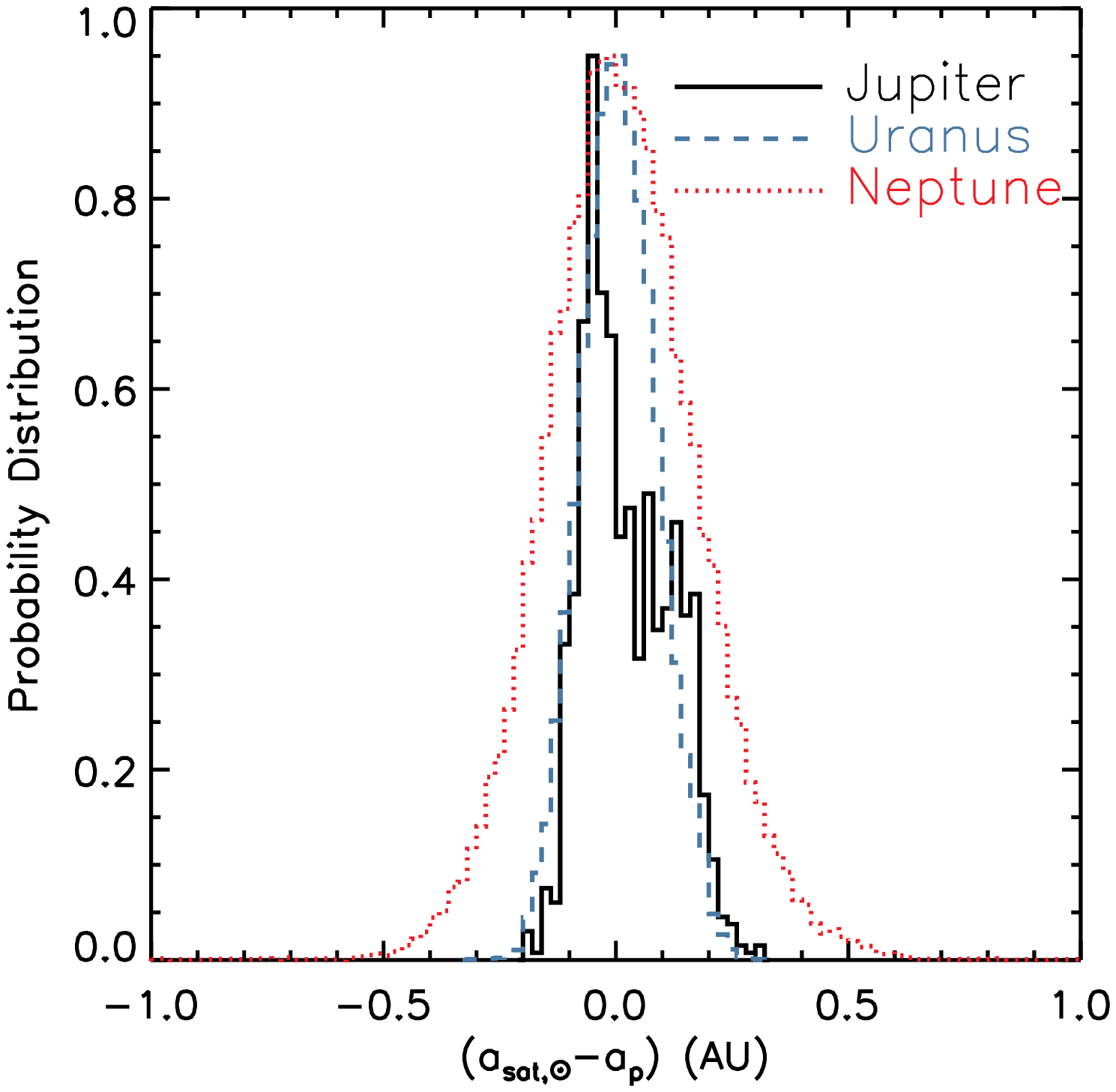}
\includegraphics[width=0.33\textwidth]{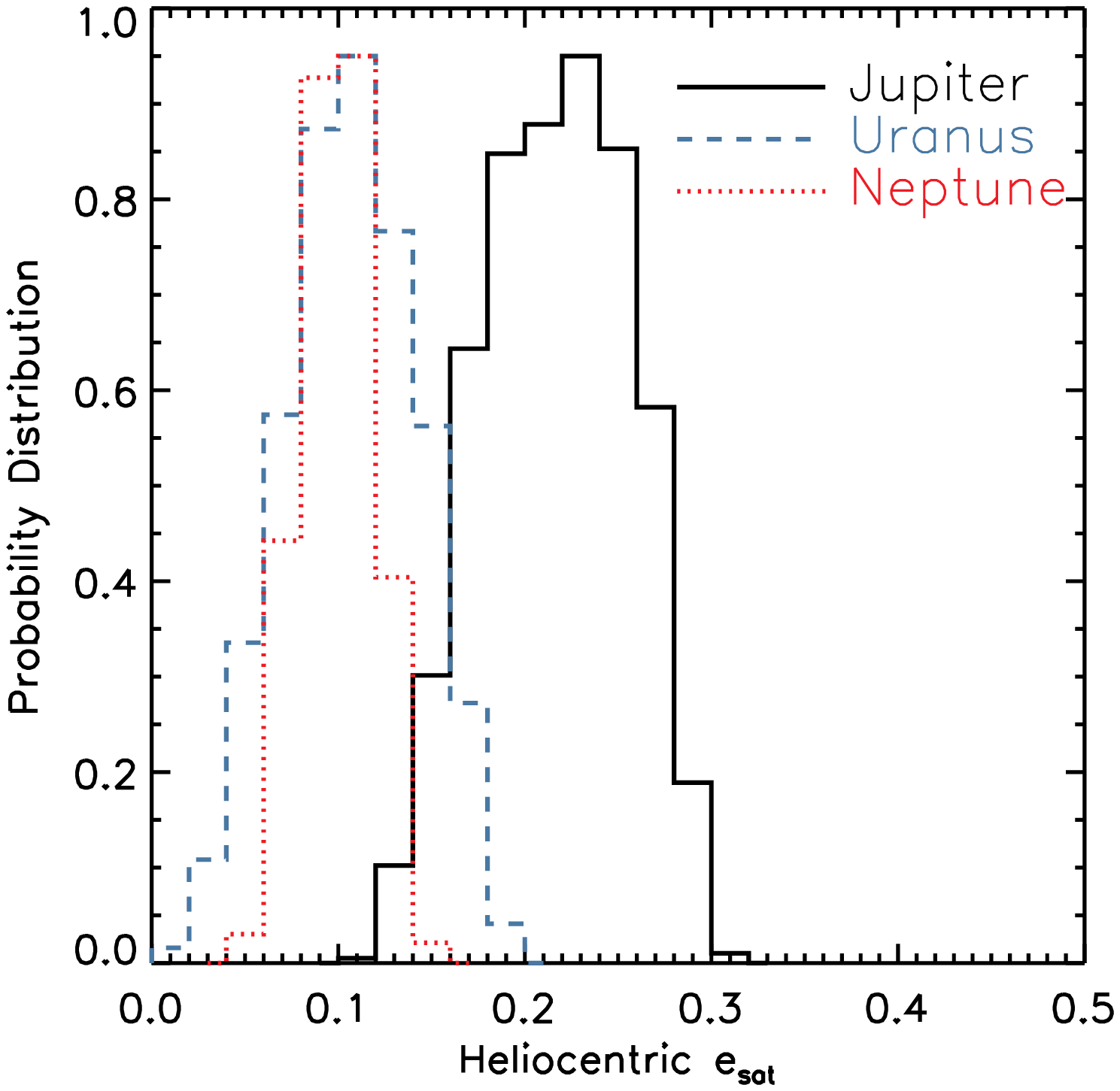}
\caption{{\em Left}: Heliocentric angular velocities of the stable
outer irregular satellites as a function of angular distance from
the planet, as viewed from the Sun in the non-rotating frame. {\em
Middle}: Histograms of the difference between the heliocentric
satellite semi-major axis and the planet semi-major axis for the
stable outer irregular satellites, where the peaks of these
distributions are arbitrarily scaled. {\em Right:} Histograms of
the heliocentric eccentricity for the stable outer irregular
satellites. The sampling of points is the same as we used to
produce the spatial stability regions in
\S\ref{sec:spatial_stab}.}\label{fig:obs_ang_vel}
\end{figure*}

\subsection{Three-dimensional H\'{e}non diagrams}\label{sec:3D_result}

We now extend the initial conditions in \S\ref{sec:2D_result} to
three dimensions by including the scaled vertical coordinate
$\zeta$. As discussed in \S\ref{subsec:Hill_equation}, we consider
a surface of section $\eta=\zeta=\dot{\xi}=0$, $\dot\eta>0$ at
$t=0$. Similar to the two-dimensional case, we sample the initial
conditions using a fine grid in the $(\Gamma,\xi)$ plane, and use
equation (\ref{eqn:3d_vel}) to generate initial velocities. We choose
a sequence of inclinations in the rotating frame,
$I=15^\circ,30^\circ,45^\circ,60^\circ,75^\circ$. Each satellite
orbit is then integrated for $10^8$ yr along with the four giant
planets and the Sun.

The general behavior when incorporating inclination is the erosion
of stable regions in the H\'{e}non diagram. As an example, we show
the results for Uranus in Fig. \ref{fig:uranus_henon_3d}. The
outer retrograde orbits quickly become unstable when the initial
inclination exceeds $\approx 20^\circ$. The inner retrograde
stable region erodes with increasing inclination and disappears at
$I\gtrsim 75^\circ$. The inner prograde stable region can survive
even at $I\approx 75^\circ$. This asymmetry between inner
retrograde and prograde orbits is due in part to the definition of
inclination in the rotating frame\footnote{When translated into
the non-rotating planetocentric frame, the inclinations of
``prograde'' (``retrograde'') orbits are actually smaller (larger)
than in the rotating frame. As we already noted in Fig.\
\ref{fig:check_Henon}, under certain circumstances retrograde
orbits in the rotating frame can even be prograde in the
non-rotating frame.}. However, even when inclination is defined in
the non-rotating frame such an asymmetry may still be present (see
\'{C}uk \& Burns 2004 for a discussion of the dynamical reasons
for the asymmetry).

To separate the destabilizing effects of inclination from the
effects of perturbations from other planets, we also ran these
three-dimensional simulations for Uranus without the other
planets. We found that for $I\lesssim 20^\circ$, perturbations
from other planets do play a major role in eroding the region of
stable outer retrograde orbits, as illustrated by comparing the
upper left and lower right panels of Fig.\
\ref{fig:uranus_henon_3d}, which show the H\'enon diagram for
$I=15^\circ$ with and without the other planets. However, for
$I=30^\circ,45^\circ,60^\circ,75^\circ$ the results for Uranus
alone are almost identical to the realistic case which includes
perturbations from the three other planets. This result suggests
that bound retrograde orbits outside the Hill sphere may not exist
at all for $I\gtrsim 20^\circ$. This is expected because the
Coriolis force, which stabilizes outer retrograde orbits by
bending their trajectory towards the planet in the rotating frame,
is reduced when the inclination angle $I$ increases. This
shrinkage of the stable outer retrograde branch with inclination
is already noticed by comparing the right panel of Fig.\
\ref{fig:uranus_only_henon2d} ($I=0$) and the bottom-right panel
of Fig.\ \ref{fig:uranus_henon_3d} ($I=15^\circ$).

Note that the inclinations $I$ are planetocentric and measured in
the rotating frame. For low-inclination orbits along the outer
retrograde branch they can be converted to heliocentric
inclinations $i$ using the approximate formula for small $e$
\begin{equation}
i\approx 2eI\ ,
\end{equation}
where $e$ is the heliocentric eccentricity. Thus our stability
region $I\lesssim 20^\circ$ corresponds roughly to $i\lesssim
4^\circ$ for $e\approx 0.1$, the typical heliocentric eccentricity
of surviving satellites for Uranus and Neptune (see Fig.\
\ref{fig:obs_ang_vel}); this is in reasonable agreement with
Wiegert et al.'s estimate that most of their long-lived orbits had
$i\lesssim 2^\circ$, especially considering that our sampling of
initial conditions is more complete than theirs. Mikkola \&
Innanen (1997) and Mikkola et al.\ (2006) estimate analytically
that the outer retrograde orbits are unstable if $i>e$ (for a
circular planet orbit), which implies instability if $I\gtrsim
30^\circ$. Our own results show that all the test particles with
initial positions outside the Hill sphere cross the escape radius
$a_p$ well before $10^4$ yr for $I\ge 30^\circ$.

As described at the end of \S\ref{subsec:Hill_equation}, a
limitation of these results is that the H\'enon diagram we have
used will not display orbits trapped in a Kozai resonance,
or other stable orbits whose argument of pericenter does not
periodically pass through 0 or $\pi$. To estimate the contribution
of such orbits, we have constructed a different set of
three-dimensional H\'enon diagrams in which the initial conditions
are changed from our usual choice $\eta=\dot\xi=\zeta=0,
\dot\eta>0$ to $\eta=\dot\xi=\dot\zeta=0, \dot\eta>0$ (i.e., when
the orbit is at its maximum height above the planet's orbital
plane, rather than crossing the orbital plane). In this case, we
define the initial inclination $I^*$ in the rotating frame by
\begin{equation}
\tan I^*=\frac{\zeta}{\xi}\bigg|_{t=0}.
\end{equation}
The results are shown in Figure \ref{fig:uranus_henon_3d_alt},
which should be compared to Figure \ref{fig:uranus_henon_3d}. Each
point in either set of H\'enon diagrams corresponds to a unique
orbit, but orbits appearing in the H\'enon diagrams of one figure
at a given value of $(\Gamma,\xi,I)$ may or may not appear in the
other figure, where they will have the same value of $\Gamma$ but
possibly different values of $\xi$ and inclination. The stable
regions are somewhat larger in Figure
\ref{fig:uranus_henon_3d_alt} at a given inclination---for
example, a few outer retrograde satellites survive for $10^8$ yr
at $I^*=30^\circ$---but the conclusions described above are not
significantly altered. In the following discussion, we neglect
stable orbits that do not appear in our fiducial
H\'{e}non diagrams (i.e., using the surface of section
$\eta=\dot\xi=\zeta=0, \dot\eta>0$); thus we may slightly underestimate the size of the stable regions. More discussion on orbits
trapped in the Kozai resonance inside the Hill sphere can be found in Carruba et al.
(2002).

\subsection{Spatial stability regions}\label{sec:spatial_stab}
We now project the phase-space volume that hosts stable orbits onto
coordinate space, to explore where stable satellites might be found.

We plot the positions of stable orbits in the two-dimensional plane
with coordinates $\left[(x^2+y^2)^{1/2},z\right]$ (non-rotating frame)
or $\left[(\xi^2+\eta^2)^{1/2},\zeta\right]$ (rotating frame).
Prograde and retrograde orbits are plotted separately on the left and
right sides of a given figure panel, where ``prograde'' and
``retrograde'' are defined in the frame used. We plot the position of
each stable (up to $10^8$ yr) point in the H\'{e}non diagram at
uniformly spaced times (every Myr) between $5\times 10^7$ and $10^8$
yr in Figs.\ \ref{fig:stability_region_Jupiter} (Jupiter) to
\ref{fig:stability_region_Neptune} (Neptune).

The stability regions within the Hill sphere are very similar to
those shown in Figs. 9-12 of Nesvorn\'{y} et al.\ (2003), though
slightly larger because we show instantaneous position rather than
semi-major axis. For Jupiter and Saturn, the stable prograde
orbits generally extend to $\sim 0.5r_{\rm H}$; the stable
retrograde orbits can extend further to $\sim 0.7r_{\rm H}$, and
nearly coplanar retrograde orbits even extend to $\sim r_{\rm H}$
for Jupiter. For Uranus and Neptune, both the prograde and
retrograde stable orbits can extend a little bit further relative
to the Hill sphere. No stable orbits exist at high latitudes,
presumably because of Kozai oscillations (Kozai 1962; Carruba et
al.\ 2002; Nesvorn\'{y} et al.\ 2003).

It is also notable that there are stable regions beyond the Hill
sphere for Jupiter, Uranus and Neptune, as we discussed in
previous sections. This is particularly the case for Uranus and
Neptune. Most of these distant stable satellites are concentrated
close to the orbital plane of the planet as for Jupiter, although
the appearance of a very thin layer in Figs.\
\ref{fig:stability_region_Jupiter} (Jupiter)-
\ref{fig:stability_region_Neptune} (Neptune) is somewhat an artifact of the
coarse sampling of inclinations in our initial conditions
($I=15^\circ,30^\circ,45^\circ,60^\circ,75^\circ$). Stable
satellites can be found as far as $\sim 5r_{\rm H}$ from Jupiter
and even $\sim 10r_{\rm H}$ for Uranus and Neptune, and as high as
$2.5r_{\rm H}$ above the orbital plane for the latter two planets.
In the following section we discuss briefly the strategy of
searches for these distant satellites.

\section{Discussion and conclusions}\label{sec:con}

We have conducted a systematic survey of the stable regions of
satellites around giant planets in the solar system, using
numerical orbital integrations that include gravitational
perturbations from the other planets. We confirm previous results
for satellites within the Hill sphere: stable retrograde
satellites can exist further out than prograde satellites (e.g.,
H\'{e}non 1970; Hamilton \& Krivov 1997; Nesvorn\'{y} et al.\
2003); and stable orbits cannot exist at high inclinations (e.g.,
Carruba et al.\ 2002; Nesvorn\'{y} et al.\ 2003).

We also confirm and extend the conclusions of Wiegert et al.\
(2000) that distant retrograde satellites (``retrograde'' as
defined in the rotating frame) can survive well beyond the Hill
sphere for at least $10^8$--$10^9$ yr, and probably for the lifetime of
the solar system. Uranus and Neptune are the most promising host
planets for such distant satellites, since their stability regions
have the largest extent (e.g., Fig.\
\ref{fig:stability_region_Uranus}-\ref{fig:stability_region_Neptune}).
Jupiter has a smaller stable region (Fig.\
\ref{fig:stability_region_Jupiter}), and Saturn appears to have no
stable regions beyond the Hill radius (Fig.\
\ref{fig:stability_region_Saturn}).

Remarkably, there is a gap between the inner and outer stability
zones for retrograde satellites, extending from about $r_{\rm H}$
to $2r_{\rm H}$, in which almost no stable orbits exist.

To check whether any of the proposed distant satellites have
already been discovered as Centaurs, we take the positions and
velocities of the known Centaurs from the IAU Minor Planet
Center\footnote{http://cfa-www.harvard.edu/iau/mpc.html} that have
planetocentric distances smaller than the semi-major axes of each
of the four giant planets at the last observed epoch. There are 31
Centaurs (Jupiter 1; Uranus 16; Neptune 14) that satisfy the
criterion. We numerically integrated these objects along with the
Sun and giant planets for $10^8$ yr, but none of them survived as
satellites according to the definition in \S\ref{sec:intro}.

Searches for satellites far beyond the Hill radius can be carried out
either with dedicated deep, wide-angle surveys around the giant
planets, or through all-sky surveys such as Pan-STARRS and LSST. The
most promising search areas are close to the orbital plane of the
planet, since only low-inclination orbits survive (e.g., compare Fig.\
\ref{fig:other_henon2d} and Fig.\ \ref{fig:uranus_henon_3d}). In Fig.\
\ref{fig:obs_ang_vel} (left panel) we show the heliocentric angular
velocity of the stable outer retrograde satellites as a function of
angular distance from the planet as viewed from the Sun, for Jupiter
(black), Uranus (blue) and Neptune (red) respectively, sampled every
Myr between $5\times 10^7$ and $10^8$ yr.


In the middle and right panels of Fig.\ \ref{fig:obs_ang_vel} we
show histograms of the difference in heliocentric semi-major axis
from their host planet and heliocentric eccentricities for the stable
outer retrograde satellites. These distributions can be used to
cull a large sample for potential satellites. Once a candidate is
identified with reliable orbit elements, a long-term orbital
integration should be run to confirm its satellite nature.

The discovery and characterization of satellites beyond the Hill
sphere would provide rich information about the early formation of
the solar system. Fabrycky (2008; also see Kortenkamp 2005)
recently performed simulations of capture of neighboring
planetesimals from the circumstellar disk during slow planet
growth, and found that such distant satellites are a natural
outcome for Uranus and Neptune. Thus an inventory of this
potential population of bodies would enhance our understanding of
the formation of planets and their satellites in the early solar
system, and the properties of the primordial planetesimal disk.

\acknowledgements

This research was supported in part by NASA grant NNX08AH83G. We thank
Dan Fabrycky and the anonymous referee for comments that greatly
improved the paper.

\end{document}